\documentclass[flalign,usenatbib]{mnras}
\usepackage{longtable}

\usepackage{amsmath,amssymb,graphicx,mathtools}
\usepackage{array}

\usepackage{newtxtext,newtxmath}
\usepackage[T1]{fontenc}
\usepackage{ae,aecompl}
\usepackage{anyfontsize}

\DeclarePairedDelimiter{\evdel}{\langle}{\rangle}
\newcommand{\ev}{\operatorname{}\evdel}

\begin{document}


\title[ACTPol: Polarized Point Sources at 148 GHz]{The Atacama Cosmology Telescope: Two-season ACTPol Extragalactic Point Sources and their Polarization properties}

\author[R. Datta et al.]{
Rahul~Datta,$^{1}$\thanks{E-mail: rahul.datta@nasa.gov}
Simone~Aiola,$^{2}$
Steve~K.~Choi,$^{2}$
Mark~Devlin,$^{3}$
Joanna~Dunkley,$^{2}$
\newauthor
Rolando~D\"unner,$^{4}$
Patricio~A.~Gallardo,$^{5}$
Megan~Gralla,$^{6}$
Mark~Halpern,$^{7}$
\newauthor
Matthew~Hasselfield,$^{8,9}$
Matt~Hilton,$^{10}$
Adam~D.~Hincks,$^{11}$
Shuay-Pwu~P.~Ho,$^{2}$
\newauthor
Johannes~Hubmayr,$^{12}$
Kevin~M.~Huffenberger,$^{13}$
John~P.~Hughes,$^{14}$
Arthur~Kosowsky,$^{15}$
\newauthor
Carlos~H.~L\'opez-Caraballo,$^{4}$
Thibaut~Louis,$^{16}$
Marius~Lungu,$^{2}$
Tobias~Marriage,$^{17}$
\newauthor
Lo\"ic~Maurin,$^{4}$
Jeff~McMahon,$^{18}$
Kavilan~Moodley,$^{10}$
Sigurd~K.~Naess,$^{19}$
\newauthor
Federico~Nati,$^{3}$
Michael~D.~Niemack,$^{5}$
Lyman~A.~Page,$^{2}$
Bruce~Partridge,$^{20}$
\newauthor
Heather~Prince,$^{2}$
Suzanne~T.~Staggs,$^{2}$
Eric~R.~Switzer,$^{1}$
and Edward~J.~Wollack$^{1}$
\\
\\
$^{1}$NASA/Goddard Space Flight Center, Greenbelt, MD 20771\\
$^{2}$Joseph Henry Laboratories of Physics, Jadwin Hall, Princeton University, Princeton, NJ 08544	\\
$^{3}$Department of Physics and Astronomy, University of Pennsylvania, 209 South 33rd Street, Philadelphia, PA 19104\\
$^{4}$Instituto de Astrof\'isica and Centro de Astro-Ingenier\'ia, Pontificia Universidad Cat\'olica de Chile, Av. Vicu\~na Mackenna 4860, Santiago, Chile\\
$^{5}$Department of Physics, Cornell University, Ithaca, NY 14953, USA\\
$^{6}$Steward Observatory, University of Arizona, 933 N Cherry Avenue, Tucson, AZ 85721, USA\\
$^{7}$Department of Physics and Astronomy, University of British Columbia, Vancouver, BC, Canada V6T 1Z4\\
$^{8}$Department of Astronomy and Astrophysics, The Pennsylvania State University, University Park, PA 16802 \\
$^{9}$Institute for Gravitation and the Cosmos, The Pennsylvania State University, University Park, PA 16802 \\
$^{10}$Astrophysics and Cosmology Research Unit, University of KwaZulu-Natal, Westville Campus, Durban 4041, South Africa\\
$^{11}$Department of Physics, University of Rome "La Sapienza", Piazzale Aldo Moro 5, I-00185 Rome, Italy\\
$^{12}$NIST Quantum Devices Group, 325 Broadway, Mailcode 817.03, Boulder, CO 80305, USA\\
$^{13}$Department of Physics, Florida State University, Tallahassee FL 32306\\
$^{14}$Department of Physics and Astronomy, Rutgers, The State University of New Jersey, Piscataway, NJ 08854-8019\\
$^{15}$Department of Physics and Astronomy, University of Pittsburgh, Pittsburgh PA 15260 USA\\
$^{16}$Laboratoire de l'Acc\'el\'erateur Lin\'eaire, Univ. Paris-Sud, CNRS/IN2P3, Universit\'e Paris-Saclay, Orsay, France\\
$^{17}$Department of Physics and Astronomy, The Johns Hopkins University, 3400 N. Charles St., Baltimore, MD 21218-2686\\
$^{18}$Department of Physics, University of Michigan, Ann Arbor, USA 48109\\
$^{19}$Center for Computational Astrophysics, 162 5th Ave, New York, NY 10003\\
$^{20}$Department of Physics and Astronomy, Haverford College, Haverford, PA, USA 19041
}
\label{firstpage}
\pagerange{\pageref{firstpage}--\pageref{lastpage}}
\maketitle
\begin{abstract}
{\fontsize{8.5}{10}\selectfont We report on measurements of the polarization of extragalactic sources at 148 GHz made during the first two seasons of the Atacama Cosmology Telescope Polarization (ACTPol) survey. The survey covered 680 deg$^{2}$ of the sky on the celestial equator. Polarization measurements of 169 intensity-selected sources brighter than 30 mJy, that are predominantly Active Galactic Nuclei, are presented. Above a total flux of 215 mJy where the noise bias removal in the polarization measurement is reliable, we detect 26 sources, 14 of which have a detection of linear polarization at greater than 3$\sigma_{p}$ significance. The distribution of the fractional polarization as a function of total source intensity is analyzed. Our result is consistent with the scenario that the fractional polarization of our measured radio source population is independent of total intensity down to the limits of our measurements and well described by a Gaussian distribution with a mean fractional polarization $p=0.028\pm$0.005 and standard deviation $\sigma_{\mathrm{p}}=0.054$, truncated at $p=0$. Extrapolating this model for the distribution of source polarization below the ACTPol detection threshold, we predict that one could get a clean measure of the E-mode polarization power spectrum of the microwave background out to $\ell=6000$ with 1 $\mu$K-arcminute maps over 10$\%$ of the sky from a future survey. We also study the spectral energy distribution of the total and polarized source flux densities by cross-matching with low radio frequency catalogs. We do not find any correlation between the spectral indices for total flux and polarized flux.  }
\end{abstract}

\section{Introduction}
\begin{table*}
\caption{Measurements of the polarization fraction of extragalactic sources from selected radio and millimeter wavelength surveys. }
\begin{center}
\begin{tabular}{| l | l | l | l | l |}
\hline
Reference & $N_\mathrm{srcs}$ & Freq (GHz) & Polarization fraction & Remarks  \\
\hline 
\cite{1998AJ....115.1693C} & 30,000 & 1.4 & SS$^{*}$: 1.1$\%$ median, 2$\%$ mean & S$_{1.4}>$ 800 mJy \\
& & & SS: 1.8$\%$ median, 2.7$\%$ mean & 200 $> S_{1.4}>$ 100 mJy \\
& & & FS$^{*}$: 1.3$\%$ median, 2$\%$ mean & \\
\hline 
\cite{2002AA...396..463M} & 8032 & 1.4 & SS NVSS sources: 1.82$\%$ median & 200 $> S_{1.4}>$ 100 mJy \\
& 3700 & & SS sources: 1.45$\%$ median & 400 $> S_{1.4}>$ 200 mJy \\
& 1438 & & SS sources: 1.37$\%$ median & 800 $> S_{1.4}>$ 400 mJy \\
& 660 & & SS sources: 0.74$\%$ median & $S_{1.4}>$ 800 mJy \\
& 6198 & & FS/IS sources: 1.84$\%$ median & 200 $> S_{1.4}>$ 100 mJy \\
& 2859 & & FS/IS sources: 1.50$\%$ median & 400 $> S_{1.4}>$ 200 mJy \\
& 1150 & & FS/IS sources: 1.32$\%$ median & 800 $> S_{1.4}>$ 400 mJy \\
& 496 & & FS/IS sources: 1.05$\%$ median & $S_{1.4}>$ 800 mJy \\
& & & 2.2$\%$ median combined & $S_{1.4}>$ 800 mJy \\
& & & & Anti-correlated with flux density \\
\hline 
\cite{2003AA...406..579K} & 106 & 2.7, 4.85, 10.5 & SS: 2--6$\%$, FS: 2.5$\%$ median & 1.4--10.5 GHz \\
\hline 
\cite{2004AA...415..549R} & 197 & 18.5 & FS: 2.7$\%$, SS: 4.8$\%$ median & Weakly correlated with $\alpha_{5-18.5}$ \\
\hline 
\cite{2006MNRAS.371..898S} & 108 & 20 & 2.3$\%$ median & Anti-correlated with flux density \\
\hline 
\cite{2009ApJ...705..868L} & 138, 122, 93, 81 & 23, 33, 41, 61 & 1.7$\%$, 0.91$\%$, 0.68$\%$, 1.3$\%$ mean & Bright WMAP sources \\
\hline 
\cite{2010MNRAS.402.2403M} & 768 & 20 & 2.6$\%$ median, 2.7$\%$ mean & FS: 2.9$\%$, SS: 3.8$\%$ \\
\hline 
\cite{0067-0049-189-1-1} & 149 & 86 & 1.5\% & Flat-radio-spectrum AGNs\\
\hline 
\cite{2011MNRAS.413..132B} & 105 & 8.4, 22, 43 & 2$\%$ median, 3.5$\%$ mean & Independent of frequency, flux density \\
\hline
\cite{2011ApJ...732...45S} & 159 & 4.86--43.34 & typically  2--5$\%$, & Slight increase with frequency,  \\	 
 &  & (4 bands) & tail extending to $\sim$15$\%$ & trend is stronger for SS and dim sources \\
\hline 	
\cite{2017MNRAS.469.2401B} & 881 & 30 & 3.05\% mean & Fractional polarization estimated by \\
&  & 44 & 3.27\% mean & applying the stacking technique on \\
&  & 70 & 2.51\% mean & 881 sources detected in the  \\
&  & 100 & 3.26\% mean &  30 GHz Planck map that are  \\
&  & 143 & 3.06\% mean & outside the Planck Galactic mask \\
&  & 217 & 3.07\% mean & \\
&  & 353 & 3.52\% mean &  \\
\hline       
\cite{2017arXiv171209639P} & 32 & 95 & FS: 2.07\% median & Independent of frequency \\ 
\hline    
\cite{2017arXiv171208412T} & 35 & 30 & 3.3\% median & Independent of frequency, flux density \\
& 9 & 44 & 2.2\% median &  Fractional polarization estimated using  \\
& 4 & 70 & 2.8\% median &  the Intensity Distribution Analysis  \\
& 14 & 100 & 1.9\% median & (IDA) method \\
& 15 & 143 & 2.9\% median &  \\
& 8 & 217 & 3.1\% median & \\
& 1 & 353 & 3.0\% median &  \\
\hline    
This work & 169 & 148 & 2.8$\pm$0.5$\%$  & Independent of flux density \\	      
\hline                   
\end{tabular}\\
\end{center}
{\footnotesize $^{*}$ SS, FS, and IS denote source populations with steep, flat, and inverted spectrum, respectively.}
\end{table*}
\label{tab:Pol_surveys}

The increasing sensitivity of millimeter wavelength telescopes has enabled detections of a large number of extragalactic sources that emit brightly in the millimeter-wavelength sky. These sources can be broadly categorized into two populations: active galactic nuclei (AGN), which are compact regions at the centre of galaxies, and dust-obscured star forming galaxies (DSFG), which represent the most intense starbursts in the universe. The radio emission from galaxies is typically dominated by synchrotron and free-free radiation. 

In AGN, relativistic jets powered by the central nuclei accelerate electrons to relativistic velocities. These electrons then emit synchrotron radiation as they interact with the galactic magnetic field. In these so-called blazars \citep{1980ARA&A..18..321A}, that are members of a larger group of active galaxies that host AGN, synchrotron self-absorption is associated with doppler-boosted structures along the jets and extends from radio to millimeter wavelengths. Free-free emission, also known as thermal bremsstrahlung, is emitted by free electrons interacting with ions in the ionized gas. The characterization of the blazar emission at millimeter (mm) and submillimeter (sub-mm) wavelengths is important for understanding the physics of these sources \citep{2011A&A...536A...7P}. 

The DSFGs are characterized by thermal radiation spanning millimeter to far infrared wavelengths emitted by dust heated by the UV and optical emission from massive stars in star-forming galaxies. This dust obscuration hides the DSFGs from optical and UV observatories. Observations at mm and sub-mm wavelengths can provide a more complete picture of the history of star formation \citep{2014PhR...541...45C}. At these wavelengths, the brightest high redshift DSFGs are detected predominantly through their gravitational lensing \citep{2013Natur.495..344V}. In the mm/sub-mm regime, the radio emission is characterized by falling spectra with frequency, while the spectra of the dust emission increases steeply with frequency. Consequently, bright sources in the sky at wavelengths longer than 1.5 mm are predominantly radio sources, whereas dusty galaxies dominate the submillimeter sky, especially at low flux density. The Second Planck Catalog of Compact Sources \citep{2016A&A...594A..26P} spanning nine frequency channels captured this transition from radio sources to dusty galaxies. They found that the change in the dominant source population occurs between 217 and 353 GHz in frequency, i.e., 1.382 and 0.850 mm in wavelength.
\begin{figure*}
\begin{center}
\includegraphics[width=0.9\linewidth,keepaspectratio]{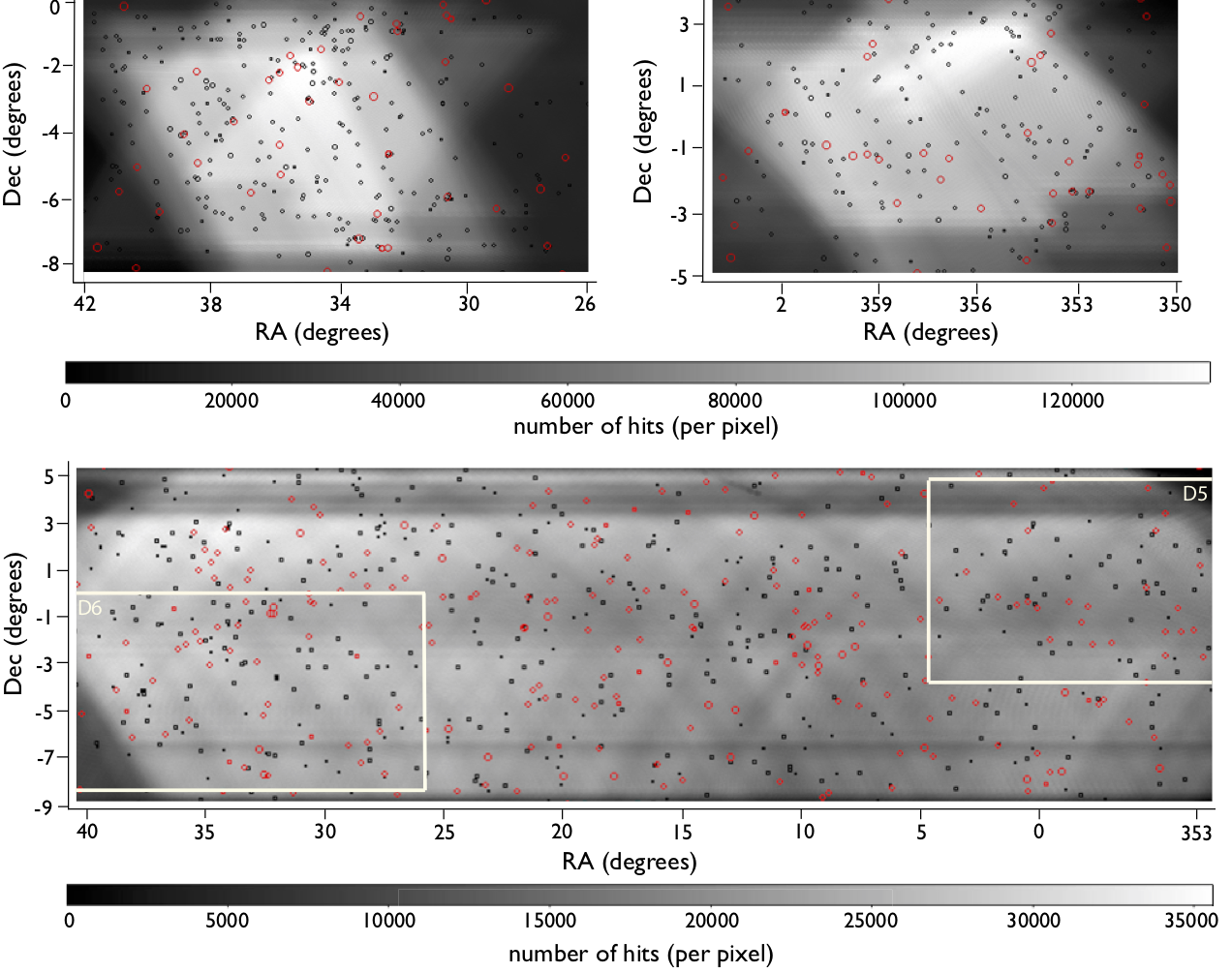}
\caption{\protect \footnotesize Sources detected in the S13+S14 D6 (left top), D5 (right top) and S14 D56 (bottom) maps shown with the hit count maps in the background. The red circles are 1$^{\prime}$ in diameter and centered at the locations of detected sources brighter than 20 mJy. The black circles are 0.5$^{\prime}$ in diameter and centered at the locations of sources detected with $\mathrm{S/N}>4.5$ that are dimmer than 20 mJy. D5 and D6 are contained within the boundaries of D56 but were observed at a different time \citep{2017JCAP...06..031L}. }
\label{fig:filtered}
\end{center}
\end{figure*}

The emission from both types of sources contaminates measurements of the CMB, which contains information on cosmological parameters. The measurement of the CMB polarization, particularly the E-mode damping tail, can help break some of the degeneracies between cosmological parameters. Measurements of the polarization damping tail are expected to become foreground-limited at a smaller angular scale (higher $\ell$) than the temperature damping tail, because of the expected low polarization of dusty point sources (see below). Further, the higher contrast of the acoustic features in EE power spectrum compared to astrophysical foregrounds \citep{2014PhRvD..90f3504G, 2014JCAP...08..010C} will ultimately provide independent and tighter constraints on the standard cosmological parameters, such as the scalar spectral index $n_\mathrm{s}$, than those from the temperature data alone. High-resolution measurements of the E-mode polarization will improve the delensing of the primordial B-modes \citep{2004PhRvD..69d3005S}, ultimately tightening the constraint on the tensor-to-scalar ratio $r$. As measurements of the small angular-scale fluctuations in the CMB are attaining higher sensitivity and finer resolution, ongoing and planned ground-based CMB surveys, such as Advanced ACTPol \citep{2016JLTP..184..772H}, SPT-3G \citep{2014SPIE.9153E..1PB}, Simons Observatory \citep{SimonsObs:2017}, CCAT-prime \citep{CCATp:2017}, and CMB Stage 4 \citep{2016arXiv161002743A, 2017arXiv170602464A} will be capable of extracting information from the E-mode damping tail out to $\ell\approx9000$. However, the contribution of the extragalactic point sources to the CMB power spectrum increases towards smaller angular scales, and it is expected to be a significant fraction of the CMB polarization power. For example, extragalactic foreground sources are expected to be the predominant contaminant for angular scales smaller than 30$\arcmin$ ($\ell\gtrsim400$) in the 70--100 GHz frequency range \citep{1998MNRAS.297..117T}. Hence, characterization of these sources in terms of their spectral and spatial distributions is essential for separating foregrounds from the CMB. 
\begin{figure*}
\begin{center}
\includegraphics[width=0.9\linewidth,keepaspectratio]{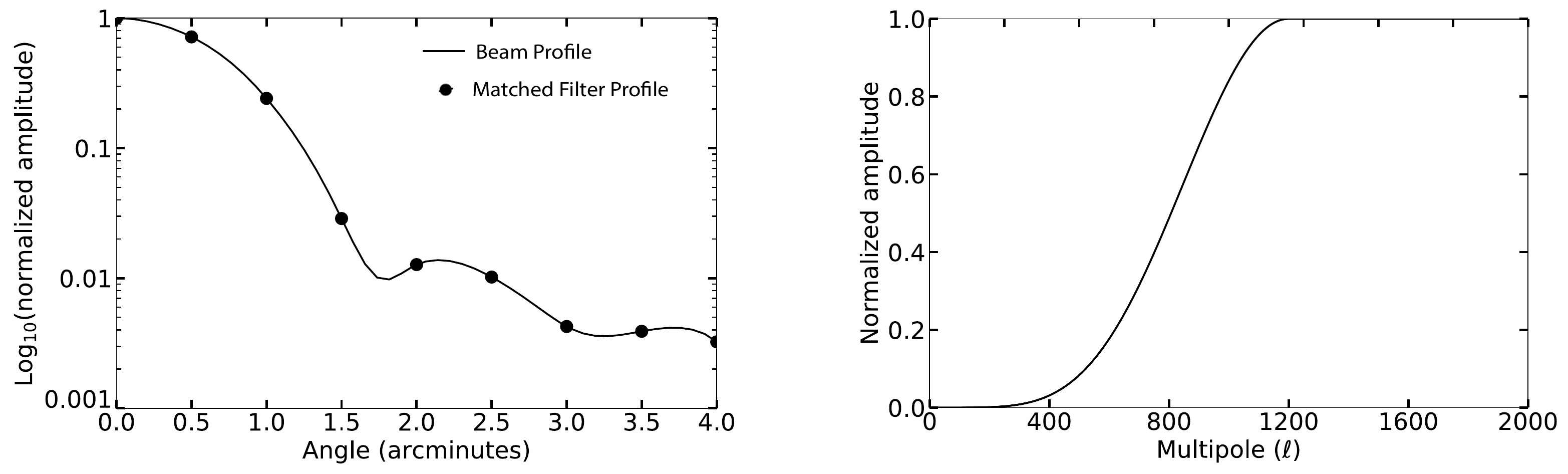}
\caption{\protect \footnotesize {\it Left:} The solid curve shows the unit normalized best-fit ACTPol beam profile. The beam FWHM (full-width at half-max) is 1.4$^{\prime}$. The dots show the unit normalized matched filter profile used in the point-source detection algorithm in the ACTPol pixelization of 0.5$^{\prime}$. {\it Right:} The profile of the high pass filter used to remove large scale modes. }
\label{fig:beams}
\end{center}
\end{figure*}

Measurement of the polarization properties of extragalactic radio sources, in itself, opens up an interesting way to study the astrophysics of the sources. Sources detected in polarization up to about 200 GHz are expected to be mostly radio sources, while the number of detected dusty galaxies increases with increasing frequency. For radio-loud AGNs, polarization data at mm/sub-mm wavelengths reveals details of the magnetic fields in the unresolved regions of their relativistic jets \citep{1998MNRAS.297..667N}. The degree of linear polarization of synchrotron emission could be intrinsically as high as 60-80$\%$ \citep{Saikia:1988cu}. However, observed polarization fractions for compact extragalactic radio sources are typically well below 10$\%$, which is believed to be the result of vector averaging along the line of sight. Nonetheless, at least some sources can have higher polarization fractions, up to 20$\%$ for individual sources \citep{2015ApJ...806..112H}.
\begin{figure*}
\begin{center}
\includegraphics[width=0.8\linewidth,keepaspectratio]{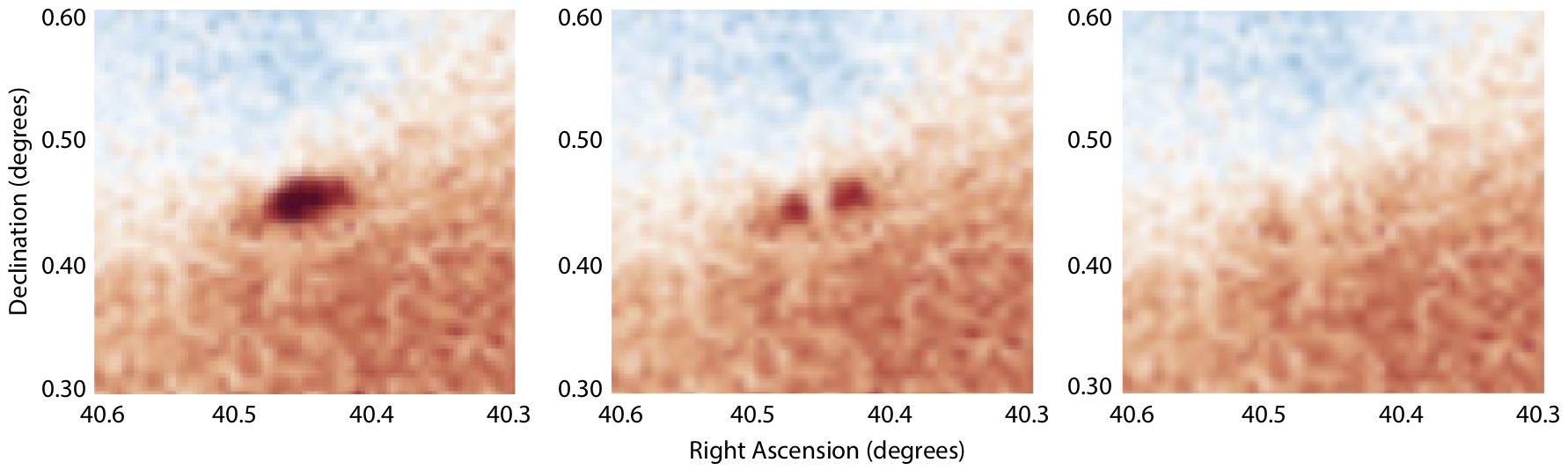}
\caption{\protect \footnotesize Example of an extended source J024145+002645, the removal of which is accomplished in a two-step matched-filtering method. The left panel shows a thumbnail from the original map, the middle panel shows the residuals after the first run of the source-removal algorithm, and the right panel shows the residuals after applying the source-removal algorithm the second time. The color scale is $\pm$400 $\mu$K.}
\label{fig:residuals}
\end{center}
\end{figure*}

Also, little is known about the degree of polarization of \mbox{DSFGs}, but it is likely to be low because the complex structure of galactic magnetic fields with reversals along the line of sight and the disordered alignment of dust grains reduce the global polarized flux when integrated over the whole galaxy. \cite{2002AIPC..609..267G} and \cite{2009ApJS..182..143M} have reported polarization measurements at 850 $\mu$m for two galaxies, M82 and M87, respectively. They found M82 to have a global net polarization of only 0.4$\%$. \cite{2007MNRAS.374..409S} placed an upper limit of 1.54\% on the polarized emission from the ultra luminous infrared galaxy (ULIRG) Arp 220 at 850 $\mu$m. \cite{2017arXiv171208412T} analyzed the fractional polarization of extragalactic sources in Planck maps and placed 90\% confidence upper limits of 2.2\% at 353 GHz and 3.9\% at 217 GHz on the polarization fraction of dusty sources.

Compared to arcsecond resolution radio telescopes with very high sensitivity, millimeter wavelength telescopes have only recently achieved the required sensitivity at arcminute level resolution to detect large numbers of compact extragalactic sources. Thousands of such sources are being detected in CMB maps as high, point-like (unresolved) fluctuations above the background fluctuation level. While the NRAO  Very Large Array - Faint Images of the Radio Sky at Twenty-centimeters (VLA-FIRST) survey \citep{1995ApJ...450..559B} and the VLA Sky survey (NVSS) \citep{1998AJ....115.1693C} at 1.4 GHz still provide the most comprehensive catalog of extragalactic radio sources both in total and polarized intensity, mm/sub-mm observations have picked up over the last decade. Table~\ref{tab:Pol_surveys} lists a selection of surveys of polarized sources between 1.4 and 353 GHz. All of them have reported low degrees of polarization, on the scale of 1--5$\%$.  

SPT (95, 150, 220 GHz); \citep{2010ApJ...719..763V, 2013ApJ...779...61M} and ACT (148, 218 GHz); \citep{2011ApJ...731..100M, 2014MNRAS.439.1556M} have reported detection of both synchrotron and dusty sources. More recently, the Planck Early Release Compact Source Catalogue \citep{2011A&A...536A...7P} and subsequent releases of the Planck Catalogue of Compact Sources \citep{2013A&A...550A.133P, 2016A&A...594A..26P} have reported on the properties of sources extracted from all sky maps at nine frequency bands spanning 30-857 GHz. \cite{2017arXiv170510603B, 2017MNRAS.469.2401B} have analyzed the fractional polarization of dusty and radio sources in the Planck maps, respectively, using stacking techniques. \cite{2017arXiv171208412T} have reanalyzed the same using a different approach. However, the polarization properties of extragalactic radio sources still remain poorly constrained at frequencies higher than 20 GHz. \cite{2012AdAst2012E..52T} and more recently, \cite{2016IJMPD..2540005G}, review the status of radio source observations at radio and mm wavelengths. \cite{2014PhR...541...45C} reviews the current status of DSFG studies. \cite{2016IJMPD..2540009M} elaborates on the role of radio sources in polarimetric cosmological studies. 

The ACTPol receiver deployed on the Atacama Cosmology Telescope (ACT) in Chile has been scanning large areas of the sky with the primary goal of measuring fluctuations in the CMB polarization signal over angular scales up to $\ell\approx9000$. In this paper, we present the polarization properties of sources detected in the CMB intensity (Stokes $I$) map made from observations during the first two seasons of the ACTPol survey covering 680 deg$^{2}$ of the sky. We model the distribution of fractional polarization as a function of total source intensity. Extrapolating this model below the ACTPol detection threshold, we predict the impact of the polarized extragalactic source population on our ability to measure the CMB EE power spectrum at high multipoles with future surveys. 

This paper is structured as follows. We introduce the ACTPol D56 survey in Section 2. A brief discussion of the source detection in the Stokes $I$ map is provided in Section 3. In Section 4, we present measurements of the polarization fraction of intensity-selected sources in the ACTPol D56 region at 148 GHz from the first two seasons of ACTPol observations. We model the distribution of fractional polarization of the sources as a function of the total source flux density using simulations. In Section 5, we discuss the spectral properties of the ACTPol sources by comparing the measured flux densities with other mm-wavelength and radio surveys. Finally, we predict the impact of the sources on the measurement of the CMB EE power spectrum in Section 6.   
\begin{figure*}
\begin{center}
\includegraphics[width=1.0\linewidth,keepaspectratio]{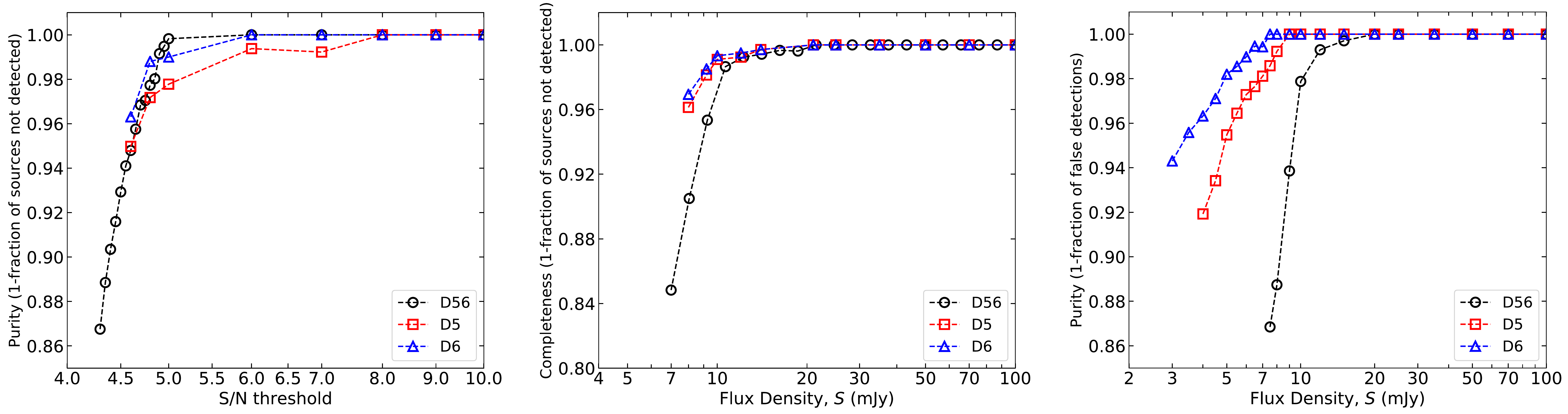}
\caption{\protect \footnotesize {\it Left:} Cumulative purity of the catalogs (D56: black, D5: red, D6: blue) as a function of the chosen detection S/N threshold. {\it Middle:} Cumulative completeness of the  catalogs as a function of source flux density. {\it Right:} Cumulative purity of the catalogs as a function of source flux density. No detections above 20 mJy are expected to be false. }
\label{fig:Compl_Pur}
\end{center}
\end{figure*}

\section{Data}
\subsection{Observations}
The Atacama Cosmology Telescope Polarimeter (ACTPol) is a millimeter-wave camera \citep{2016ApJS..227...21T} which has surveyed thousands of square degrees of sky with 1.4$^{\prime}$ resolution and sensitivity to sources at the level of a few milli-Jansky (mJy). Achieving first light in 2013, ACTPol observed in a band centered at 149 GHz ($\sim$2.0 mm) for the first two seasons. Though the effective central frequency of the broadband receiver to the CMB was 149 GHz, the effective band central frequency of a non-CMB source depends on its spectrum, characterized by spectral indices. These are defined as:
\begin{equation}
	\alpha_{\nu_{b}-\nu_{a}} = \frac{\log(S_{\nu_{a}})- \log(S_{\nu_{b}})} {\log(\nu_{a})- \log(\nu_{b})}
\label{eq:spec}
\end{equation}
According to the conventional single power law model, the distribution of source flux densities $S$ as a function of frequency $\nu$ is given by $S_{\nu}\propto \nu^{\alpha}$. At radio frequencies, a negative $\alpha$ is associated with synchrotron-dominated sources. Sources with free-free as the dominant emission mechanism are characterized by an index close to zero. Sources dominated by graybody re-emission from warm dust, including many of those detected at high redshift, typically have $\mathrm{\alpha> 2}$. We follow the method outlined in Appendix A which follows from \cite{0004-637X-585-1-566} for computing the effective central frequencies of the broadband ACTPol receiver to sources with different spectra.  We do not know the spectral indices of the sources cataloged here, but we assume most are synchrotron dominated.  Hence to calculate flux densities, we assume a nominal --0.7 synchrotron index and compute the effective central frequency, 147.6 GHz.  We show the results of similar calculations for sources dominated by free-free emission ($\alpha = -0.1$) and dust re-emission ($\alpha = +3.7$) in Appendix A. 

In its first season (S13) from September 11, 2013 to December 14, 2013, ACTPol observed four regions of the sky, including the two deep regions D5 and D6 (see Fig.~\ref{fig:filtered}), with a single receiver array. In its second season (S14) from August 20, 2014 to December 31, 2014, ACTPol observed two wider regions, one of which was the D56 region which included both the D5 and D6 sub-regions, with two receiver arrays. In this work, we present the properties of extragalactic sources in the night-time data of the wider D56 region including the deeper D5 and D6 sub-regions \citep{2017JCAP...06..031L} mapped in both S13 and S14. The D56 region covers 680 $\mathrm{deg^{2}}$ of the sky with boundaries $-8.8^\circ < \mathrm{Dec} < 5.2^\circ$, $352.0^\circ < \mathrm{RA} < 40.6^\circ$. The D6 region covers 160 $\mathrm{deg^{2}}$ with boundaries $-8.4^\circ < \mathrm{Dec} < 0.0^\circ$, $26.0^\circ < \mathrm{RA} < 42.0^\circ$, and the D5 region covers 181 $\mathrm{deg^{2}}$ with boundaries $-3.8^\circ < \mathrm{Dec} < 4.8^\circ$, $350.0^\circ < \mathrm{RA} < 4.6^\circ$. 

\subsection{Maps} 
\label{sec:MapMake}
The mapmaking procedure is described in \cite{2014JCAP...10..007N} and \cite{2013ApJ...762...10D}. We combine maps made from both arrays and both seasons, inverse-variance weighted by their white noise level. These maps are publicly available on Legacy Archive for Microwave Background Data Analysis (LAMBDA) \footnote{https://lambda.gsfc.nasa.gov/product/act/} and presented in \cite{2017JCAP...06..031L}. Combining the S14 data from both arrays for D56, and the S13 data for D5 and D6, the resulting Stokes $I$ map sensitivity is 18, 12, and 11 $\mu$K-arcmin for D56, D5, and D6, respectively. The Stokes $Q$ and $U$ sensitivities are the $I$ sensitivities multiplied by a factor of $\sqrt{2}$. The average white noise level, when matched filtered with the ACTPol beam, results in typical sensitivities to point source flux densities of 8 mJy in D56, 4 mJy in D5 and 3 mJy in D6, respectively, with a signal-to-noise ratio (S/N) threshold of 4.5. 

\section{Source Detection}
\label{sec:srcDet}
We use the matched filtering method described in \cite{2011ApJ...731..100M} to detect sources in the intensity map and estimate their amplitudes. Here, we summarize the method and discuss some additional aspects relevant for this work. 

We first multiply the map, pixel by pixel, by the square-root of the number of observations per pixel $N_\mathrm{obs}$, where an observation represents data acquired over 0.0025 seconds. This accounts for the local variation in the map white-noise level and results in a noise-weighted map with approximately uniform white noise level. The average number of observations per pixel is 5534, and the peak number of observations is 35596. We mask the noisy edges of the survey region, where $N_\mathrm{obs}$ is below a threshold of 2000; these regions are prone to false detections. The remaining map region is then used for source detection. The matched filter applied to the Fourier transformed map is given by:
\begin{flalign}
	\begin{gathered}
	\Phi_{MF}(\mathrm{\textbf{k}}) = \frac{F_{k_{0}}(\mathrm{\textbf{k}}) \tilde{B}^{*}(\mathrm{\textbf{k}}) |\tilde{T}_\mathrm{other}(\mathrm{\textbf{k}})|^{-2}}{\int{\tilde{B}^{*}(\mathrm{\textbf{k}}') F_{k_{0}}(\mathrm{\textbf{k}}') |\tilde{T}_\mathrm{other}(\mathrm{\textbf{k}}')|^{-2} \tilde{B}(\mathrm{\textbf{k}}') d\mathrm{\textbf{k}}'}}
	 \end{gathered}	 
\label{eq:stokesQU}
\end{flalign} 
\noindent where $\mathrm{\textbf{k}}$ refers to the wavenumber and $\tilde{B}(\mathrm{\textbf{k}})$ is the Fourier transform of the ACTPol beam. The quantity $\tilde{T}_\mathrm{other}$ is the Fourier transform of the combination of all expected signal components other than point sources, i.e., atmospheric noise, detector noise, SZ signal, primary CMB, and undetected point sources. The function $F_{k_{0}}$ represents the same high-pass filter as was used in \cite{2011ApJ...731..100M}, which tapers off from unity above multipole $\ell=1200$ to zero at $\ell=0$ as shown in the right panel of Fig.~\ref{fig:beams}. This is required to remove large scale noise due to the atmosphere. The effective instrument beam \citep{2017JCAP...06..031L} is derived from planet and bright point source observations as described in \cite{2013ApJS..209...17H} for the previous generation ACT beam. To arrive at the most accurate and refined representation of the beam, the profiles of the brightest sources in the ACTPol map were examined and fit with the beam profile and a Gaussian broadening term in order to estimate the effect of pointing variance in each region. The azimuthally symmetric beam profile is shown in the left panel of Fig.~\ref{fig:beams} along with the matched filter profile. A robust estimate of the power spectrum of $\tilde{T}_\mathrm{other}$ used for this analysis was constructed from the power spectrum of the map itself after masking the brightest sources with $\mathrm{S/N>50}$. This works because the residual source power spectrum is known to be significantly lower than that of the other components. 

The source detection proceeds in two steps: first, sources with $\mathrm{S/N>50}$ are identified in an initial run of the source-finder, and masked. This is done to prevent ringing around the very bright sources in the filtered map which could impact detection of the dimmer sources. Next, the matched filter is applied to detect sources with $\mathrm{S/N>4.5}$. The S/N is defined as the temperature at a location in the filtered map divided by the square-root of the variance of the filtered map itself after masking the sources with $\mathrm{S/N>50}$. The finite ACTPol map pixelization of 0.5$^\prime$ means that unless a detected source lies exactly at the center of the pixel, the flux estimation would be erroneous. In order to correct for this, we use the method of Fourier interpolation to achieve a finer pixel spacing in map space by a factor of 2$^{4}$ as described in \cite{2011ApJ...731..100M}. This enables a more precise estimation of the location of the source peak amplitude after the pixel window function is accounted for. The raw flux densities (in Jy) associated with the source detections are computed from the amplitudes (in $\mu$K) following the method described in Appendix A. The flux estimates obtained from the matched filtering method are accurate under the assumption that the sources are intrinsically unresolved. Distant extragalactic sources meet this criterion for ACTPol. However, objects at very low redshifts or with elongated structures, such as highly elliptical galaxies and AGNs with jets or lobes, would appear extended in our maps. The detection and template-based source-removal process does not work well for extended sources. In order to accurately remove these extended sources from the CMB map, we adopt a two-pass detection and removal method. In the first pass, the matched-filter source-finding algorithm is run on the original map and a source catalog is generated. We then subtract a template map (made from this catalog) from the original map and search for residual bright spots (from the initial subtraction) in these template-subtracted maps. We identify nine extended sources using this method. We then run the source-finding algorithm again on a $1^{\circ}\times 1^{\circ}$ patch of the template-subtracted map centered around each of these bright spots and perform a second pass of source subtraction. An example is shown in Fig.~\ref{fig:residuals}. The total flux for each extended source is obtained by adding together the flux from the two steps. 

\subsection{Source Catalog} 
With a S/N threshold of 4.5, 665, 324, and 213 sources were detected in the D56 (S14), D6 (S13), and D5 (S13) regions respectively. The bottom panel of Fig.~\ref{fig:filtered} shows the hit count map of the S14 D56 region used in this work in the background with circles centered at the locations of the sources brighter than 10 mJy. The red circles correspond to sources brighter than 20 mJy. The top left and right panels show similar images for the D6 and D5 regions respectively. Combining the detections in the D56, D5 and D6 maps, 169 sources brighter than 30 mJy in total intensity are selected for the polarization analysis presented in Section 4. This threshold of 30 mJy is chosen because the measurement of polarized flux for dim sources is heavily biased by noise as discussed in Section 4. A catalog of these sources is presented in Table~\ref{tab:cat}{\color{blue}4} and the full catalog is being released on LAMBDA \footnote{https://lambda.gsfc.nasa.gov/product/act/actpol\_ps\_cat\_info.cfm}. The extended sources are marked with an asterisk in the catalog table. The differential source number counts are derived in Section 5.1 separately from two catalogs: the S14 D56 catalog and from the S13+S14 D5+D6 catalog. 
\begin{figure*}
\begin{center}
\includegraphics[width=1.0\linewidth,keepaspectratio]{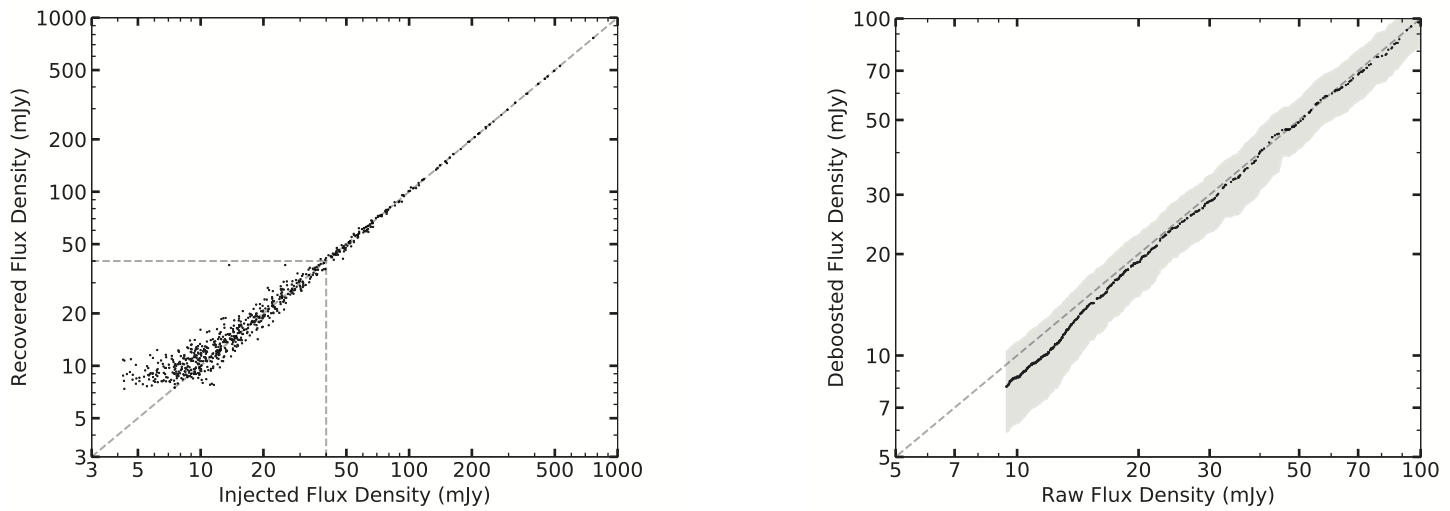} 
\caption{\protect \footnotesize {\it Left:} Source injection simulations: injected versus recovered source flux densities. The ratio is approximately consistent with unity (dashed line) at the 1$\%$ level except for the low flux end ($S$ $<$ 40 mJy), where the effect of flux boosting becomes significant. {\it Right:} Raw versus deboosted source flux densities shown only for sources dimmer than 100 mJy. For e.g., a measured source flux of 10 mJy is deboosted to a true flux of $\sim$8.5 mJy. The grey band encompasses the 68$\%$ confidence interval derived from the kernel density estimation. }
\label{fig:deb1}
\end{center}
\end{figure*}  

\subsection{Purity and Completeness of the Source Catalog}
The S13+S14 D5 and D6 maps are significantly deeper than the D56 map. In order to estimate the completeness of each of the source catalogs, we inject simulated sources in the source-subtracted original map and run the source detection algorithm on it just as with the original maps. We first generate a catalog of sources with S/N $>$ 4.5 by running the source detection algorithm on the original map as described in Section 3. From this catalog, we construct a template map and subtract this from the original map. We then construct a simulated source-only template map the same size as the original map which has 1500 simulated sources at randomly chosen pixels with flux densities down to 5 mJy, drawn from the C2Co model for source counts proposed in \cite{2012AdAst2012E..52T} (T2012). This map is convolved with the best-fit beam and the resulting map is added to the source-subtracted map. The source detection algorithm is then run on this map and the resulting catalog is searched for matches to the list of injected sources. The cumulative completeness curve plotted in the middle panel of Fig.~\ref{fig:Compl_Pur} shows the fraction of injected sources that were detected above a given flux threshold. Based on this test, the expected completeness of the S/N $>$ 4.5 D56 catalog is 87$\%$ at 8 mJy, 97$\%$ at 10 mJy, and 100$\%$ at 20 mJy. The completeness of the S/N $>$ 4.5 D5/D6 catalog is 96$\%$/97$\%$ at 8 mJy, 99$\%$ at 10 mJy, and 100$\%$ at 20 mJy. There are no D5/D6 sources with flux greater than 20 mJy in the overlapping area of D56 with D5/D6 that were not also detected in D56. This is consistent with the completeness of the D56 catalog being 100$\%$ above 20 mJy.
\begin{figure*}
\begin{center}
\includegraphics[width=1.0\linewidth,keepaspectratio]{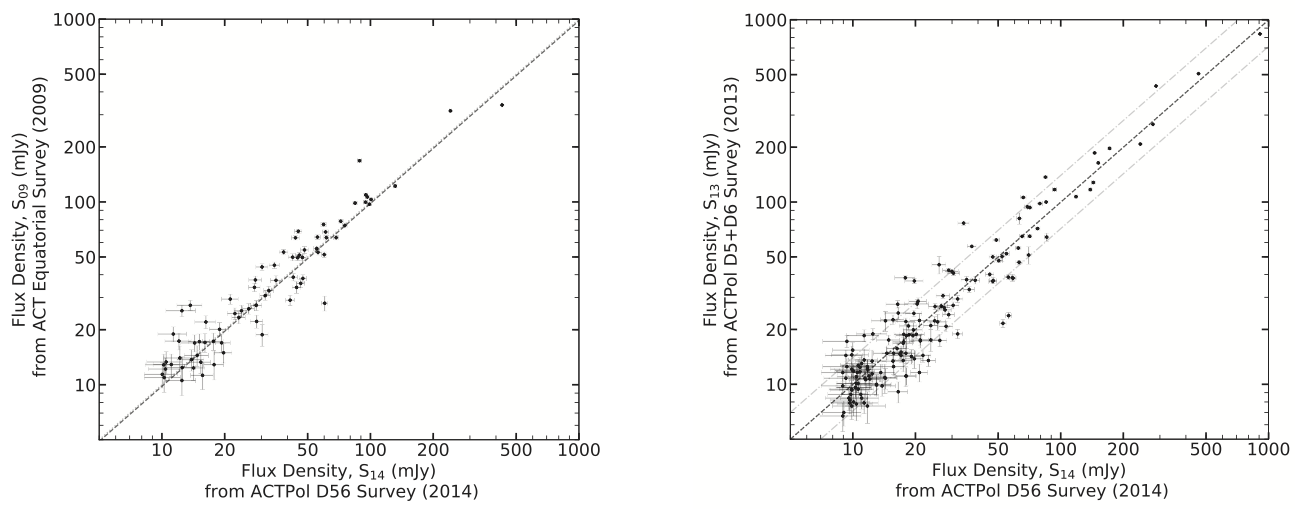} 
\caption{\protect \footnotesize {\it Left:} Comparison of flux densities of cross-matched sources (with $S_\mathrm{14}$>10 mJy)  between the ACTPol D56 survey (2014) and the ACT equatorial survey (2009). The dashed line is the best-fit straight line and corresponds to $S_\mathrm{09}$ = 0.98$\pm$0.03 $S_\mathrm{14}$ and the dotted line corresponds to $S_\mathrm{09}$ = $S_\mathrm{14}$. {\it Right:} Comparison of flux densities of cross-matched sources (with $S_\mathrm{14}>10$ mJy) between the ACTPol D56 survey (2014) and the ACTPol D5 and D6 surveys (2013). The dashed line is the best-fit straight line and corresponds to $S_\mathrm{13}$ = 1.00$\pm$0.01 $S_\mathrm{14}$. These sources are predominantly synchrotron and hence expected to be variable with no preference for one direction over another. The fits are forced to pass through zero.} 
\label{fig:Var1}
\end{center}
\end{figure*} 

The number of false detections in each catalog is also estimated from simulations. We start with simulated temperature and polarization maps of the CMB, and add noise to these maps. The noise model is constructed so as to replicate the combined effect of instrumental white noise, atmospheric noise and detector 1/$f$ noise in the ACTPol maps. In order to model the noise, we start with ``4-way maps" made from four data splits where each split is allocated two consecutive nights of data. The cycle repeats every eight nights. Next, we construct a pair of 2-way maps, each made by co-adding two of the 4-way maps. Then, we compute the noise power spectra from maps obtained by differencing the 2-way maps in $I$, $Q$, and $U$. Noise maps are then constructed given by a Gaussian realization of the noise spectrum of the difference maps. These noise maps are then added to the simulated $I$, $Q$ and $U$ maps. We then inject sources with flux densities down to 10$^{-4}$ Jy, drawn from the C2Co source counts model, into the simulated $I$ map. The source detection algorithm is then run on this map. We then counted the number of sources detected above a given S/N or flux threshold and look for matches to the list of injected sources. The cumulative purity curves shown in the left and right panels of Fig.~\ref{fig:Compl_Pur} plot the quantity (1-${N_\mathrm{false}}/{N_\mathrm{total}}$) as a function of the S/N and flux threshold, respectively, where $N_\mathrm{false}$ is the number of unmatched detections and $N_\mathrm{total}$ is the total number of detections above the corresponding S/N or flux threshold. The number of false detections in the D56/D5/D6 catalogs are less than 7$\%$/5$\%$/4$\%$ for a S/N threshold of 4.5. The expected purity of the S/N $>$ 4.5 D56 catalog is thus estimated to be 87$\%$ at 8 mJy, 97$\%$ at 10 mJy, and 100$\%$ at 20 mJy. The expected purity of the S/N $>$ 4.5 D5/D6 catalog is estimated to be 94$\%$/98$\%$ at 5 mJy, 99$\%$/100$\%$ at 8 mJy, and 100$\%$ at 10 mJy. We use these estimates of completeness and purity of the catalogs to correct the differential source number counts presented in Section 5.1. There are no D56 sources with flux greater than 20 mJy in the overlapping area of D56 with D5/D6 that were not also detected in the deeper D5/D6 maps. This is consistent with the purity of the D56 catalog being 100$\%$ above 20 mJy. 
\begin{figure*}
\begin{center}
\includegraphics[width=0.85\linewidth,keepaspectratio]{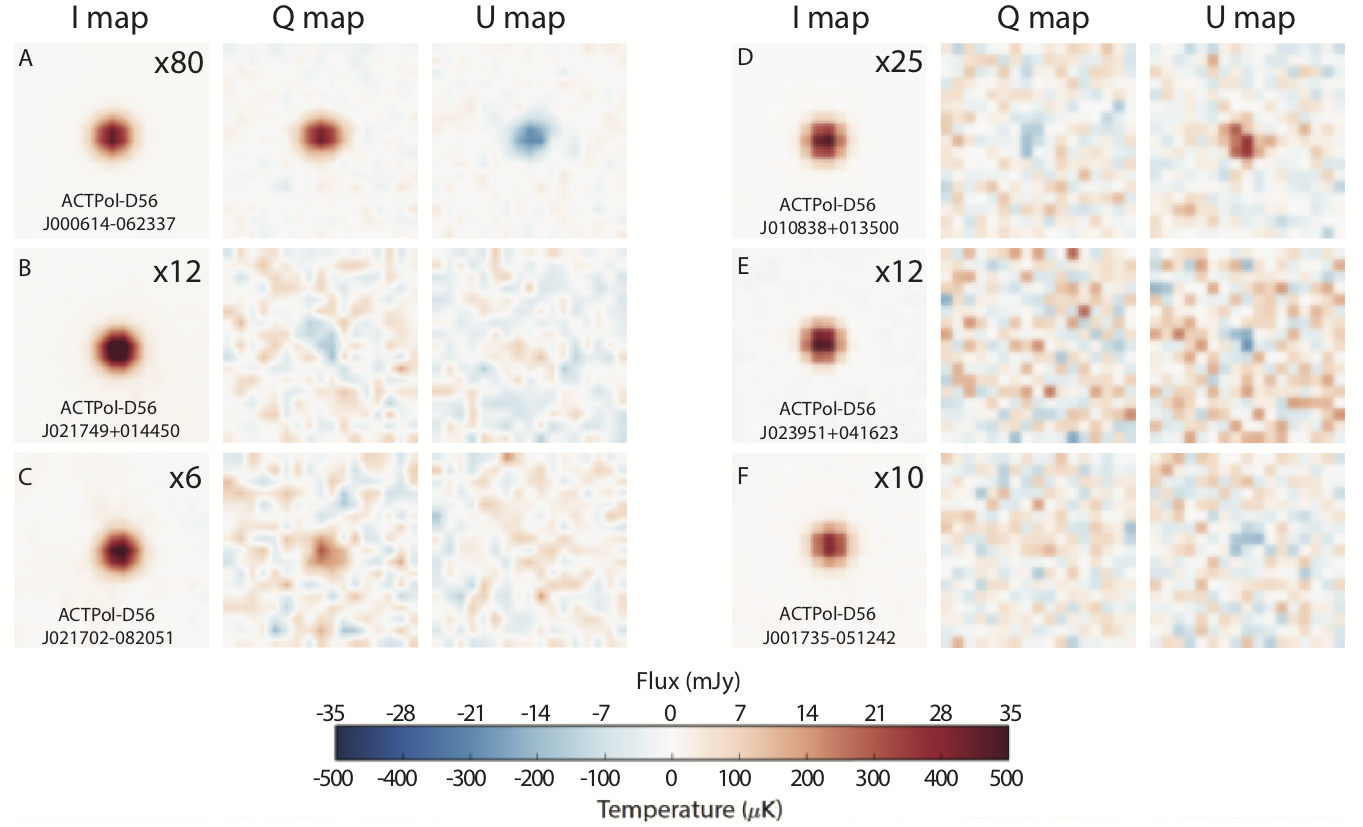}  
\caption{\protect \footnotesize $I$, $Q$, and $U$ thumbnails (0.02 deg$^{2}$ in area) of six intensity-selected sources that have the strongest signal in polarization. The intensity map temperatures have been scaled down by a factor noted in the top right corner of the $I$ thumbnails so as to keep the color scale same for $I$, $Q$, and $U$. The color scale spans $\pm$500 $\mu$K for all but the brightest source in the top left panel ($\pm$2000 $\mu$K). }
\label{fig:iqustamps}
\end{center}
\end{figure*}

\subsection{Deboosting of Source Fluxes}
\begin{figure}
\begin{center}
\includegraphics[width=0.85\linewidth,keepaspectratio]{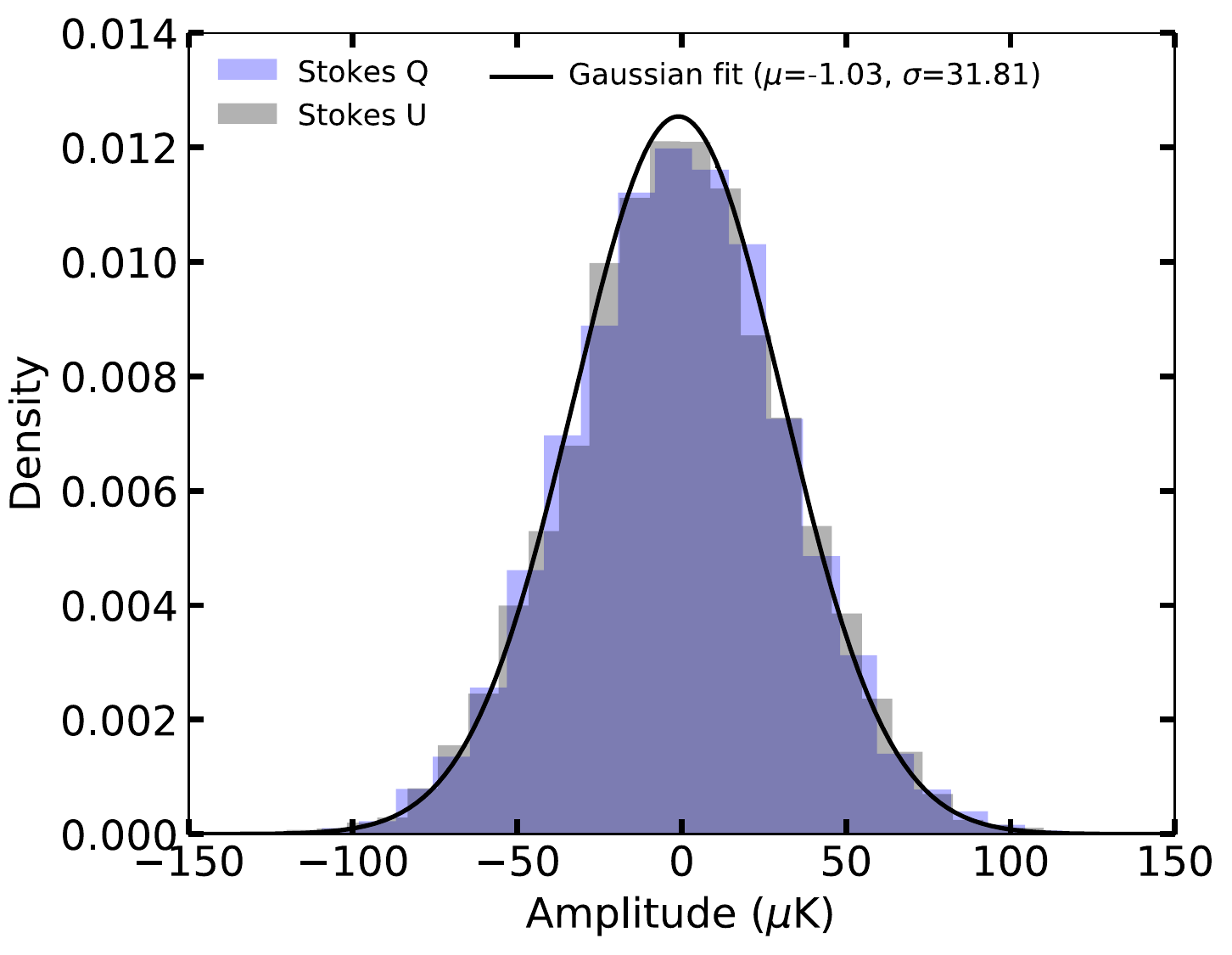} 
\caption{\protect \footnotesize Histograms of pixel values for the Stokes $Q$ (purple) and $U$ (grey) D56 maps. The black line is a Gaussian with mean equal to the average of the $Q$ and $U$ values and standard deviation equal to the average of the $\sigma_{\mathrm{Q}}$ and $\sigma_{\mathrm{U}}$ values. }
\label{fig:QU_dist}
\end{center}
\end{figure} 
To accurately estimate the intrinsic flux densities of sources given their measured values, we have used a method adapted from \cite{2010ApJ...718..513C} to correct for the flux boosting effect due to the bias for selecting intrinsically dim sources that coincide with noise fluctuations. We start with the D56 map and subtracted from it a template map constructed from sources detected down to a S/N threshold of 4. We then populate a zero map the same size as the D56 map with sources at random locations, drawn from the C2Co source counts model, down to a flux threshold of 7 mJy. The C2Co model was selected because it best describes the source counts from this work (see Section 5.1). This map is convolved with the best-fit ACTPol beam and the resulting template map was added to the source-subtracted D56 map. We then run the source finder on this simulated map and verify that the recovered source flux densities are consistent with the injected flux densities at the 1$\%$ level (see left panel of Fig.~\ref{fig:deb1}) for sources above a flux density threshold of 40 mJy. Hence, the effect of flux boosting on these sources is expected to be negligible. While selecting the lower flux limit for the injected sources, we chose a flux cut of 7 mJy that is sufficiently lower than the dimmest source whose flux we are trying to deboost, i.e., 10 mJy; this is done to minimize the effect of incompleteness in the low flux end. 

Using this simulated map, we extract the true underlying posterior probability distribution $\mathcal{P}(S_\mathrm{max}|S_\mathrm{p,m})$ for the intrinsic flux density of the brightest source within a pixel as:
\begin{flalign}
	\begin{gathered}
	\mathcal{P}({S_\mathrm{max}}|{S_\mathrm{p,m}}) \propto \mathcal{P}({S_\mathrm{p,m}}|{S_\mathrm{max}}) \mathcal{P}({S_\mathrm{max}}) \\
	\end{gathered}	 
\label{eq:posterior}
\end{flalign} 
\noindent where $S_\mathrm{p,m}$ is the measured source flux in a pixel, and $\mathcal{P}(S_\mathrm{p,m}|S_\mathrm{max}$) is the likelihood of measuring flux $S_\mathrm{p,m}$ in a pixel given the brightest source in that pixel has flux $S_\mathrm{max}$. The prior $\mathcal{P}(S_\mathrm{max})$ is given by: 
\begin{flalign}
	\begin{gathered}
	\mathcal{P}(S_\mathrm{max}) \propto  \frac{dN}{dS} \exp \bigg(-\Delta \Omega_{\mathrm{p}} \int_{S_\mathrm{max}}^{\infty} \frac{dN}{dS'} dS'\bigg)\\
	\end{gathered}	 
\label{eq:prior}
\end{flalign} 
\noindent where $\Delta \Omega_{\mathrm{p}}$ is the solid angle corresponding to each pixel. The effect of the d$N$/d$S$ source count distribution is essentially implemented in the simulations by drawing sources from the C2Co d$N$/d$S$ model. In order to compute the posterior given by equation~(\ref{eq:posterior}) for each detection with measured flux $S_\mathrm{p,m}$, we started by noting each source detected in the simulated map with flux in the range 0.85 $S_\mathrm{p,m}$ to 1.15 $S_\mathrm{p,m}$. We find every matching injected source (within a matching radius of 0.01$^{\circ}$), and compare the recovered and injected flux densities. We then apply a kernel density estimator (KDE) to the distribution of injected source flux densities. The bandwidth parameter for the KDE is chosen to be 0.15 $S_\mathrm{p,m}$. The true, underlying posterior $\mathcal{P}(S_\mathrm{max}|S_\mathrm{p,m})$ for a measured flux $S_\mathrm{p,m}$ is then simply obtained by normalizing the resulting probability density distribution multiplied by the exponential term in equation~(\ref{eq:prior}). The right panel of Fig.~\ref{fig:deb1} shows the deboosted flux densities plotted against the raw measured flux densities for sources detected with total flux less than 100 mJy. As expected, the correction due to deboosting becomes more significant for dimmer sources. We report the deboosted flux densities for sources with measured total flux below 100 mJy in Table~\ref{tab:cat}{\color{blue}4}.

\subsection{Source Variability}
The majority of the sources in this study are synchrotron dominated, presumably AGN-powered radio galaxies. These are likely to be variable, attributable to the birth and expansion of new components and shocks forming in relativistic flows in parsec-scale regions. The variations may be on times scales of hours to months or even years. We examine variability of the sources treated here by cross-matching with the ACT equatorial catalog \citep{Gralla:2018} composed of sources observed in 2009. The left panel of Fig.~\ref{fig:Var1} shows a comparison of the measured flux densities between 2009 and 2014. The best fit straight line through these points has a slope of 0.98$\pm$0.03. We also compare measured flux densities of 158 sources observed both during the 2013 (D5+D6) and 2014 (D56) seasons of ACTPol observations in the right panel of Fig.~\ref{fig:Var1}. The best fit straight line through these points has a slope of $1.00\pm0.01$. The slope of the best fit line being close to unity in both comparisons implies that on average, there is no bias introduced by source variability or calibration of our data across seasons. In other words, there is no preference of the flux densities to vary in one direction over another, on average. On the other hand, the obvious scatter about the best fit line seen in the form of a significant number of data points lying well away from the unity slope line is an indication of source variability. A detailed study of source variability across seasons of ACTPol observations is underway and will be the topic of a future paper. Comparing the measured per-season flux densities of the 158 sources observed during both seasons, we find 36 sources with flux densities varying by more than 40$\%$ between the two seasons. These are the data points lying outside the two dash-dot lines in Fig. 6 and are flagged as variable in the catalog table. 
\begin{figure*}
\begin{center}
\includegraphics[width=0.85\linewidth,keepaspectratio]{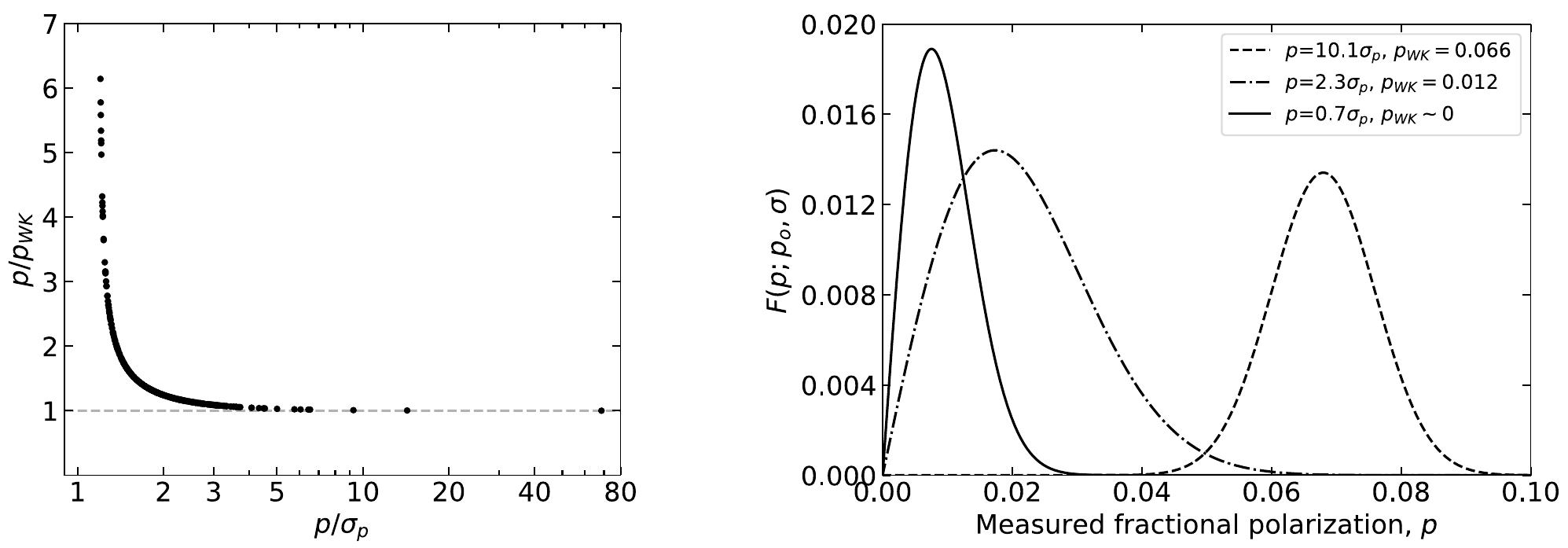} 
\caption{\protect \footnotesize {\it Left:} Ratio of measured $p$ to the noise bias removed estimate, $p_{\mathrm{WK}}$ as a function of $p$/$\sigma_{\mathrm{p}}$. For high values of $p$/$\sigma_{\mathrm{p}}$, $p_{\mathrm{WK}}$ approaches $p$ and as $p$ approaches $\sigma_{\mathrm{p}}$, $p_{\mathrm{WK}}$ approaches zero. {\it Right:} Rice distribution for the measured $p$ of three example sources with $p=0.7\sigma_{\mathrm{p}}$, $p=2.3\sigma_{\mathrm{p}}$, and $p=10.1\sigma_{\mathrm{p}}$. The peak of the distribution corresponds to the most probable observed value of $p$. Here $p_{\mathrm{WK}}$ is the noise bias subtracted estimate of the intrinsic fractional polarization. }
\label{fig:rice}
\end{center}
\end{figure*}

\section{Fractional Polarization of intensity-selected Sources}
\label{sec:Polfrac}
While hundreds of sources are detected in the total intensity map, the situation is very different in the case of the polarization maps, where only eight sources are detected with signal-to-noise ratio greater than 5. Fig.~\ref{fig:iqustamps} shows thumbnails of six sources having the strongest signal in polarization. We investigate the polarization properties of the 169 intensity-selected sources in the D56 map brighter than 30 mJy in total flux density. Polarization is quantified with the Stokes $Q$ and $U$ components:
\begin{flalign} 
	\begin{gathered}
	 Q = p I \mathrm{\cos(2\psi)}, \\
	 U = p I \mathrm{\sin(2\psi)}
	 \end{gathered}	 
\label{eq:stokesQU}
\end{flalign} 
The polarization fraction, $p$, and polarization angle, $\psi$, can then be derived as:
\begin{flalign} 
	\begin{gathered}
	p = \sqrt{q^{2} + u^{2}}, \; \textnormal{where} \; q = \frac{Q}{I}, \; \textnormal{and} \; u = \frac{U}{I} \\
	\psi = \frac{1}{2}\arctan(U/Q)
	 \end{gathered}	 
\label{eq:polarized}
\end{flalign} 
\noindent The estimates of the Stokes parameters from the D56 $Q$ and $U$ maps are unbiased and approximately uncorrelated, but computing the polarized flux introduces a noise bias because of the squaring, which is significant when the signal-to-noise ratio is low. In the presence of measurement noise, a hypothetical source with a true polarization fraction $p_{0} = 0$ can appear to have a $p > 0$, unless the noise bias is accurately subtracted. This bias was first discussed in \cite{1958AcA.....8..135S} and has since been treated extensively in the literature: \cite{1985A&A...142..100S} account for the biases when Stokes parameters are uncorrelated and have the same errors; an analytic, approximate distribution of $p$ for the general case of correlated errors was developed by \cite{2014MNRAS.439.4048P}; \cite{2012A&A...538A..65Q} and \cite{2014PASP..126..459M} adopted a Bayesian approach based on the assumption of Gaussian error; \cite{2011ApJ...729...25L}, \cite{2012AdAst2012E..40R}, and \cite{0004-637X-787-2-99} corrected for polarization bias through a Monte Carlo analysis; and \cite{2015A&A...574A.135M, 2015A&A...574A.136M} compared several estimation methods. This paper takes a new approach of using maps with uncorrelated noise constructed from data splits to avoid the noise bias.
\begin{figure*}
\begin{center}
\includegraphics[width=0.75\linewidth,keepaspectratio]{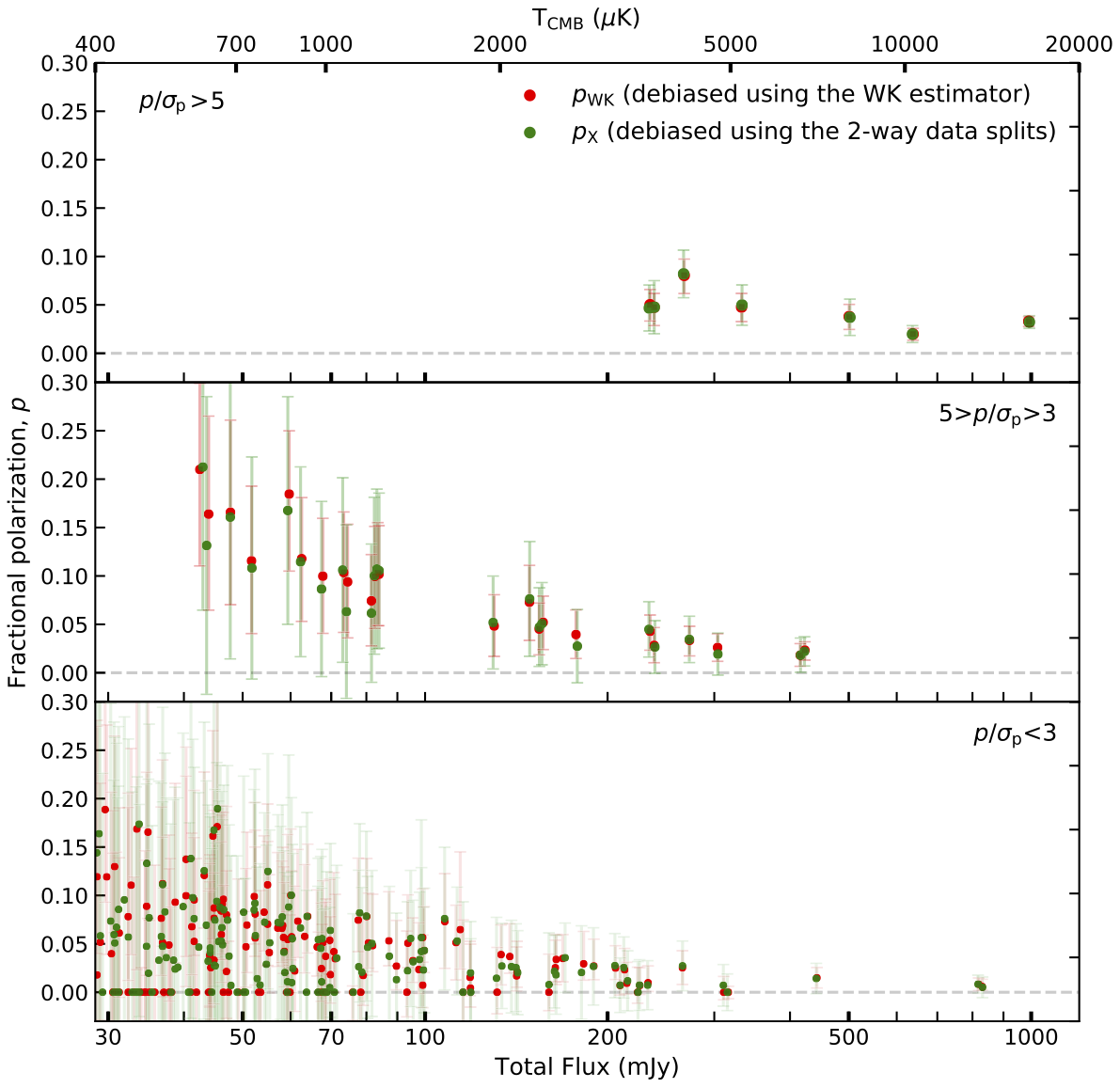} 
\caption{\protect \footnotesize Measurements of fractional polarization plotted against total flux density. The raw measurements are debiased using (a) the Wardle-Kronberg estimator (red), and (b) 2-way data splits (green). {\it Top:} Sources detected in polarization with $p\mathrm{>5\sigma_{\mathrm{p}}}$. The error bars represent the 95$\%$ confidence interval. {\it Middle:} Sources with $\mathrm{5\sigma_{\mathrm{p}}>p>3\sigma_{\mathrm{p}}}$. {\it Bottom:} Sources with $p\mathrm{<3\sigma_{\mathrm{p}}}$. The green error bars are uncertainties in estimates of $p$ obtained using method (b), where each of the 2-way maps contains only half the data, and hence the error bars are larger. The green data points at $p=0$ correspond to sources for which the quantity $q_{\mathrm{1}}q_{\mathrm{2}} + u_{\mathrm{1}}u_{\mathrm{2}}$ is negative. }
\label{fig:polfrac}
\end{center}
\end{figure*}
Our goal is to estimate the true values $I_\mathrm{o}$, $Q_\mathrm{o}$, and $U_\mathrm{o}$ of the Stokes parameters, given measured values $I$, $Q$, and $U$. The fundamental assumption we make is that the measurements of $I$, $Q$, and $U$ are described by a three-dimensional Gaussian with standard deviations $\sigma_{\mathrm{I}}, \sigma_{\mathrm{Q}}$, and $\sigma_{\mathrm{U}}$, respectively. Fig.~\ref{fig:QU_dist} shows the noise distribution for Stokes $Q$ and $U$ maps, where sources detected in total intensity are masked out of the $Q$ and $U$ maps leaving pixels that are free of detectable polarized emission. We then use our source finding algorithm to pick out the $Q$ and $U$ amplitudes at 10$^{4}$ randomly selected locations on the maps, just as we would do for extracting the $Q$ and $U$ amplitudes of the intensity-selected sources. We calculate the average $Q$, $U$, $\sigma_{\mathrm{Q}}$, and $\sigma_{\mathrm{U}}$ values to be --1.03, --1.03, 31.8, and 31.9 $\mu$K, respectively. The solid curve in Fig.~\ref{fig:QU_dist} represents a Gaussian distribution with mean equal to the average of the $Q$ and $U$ values and standard deviation equal to the average of the $\sigma_{\mathrm{Q}}$ and $\sigma_{\mathrm{U}}$ values. This describes well the $Q$ and $U$ probability density functions and thus validates our assumption. We get the measured value for the total intensity $I$ and the uncertainty in its measurement $\sigma_{\mathrm{I}}$ from the matched filtering method as described in Section 3. The Stokes $Q$ and $U$ measurements for these intensity-selected sources and their respective uncertainties $\sigma_{\mathrm{Q}}$ and $\sigma_{\mathrm{U}}$ are then obtained using the same matched filtering method on the $Q$ and $U$ maps at the locations of the detections in the $I$ map.
\begin{figure*}
\begin{center}
\includegraphics[width=1.0\linewidth,keepaspectratio]{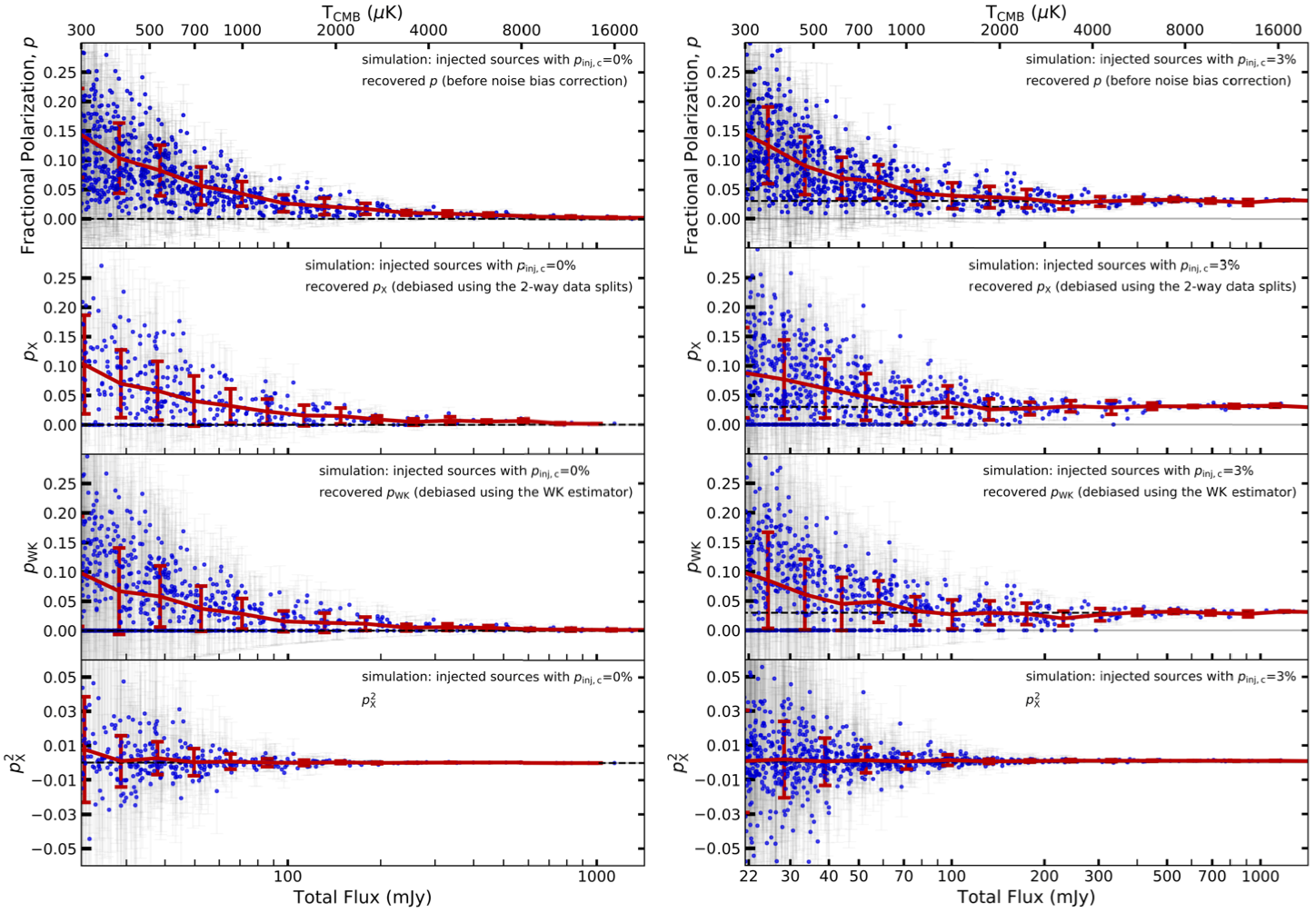} 
\caption{\protect \footnotesize ``C'' simulations: injected sources with a constant polarization fraction, $p_\mathrm{inj,c}=0$ (left) and 0.03 (right). The polarization angles of the sources are randomly drawn from a flat distribution. The blue dots represent the polarization fractions of the injected sources recovered from the simulated maps along with the 95$\%$ confidence intervals and the red lines connect the average fractional polarization values binned in total intensity with the error bars denoting the 1$\sigma$ standard deviation $\sigma_{p_\mathrm{rec,c}}$ within each bin. The apparent lack of data points near $p$ = 0 at the low flux end in the top left panel is a consequence of the noise bias which is significant at the low flux end and biases the recovered $p$ away from zero. The residual bias in the recovered $p$ from the 2-way maps at low total flux comes from the fact that $p$ is only defined if $p^\mathrm{2}$ is positive. The plot of $p^\mathrm{2}$ versus total flux from the 2-way maps shows that the de-biasing works well on average. (continued in Fig.~\ref{fig:psim2}) }
\label{fig:psim1}
\end{center}
\end{figure*}
\begin{figure*}
\begin{center}
\includegraphics[width=1.0\linewidth,keepaspectratio]{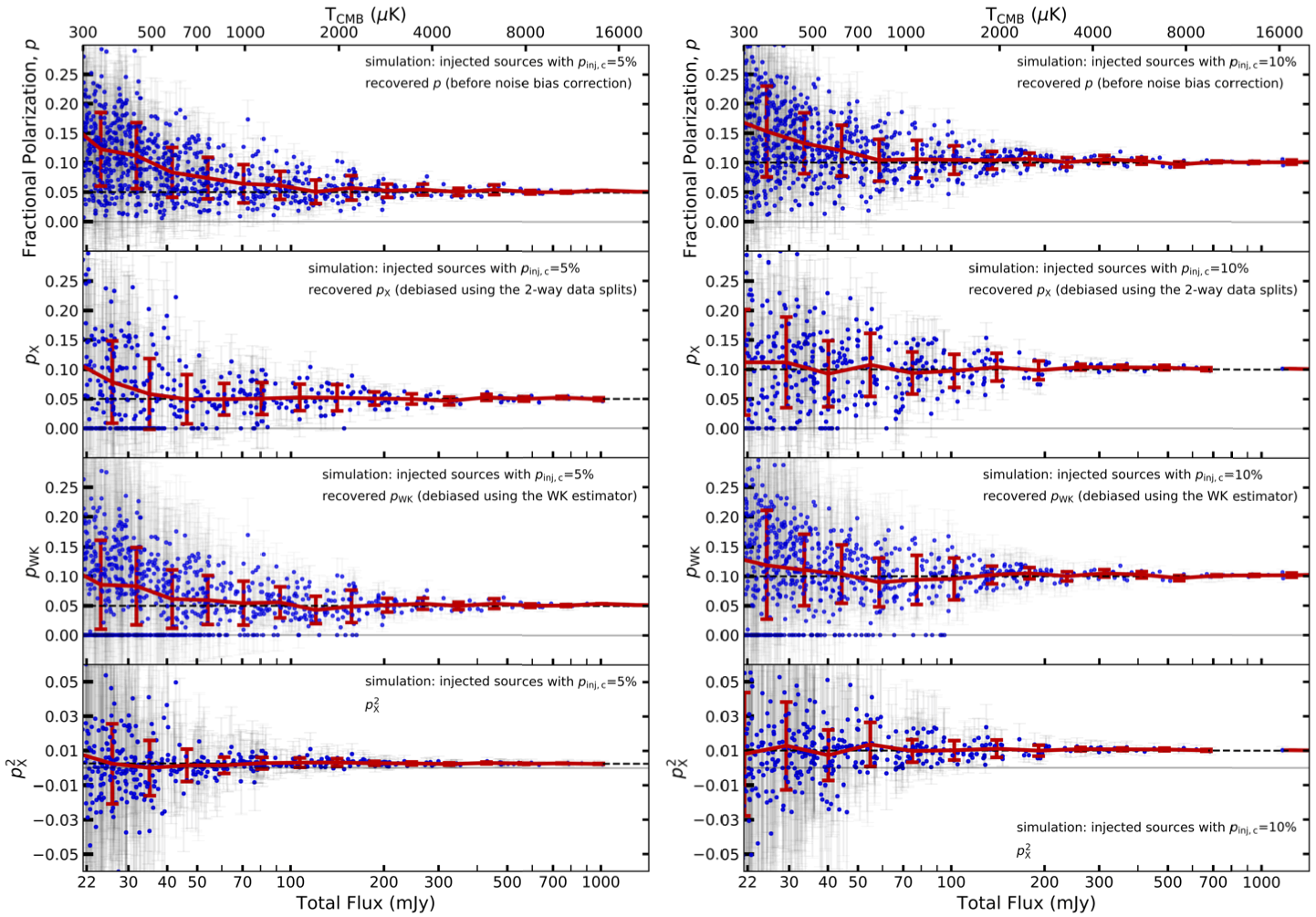} 
\caption{\protect \footnotesize (continued from Fig.~\ref{fig:psim1}) ``C'' simulations: injected sources with a constant polarization fraction, $p_\mathrm{inj,c} = 0.05$ (left) and 0.10 (right). The polarization angles of the sources are randomly drawn from a flat distribution. The blue dots represent the polarization fractions of the injected sources recovered from the simulated maps along with the 95$\%$ confidence intervals and the red lines connect the average fractional polarization values binned in total intensity with the error bars denoting the 1$\sigma$ standard deviation $\sigma_{p_\mathrm{rec,c}}$ within each bin. The residual bias in the recovered $p$ from the 2-way maps at low total flux comes from the fact that $p$ is only defined if $p^\mathrm{2}$ is positive. The plot of $p^\mathrm{2}$ versus total flux from the 2-way maps shows that the de-biasing works well on average. }
\label{fig:psim2}
\end{center}
\end{figure*}

\subsection{Systematics: Temperature-to-Polarization leakage}
We consider the simplest of a range of systematic effects that can lead to power from the Stokes $I$ map contaminating the Stokes $Q$ and $U$ maps. As we are dealing with faint polarization signals, it is essential to correct for any such I-to-P leakage. An approximate estimate of this instrument-induced systematic effect is obtained by performing a combined fit of the $Q (U)$ pixel values at each location of the detections in the $I$ map to the inverse variance weighted average $Q (U)$ value plus a variable temperature leakage term $\alpha_\mathrm{I-Q}I$ ($\alpha_\mathrm{I-U}$I). The 336 sources with total flux between 10 and 20 mJy were used for this fitting. The percentage leakages are found to be $\alpha_{\mathrm{(I-Q)}}=0.13\pm0.15\%$ and $\alpha_{\mathrm{(I-U)}} = 0.24\pm0.16\%$ and are corrected for in the calculation of the polarized flux. 

\subsection{Noise Bias Corrected Fractional Polarization}
In the presence of random noise, the distribution of the degree of linear polarization can be approximated by a Gaussian only for sources with a strong $Q$ or $U$ signal. In the measurement of the fractional polarization $p$, for a source with an intrinsic fractional polarization $p_{\mathrm{o}}$, the statistical distribution of noise is a Rice distribution \citep{1958AcA.....8..135S}:
\begin{flalign}
	\begin{gathered}
	 F(p;p_{\mathrm{o}},\sigma_{\mathrm{p}}) = \frac{p}{\sigma_{\mathrm{p}}^{2}} J_{o} \Bigg(i\frac{pp_{\mathrm{o}}}{\sigma_{\mathrm{p}}^{2}}\Bigg) \exp\Bigg(-\frac{p^{2}+p_{\mathrm{o}}^{2}}{2\sigma_{\mathrm{p}}^{2}}\Bigg)
	 \end{gathered}	 
\label{eq:rice}
\end{flalign}
\noindent where the uncertainty, $\mathrm{\sigma_{\mathrm{p}} \approx \sigma_{\mathrm{q}} \approx \sigma_{\mathrm{u}}}$, and $\mathrm{\sigma_{q,u} = \sigma_{Q,U}}/I$. Fig.~\ref{fig:rice} shows the Rice distribution $F(p$;$p_{o},\sigma_{\mathrm{p}}$) for three sources: one each with $p\mathrm{<\sigma_{\mathrm{p}}}$, $p\mathrm{>\sigma_{\mathrm{p}}}$, and $p\gg\sigma_{\mathrm{p}}$. For sources with $p\gg\sigma_{\mathrm{p}}$, the Rice distribution is close to Gaussian, and $p_{o}\approx p$. \cite{1985A&A...142..100S} have critiqued and compared various methods for noise bias correction and shown that all of them leave some residual bias, especially when $p/\sigma_{\mathrm{p}}$ is small. They further showed that the method developed by \cite{1974ApJ...194..249W} is one of the most effective in removing noise bias from measurements of fractional polarization of point sources. We perform the noise bias correction in two ways: using the Wardle-Kronberg (which we will call ``WK") estimator and using 2-way maps (defined in Section 3.2) constructed from 4-way data splits. 
\begin{figure}
\begin{center}
\includegraphics[width=0.95\linewidth,keepaspectratio]{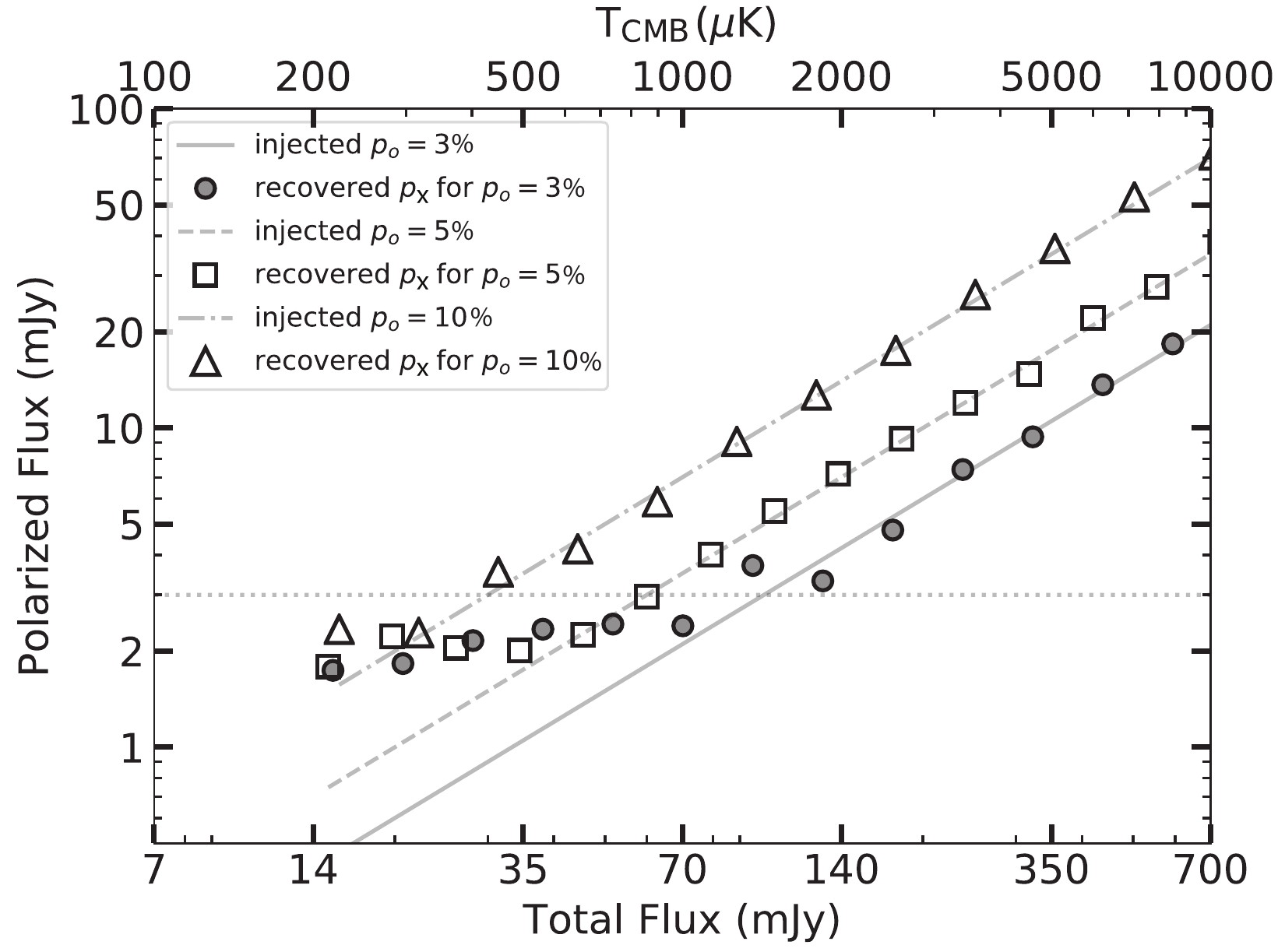} 
\caption{\protect \footnotesize The binned average values of the recovered polarized flux as a function of the total flux for injected source  populations having constant $p_{o} = 0.03$ (circle), 0.05 (square), 0.10 (triangle). The measured polarized flux values were de-biased using the 2-way maps and this plot shows that the de-biasing works well when the polarized flux $P\gtrsim3$ mJy as evident from the fact that the recovered values shown by the dots follow the injected values shown by continuous lines, but diverge for $P\lesssim3$ mJy indicating residual bias. } 
\label{fig:polflux}
\end{center}
\end{figure}

The ``WK" estimator is defined as the value of the intrinsic fractional polarization $\hat{p}_\mathrm{WK}$, for which the observed value $p$ is equal to the maximum of the Rice distribution. This is obtained by solving the equation that we get by differentiating equation~(\ref{eq:rice}) with respect to $p$ and setting it to zero as below.
\begin{flalign}
	\begin{gathered}
	 J_{o} \Bigg(i\frac{p {\hat{p}_{\mathrm{WK}}}}{\sigma_{\mathrm{p}}^{2}}\Bigg) \Bigg(1 - \frac{p^{2}}{\sigma_{\mathrm{p}}^{2}}\Bigg) -i \frac{p {\hat{p}_{\mathrm{WK}}}}{\sigma_{\mathrm{p}}^{2}} J_{1} \Bigg(i\frac{p {\hat{p}_{\mathrm{WK}}}}{\sigma_{\mathrm{p}}^{2}}\Bigg) = 0
	 \end{gathered}	 
\label{eq:WK}
\end{flalign}

We obtain the noise bias corrected value of the source fractional polarization $p_{\mathrm{WK}}$ as:
\begin{flalign}
	\begin{gathered}
		p_{\mathrm{WK}} = \hat{p}_{\mathrm{WK}} \; \textnormal{when} \; p < 4\sigma_{\mathrm{p}} \textnormal{,} \; \textnormal{and} \\
		p_{\mathrm{WK}} = \sqrt{p^{2} - \sigma_{\mathrm{p}}^{2}} \; \textnormal{when} \; p > 4\sigma_{\mathrm{p}} \\
	 \end{gathered}	 
\label{eq:po_1}
\end{flalign}
It follows from equation~(\ref{eq:WK}) that for any measured value $p$ less than or equal to $\sigma_{\mathrm{p}}$, the intrinsic fractional polarization $p_{\mathrm{WK}}$ is equal to zero. Fig.~\ref{fig:rice} shows a plot of the ratio $p/p_{\mathrm{WK}}$ as a function of $p/\sigma_{\mathrm{p}}$, where $p_{\mathrm{WK}}$ approaches 0 as $p/\sigma_{\mathrm{p}}$ approaches 1. The 95$\%$ upper and lower confidence interval limits on the estimate of $p_{\mathrm{WK}}$ for any measured $p$ is obtained using the Rice distribution as the probability density function of the estimator. Given the shape of the Rice distribution, these error bars are asymmetric in general, except when $p \gg \sigma_{\mathrm{p}}$.
\begin{figure}
\begin{center}
\includegraphics[width=1.0\linewidth,keepaspectratio]{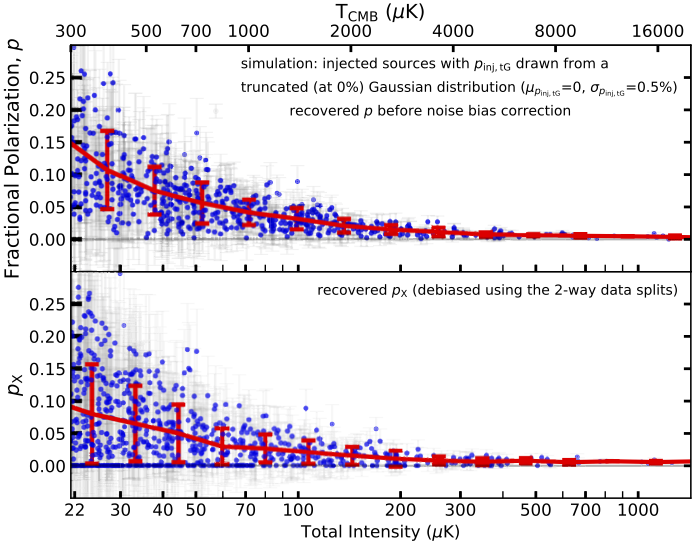} 
\caption{\protect \footnotesize ``tG'' simulations: injected sources that are assigned a polarization fraction drawn from a Gaussian distribution with mean $\mu_{p_\mathrm{inj,tG}} = 0$ and standard deviation $\sigma_{p_\mathrm{inj,tG}} = 0.005$. The polarization angles of these sources are randomly drawn from a flat distribution. The blue dots represent the polarization fractions of the injected sources recovered from the simulated maps along with the 95$\%$ confidence intervals and the red lines connect the average fractional polarization values binned in total intensity with the error bars denoting the 1$\sigma$ standard deviation $\sigma_{p_\mathrm{rec,tG}}$ within each bin. }
\label{fig:psim3}
\end{center}
\end{figure}

In the second method, we use the 2-way maps with uncorrelated noise to naturally de-bias the measurement of $p$. We get the amplitudes of the Stokes $I$, $Q$, and $U$ at the locations of the intensity-selected sources independently from the two maps. Then, the intrinsic fractional polarization $p_{\mathrm{\times}}$ is computed as below.
\begin{flalign}
	\begin{gathered}
	p_{\mathrm{\times}} = \sqrt{q_{1} q_{2} + u_{1} u_{2}}
	\end{gathered}	 
\label{eq:polarized}
\end{flalign}
\noindent where $q_{1,2} = Q_{1,2}/I_{1,2}$ and $u_{1,2} = U_{1,2}/I_{1,2}$. If the quantity under the square root $q_{1}q_{2}+u_{1}u_{2}$ is negative, the $Q$ and $U$ amplitudes could be attributed to random fluctuations and we set $p_{\mathrm{\times}}$ equal to zero. This method naturally de-biases the measurement of $p$. 
\begin{figure}
\begin{center}
\includegraphics[width=0.9\linewidth,keepaspectratio]{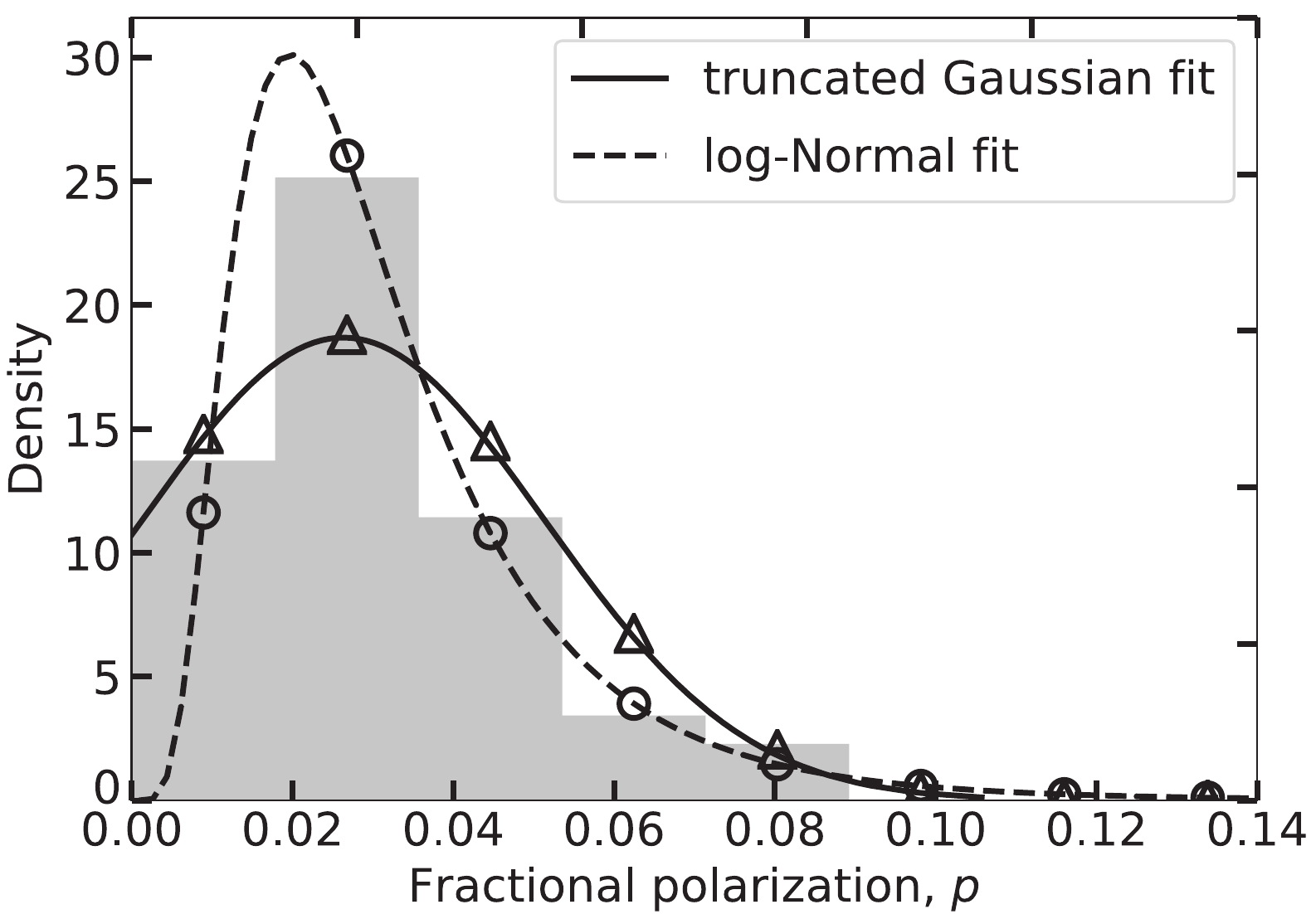}
\caption{\protect \footnotesize Histogram of the measured fractional polarization $p_\mathrm{\times}$ for sources detected with total flux density $S>30$ mJy and having a non-zero $p_\mathrm{\times}$. The dashed line shows the best-fit log-normal distribution to this data described by $\mu_\mathrm{lN}$ = 0.027 and $\sigma_\mathrm{lN} = 0.572$ and the solid line shows the best-fit truncated Gaussian distribution with $\mu_{p,\mathrm{tG}} = 0.027$ and $\sigma_{p,\mathrm{tG}} = 0.025$.}
\label{fig:src_model}
\end{center}
\end{figure} 
\begin{figure*}
\begin{center}
\includegraphics[width=1.0\linewidth,keepaspectratio]{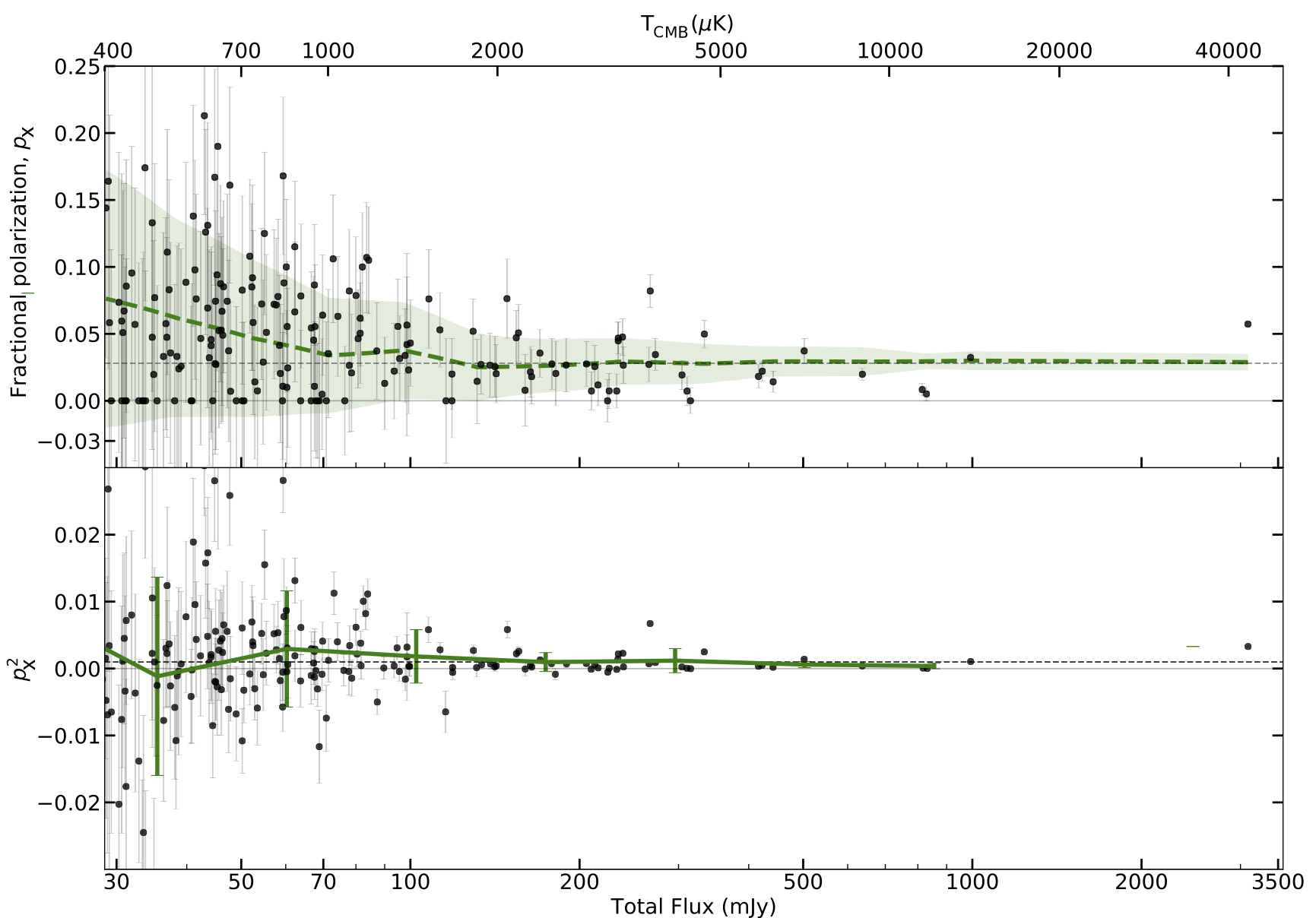} 
\caption{\protect \footnotesize {\it Top:} The green dashed line ($p$ = 0.028) shows the best fit model for the distribution of fractional polarization with the green band representing the $\pm1\sigma_{\mathrm{p}_\mathrm{m}}$ ($\sigma_{\mathrm{p}_\mathrm{m}} = 0.054$) uncertainty band. The data points represent sources with total intensity above 410 $\mu$K or total flux density above 30 mJy that were used in the fitting with individual error bars representing the 68$\%$ confidence interval. The data points at $p = 0$ correspond to sources for which the quantity $q_{\mathrm{1}}q_{\mathrm{2}} + u_{\mathrm{1}}u_{\mathrm{2}}$ is negative. The gray lines are at $p = 0$ (solid) and 0.028 (dashed). {\it Bottom:} Distribution of $p^\mathrm{2} = q_{\mathrm{1}}q_{\mathrm{2}} + u_{\mathrm{1}}u_{\mathrm{2}}$ from the 2-way maps as a function of total flux. The green line shows the data binned in total intensity with the error bars denoting the 1$\sigma$ standard deviation. The grey dashed line corresponds to $p^\mathrm{2} = 0.028^{2}$. }
\label{fig:p_best_fit2}
\end{center}
\end{figure*} 

Informed by simulations described in Section 4.3 below, we trust the accuracy of the de-biasing method when the total flux density of a source is greater than 215 mJy ($\sim$2900 $\mu$K). We have significant detection, defined by $p>3\sigma_{\mathrm{p}}$, of the polarization signal in 14 out of 26 such sources shown in the center ($5>p$/$\sigma_{\mathrm{p}}>3$) and top ($p$/$\sigma_{\mathrm{p}}>$5) panels of Fig.~\ref{fig:polfrac} with error bars that encompass the $95\%$ confidence interval. The red points represent measurements of fractional polarization de-biased using the first method, and green points correspond to estimates obtained using the second method. For sources with $p<3\sigma_{\mathrm{p}}$, we show the estimates of their fractional polarization in the bottom panel of Fig.~\ref{fig:polfrac}. The errors on the estimate of $p_{\mathrm{\times}}$ are larger than the errors on the estimate of $p_{\mathrm{WK}}$ because the $Q$ and $U$ errors are larger when using half the data as in the case of the 2-way maps compared to the full map. The error in the measurement of each source polarization fraction is obtained from a thousand realizations of Gaussian distributed $Q_\mathrm{1}$, $Q_\mathrm{2}$, $U_\mathrm{1}$, $U_\mathrm{2}$, $I_\mathrm{1}$, and $I_\mathrm{2}$ about the measured values giving a distribution for $p_{\mathrm{\times}}^\mathrm{2}$. The 95\% confidence interval for $p_{\mathrm{\times}}^\mathrm{2}$ is then obtained from a kernel density estimator applied to the distribution of $p_{\mathrm{\times}}^\mathrm{2}$. The errors are then propagated to errors on $p_{\mathrm{\times}}$. Table~\ref{tab:cat}{\color{blue}4} reports the fractional polarization of sources brighter than 74 mJy ($\sim$1000 $\mu$K) in total intensity, de-biased using the two methods. It should be noted however, that for the sources with total flux density between 74 and 215 mJy, the reported $p_{\mathrm{WK}}$, $p_{\times}$ values could potentially be overestimates, as discussed below. 
 
\subsection{Simulations}
Simulations of the measured quantities are used to statistically analyze the calculated fractional polarization and any residual bias after our noise-bias removal. First, sources detected in total intensity down to a flux density level of 5 mJy are masked out of the D56 Stokes $I$, $Q$ and $U$ maps. We again create a template map the same size as the D56 map where all pixels are set to zero. Random pixels are selected from this template map and assigned amplitudes drawn from the C2Co source counts model down to a flux level of 5 mJy. We call this the total intensity template map for simulated sources. Each source is assigned a known polarization fraction and a random polarization angle and the corresponding $Q$ and $U$ amplitudes are computed. With these, similar template maps are created for $Q$ and $U$. These three template maps are then convolved with the ACTPol beam and added to the source subtracted $I$, $Q$, and $U$ maps respectively. We then take these maps with simulated sources and estimate the polarization fractions for the sources in the same way as for the real maps. We run multiple realizations for a few different hypothetical source populations which we refer to as ``C'' (see Figs.~\ref{fig:psim1}--\ref{fig:psim2}) for sources with a constant polarization fraction $p_\mathrm{inj,c}$, and ``tG'' (see Fig.~\ref{fig:psim3}) for sources that are assigned a polarization fraction drawn from a Gaussian distribution with mean $p_\mathrm{inj,tG}$ and standard deviation $\sigma_{p_\mathrm{inj,tG}}$ truncated at $p_\mathrm{inj,tG}$ = 0 (see Figs.~\ref{fig:psim1}--\ref{fig:psim3}). We compare the two methods for noise bias correction and find that both methods work well for estimating the average polarization properties of a source population when either $I_{\mathrm{inj}}$ is greater than $\sim$215 mJy or both $p_{\mathrm{inj}}$ is greater than 3$\%$ and the polarized flux given by $P_{\mathrm{inj}} = p_{\mathrm{inj}}I_{\mathrm{inj}}$ is greater than $\sim$3 mJy. When $P < 3\ \mathrm{mJy}$, both methods leave a residual bias as shown in Fig.~\ref{fig:polflux}. We consider our measurement of the fractional polarization of individual sources to be unbiased only for the 26 sources with total flux density above 215 mJy. 
\begin{figure*}
\begin{center}
\includegraphics[width=0.7\linewidth,keepaspectratio]{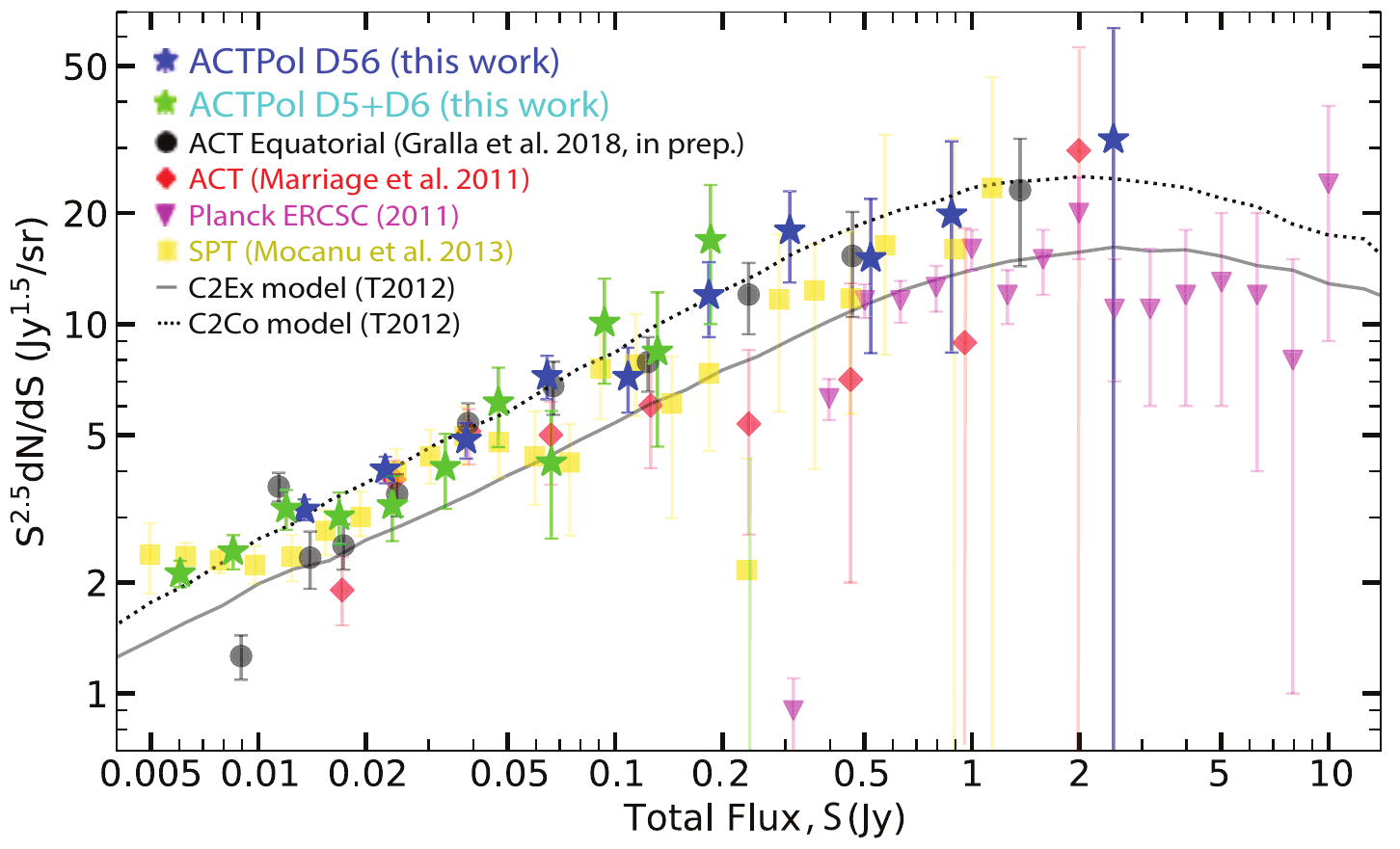} 
\caption{\protect \footnotesize Differential Source Number Counts in Total intensity. The ACTPol 148 GHz differential source number counts (corrected for incompleteness and purity, D56: blue stars, D5+D6: green stars) are plotted together with measurements from the ACT equatorial survey \citep{Gralla:2018} at 148 GHz (black circles), ACT southern survey \citep{2011ApJ...731..100M} at 148 GHz (red diamonds), Planck ERCSC \citep{2011A&A...536A...7P} at 143 GHz (magenta triangles), SPT \citep{2013ApJ...779...61M} at 148 GHz (gold squares), and the T2012 C2Ex model (solid gray line) and C2Co model (dotted black line) for radio and infrared source populations at 143 GHz. The error bars on the measurements are Poissonian. } 
\label{fig:ACTPol_dNdS}
\end{center}
\end{figure*}

\subsection{Distribution of Fractional Polarization}
In addition to measuring the polarization of the extragalactic point sources, we predict the distribution of their polarization fraction as a function of total intensity that is valid not only down to the sensitivity level of the D56 map but also extendable to lower flux limits. This helps predict the contribution of unsubtracted point sources to the E-mode polarization power spectrum at high-$\ell$. Earlier work by \cite{2002AA...396..463M} and \cite{2004MNRAS.349.1267T} presented fractional polarization distributions for NVSS sources, exhibiting a log-normal form. They further concluded that NVSS sources exhibited an anti-correlation between fractional polarization and total flux as did 1.4 GHz polarimetric studies of the ELAIS-N1 field by \cite{2007ApJ...666..201T}, \cite{2010ApJ...714.1689G}, and ATLBS fields by \cite{2010MNRAS.402.2792S}. However, \cite{2014MNRAS.440.3113H} argued that this apparent anti-correlation represented a selection bias. Since it is very hard to detect low levels of fractional polarization for the faint sources in total intensity, the average fractional polarization of detected polarized sources will always appear to increase with decreasing total flux. \cite{2014MNRAS.440.3113H} also found the distribution to be log-normal and that the maximum level of fractional polarization exhibited by ATLAS sources was not correlated with total flux density. \cite{2017MNRAS.469.2401B} implemented a stacking technique to estimate the average fractional polarization from 30 to 353 GHz of a primary sample of 1560 compact sources detected in the 30 GHz Planck all-sky map \citep{2016A&A...594A..26P} and described the distribution of fractional polarization as log-normal and independent of frequency. \cite{2017arXiv171208412T} extract polarization degrees of radio sources from measurements of signals in a map at the positions of a given source catalog and comparing with that for the blank sky, measured at random positions, away from sources. They find a median fractional polarization of 2.83\% across the Planck frequencies 30--353 GHz. \cite{2017arXiv171209639P} present a compendium of previous polarization measurements and using a log-normal fit to the distribution of polarization fractions, they show that most experiments are consistent with a level of polarization around 1--5\% largely independent of frequency. Our data is consistent with this level of polarization as described below.
\begin{figure}
\begin{center}
\includegraphics[width=1.0\linewidth,keepaspectratio]{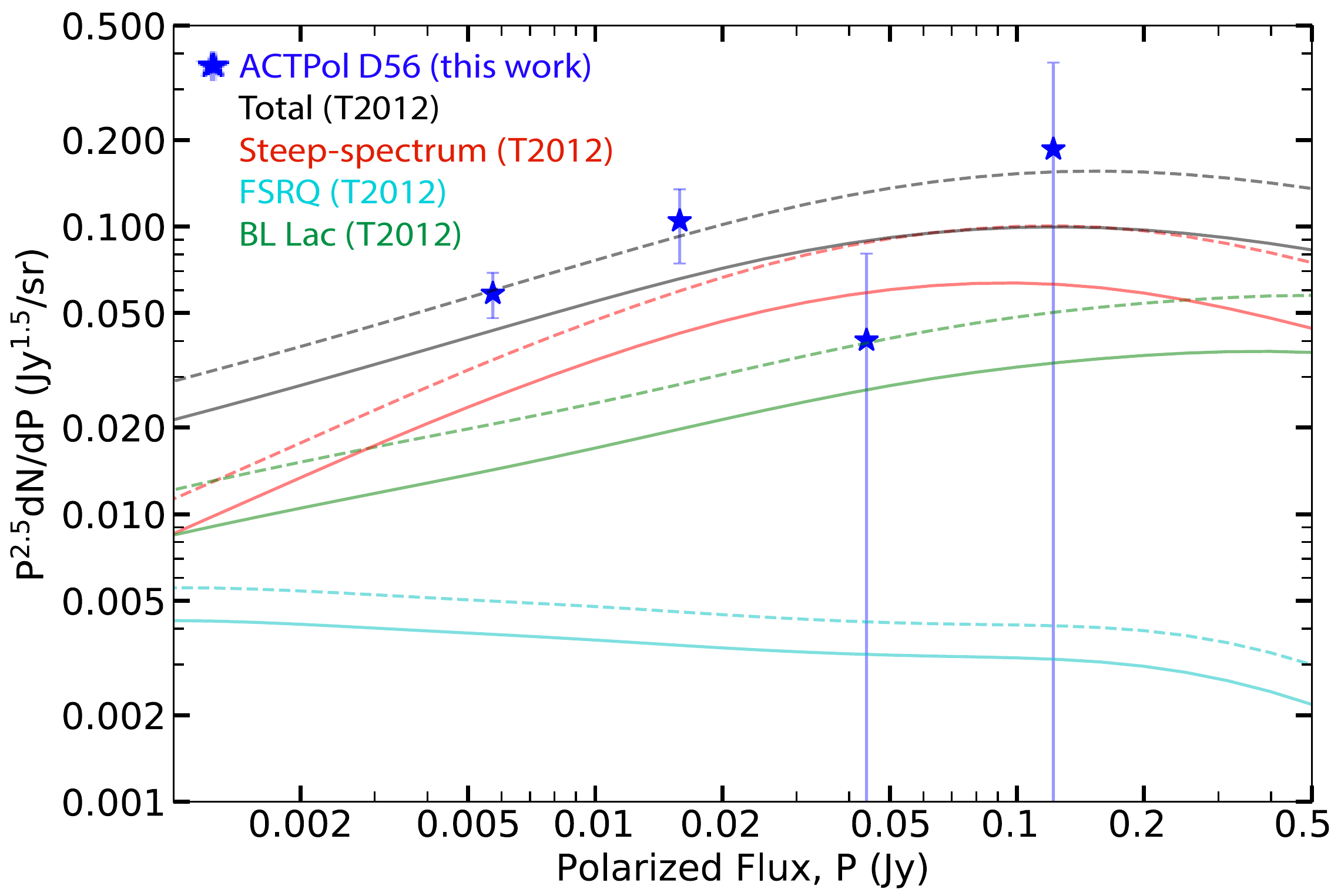} 
\caption{\protect \footnotesize Differential Source Number Counts in Polarized intensity. Comparison of the measured ACTPol source number counts with predictions from T2012 for different source populations: the solid lines represent their ``optimistic" model and the dashed lines represent their ``conservative" model. The error bars on the measurements are Poissonian. }
\label{fig:ACTPol_dNdP}
\end{center}
\end{figure}
In order to model the distribution of the fractional polarization $p$ of the AGN source population, we use simulations to statistically analyze the measured distribution of $p$ as a function of intensity. \cite{2011MNRAS.413..132B} argued that the distribution of fractional polarization $p$ could be modeled as a Gaussian in log($p$). In other words, the probability density function would follow a log-normal distribution (T2012) given by: 
\begin{flalign}
	\begin{gathered}
	 \ f(p;\mu_\mathrm{lN},\sigma_\mathrm{lN}) = \frac{1}{p \sqrt{2 \pi \sigma_\mathrm{lN}^{2}}} \exp \bigg[-\frac{(\mathrm{ln}(p/\mu_\mathrm{lN}))^{2}}{2 \sigma_\mathrm{lN}^{2}}\bigg]
	 \end{gathered}	 
\label{eq:logN}
\end{flalign}
\noindent where we fit for the values of $\mu_\mathrm{lN}$ and $\sigma_\mathrm{lN}$, the mean and the standard deviation in log, respectively. One can then obtain a good approximation of the fractional polarization using a combination of the log-normal parameters as:
\begin{center}
\begin{flalign}
	\begin{gathered}
	\ev{p_\mathrm{lN}} \approx \mu_\mathrm{lN} e^{\frac{1}{2}\sigma_\mathrm{lN}^{2}} \\  
	\ev{p_\mathrm{lN}^{2}} \approx \mu_\mathrm{lN}^{2} e^{2\sigma_\mathrm{lN}^{2}} \\
	p_\mathrm{lN,med} \approx \mu_\mathrm{lN} \\
	\end{gathered}
\label{eq:LogN}
\end{flalign}
\end{center}
Fig.~\ref{fig:src_model} shows a comparison of the best-fit log-normal distribution with parameters $\mu_\mathrm{lN}=0.027$ and $\sigma_\mathrm{lN}=0.572$ and the best-fit truncated Gaussian distribution with $\mu_{p,\mathrm{tG}}=0.027$ and $\sigma_{p,\mathrm{tG}}=0.025$ along with a histogram of the measured fractional polarization $p_\mathrm{x}$ for sources detected with total flux density $S>30$
 mJy and having a non-zero $p_\mathrm{\times}$. Both the fits have a $\chi^{2}$ per degree of freedom less than 1 and seem to describe our observations well. The estimates for $\ev{p_\mathrm{lN}}$, $\ev{p_\mathrm{lN}^{2}}^{1/2}$, and  $p_\mathrm{lN,med}$ from the log-normal parameters are 0.032, 0.038, and 0.027 respectively. In this work, we choose a truncated Gaussian model to describe our observed fractional polarization because it appears to capture the sources with very low measured levels of polarization better than the log-normal distribution.

We use the method described in Section 4.2 to recover the fractional polarization of injected sources from the ``C'' and ``tG'' simulations. The results for a few of these simulations are shown in Figs.~\ref{fig:psim1},~\ref{fig:psim2},~and~\ref{fig:psim3}. These simulations are then interpolated over values between $p_\mathrm{inj,c}$ = 0 and 0.10 and $\sigma_{p_\mathrm{inj,tG}}$ = 0.005 and 0.10. Our model source population consists of a constant polarization fraction component $p_\mathrm{m,c}$, and a variance $\sigma_{p_\mathrm{m}}^{2}$ as model parameters. While only the ``tG'' simulations have a standard deviation $\sigma_{p_\mathrm{inj,tG}}$ as a simulated parameter, both simulations have a variance in the distribution of the data points corresponding to the recovered $p$ values (see Figs.~\ref{fig:psim1},~\ref{fig:psim2},~and~\ref{fig:psim3}). The model parameter $\sigma_{p_\mathrm{m}}^{2}$ is the sum of these data space variances $\sigma_{p_\mathrm{rec,c}}^{2}$ and $\sigma_{p_\mathrm{rec,tG}}^{2}$ following the convention of adding uncorrelated errors in quadrature. The best fit to the measured fractional polarization is determined by optimizing over the parameters $p_\mathrm{m,c}$ and $\sigma_{p_\mathrm{m}}^{2}$ to minimize the $\chi^{2}$ as given by equation~(\ref{eq:chi_sq}). Since $\sigma_{p_\mathrm{m}}^{2}$ is a free parameter in our fits, we can lower the $\chi^{2}$ of our fit by increasing $\sigma_{p_\mathrm{m}}^{2}$ indefinitely, and hence we decide to stop when a $\chi^{2}$ of unity is reached. 
\begin{center}
\begin{flalign}
	\begin{gathered}
	\chi^{2} = \frac{1}{n}\sum_{i=1}^{n}\Bigg(\frac{p_{\mathrm{meas,}i} - p_{\mathrm{m,}i}}{\sigma_{p_{\mathrm{m,}i}}}\Bigg)^{2}\\
	\textnormal{where,} \; p_{\mathrm{m,}i} = p_{\mathrm{inj,c,}i} \\ 
	\textnormal{and,} \; \sigma_{p_{\mathrm{m,}i}} = \sqrt{\sigma_{p_{\mathrm{rec,tG,}i}}^{2} + \sigma_{p_{\mathrm{rec,c,}i}}^{2}} \\
	\end{gathered}
\label{eq:chi_sq}
\end{flalign}
\end{center}

We are essentially fitting for the intrinsic scatter in the fractional polarization of the source population we are investigating. We perform the fitting with sources whose total flux is greater than 30 mJy and measurement of $p$ de-biased using the 2-way maps. The best fit model is described by $p_\mathrm{m,c}$ = 0.028$\pm$0.005 (where the quoted uncertainty is the standard error on the mean) and $\sigma_{p_\mathrm{m}}$ = 0.054, the $\pm1\sigma_{\mathrm{p}_{\mathrm{m}}}$ error band is shown in the top panel of Fig.~\ref{fig:p_best_fit2}. This value of the mean fractional polarization is consistent with the best-fit value in Fig.~\ref{fig:src_model} and with $p_\mathrm{m} = 0.027\pm0.004$ for the 26 sources with total flux density greater than 215 mJy, whose estimates of fractional polarization are expected to be free of residual noise bias. Hence our findings are consistent with the scenario that the fractional polarization of the radio source population we are probing is independent of total intensity or total flux density down to the limits of our measurements. Our estimate for the average fractional polarization is consistent with the numbers typically in the range 1--5$\%$ reported by other mm-wave and radio surveys listed in Table~\ref{tab:Pol_surveys}. Deeper maps at 148 GHz together with maps at five frequencies spanning 24--280 GHz from subsequent seasons of the ACTPol and Advanced ACTPol (AdvACT) surveys will significantly improve the statistics on the polarization properties of extragalactic sources at millimeter wavelengths.  

\section{Comparison to other Source Catalogs} 
\label{sec:Pub_Cats}
\subsection{Differential source number counts}
The differential source number counts per steradian in total intensity d$N$/d$S$ from the ACTPol two-season catalogs are shown in Fig.~\ref{fig:ACTPol_dNdS} along with counts from other surveys at similar frequencies. The data are presented in Table~\ref{tab:dNdS}. The number counts are plotted separately for the S14 D56 and the S13+S14 D5+D6 maps, the latter being much deeper. For flux densities lower than 40 mJy, the ACTPol counts shown are deboosted and corrected for incompleteness and expected number of false detections. We compare our counts to the T2012 C2Ex and C2Co models, which are constructed based on extrapolations of number counts from high radio frequencies (5 GHz) and observed counts from SPT \citep{2010ApJ...719..763V, 2013ApJ...779...61M} and ACT \citep{2011ApJ...731..100M}, building on the earlier models proposed in \cite{2005A&A...431..893D}. These models consider the spectral behavior of the different source populations, flat-spectrum (FSRQs and BL Lac), steep-spectrum and inverted spectrum, in a statistical way and feature different distributions of spectral break frequencies for FSRQs and BL Lacs, and the effects of spectral steepening. \cite{2012AdAst2012E..52T} found that the C2Ex model was the best fit to then available observational counts above 100 GHz. Both the ACTPol (this work) and the ACT equatorial \citep{Gralla:2018} counts favor the C2Co model over the C2Ex model favored by the Planck counts \citep{2011A&A...536A...7P} at 148 and 220 GHz for sources above 500 mJy. The reduced $\chi^{2}$ of the fit of the C2Ex and C2Co models to our D56/(D5+D6) counts is 5.0/1.9 and 2.8/0.4, respectively. \cite{2013ApJ...779...61M} reported that the C2Ex model is a good fit to the SPT counts above 80 mJy and below 20 mJy in all bands (95, 150, and 220 GHz), but under-predicts the counts in the intermediate flux range at 95 and 150 GHz. 

The differential number counts per steradian in polarized intensity d$N$/d$P$ as a function of the polarized flux $P$ are compared with the T2012 models for different source populations in Fig.~\ref{fig:ACTPol_dNdP}. Sources detected in total intensity with S/N greater than 5 and total flux density greater than 74 mJy with noise-debiased polarized flux greater than 3 mJy were included to derive the d$N$/d$P$ here. \cite{2012AdAst2012E..52T} proposed two models (a ``conservative" case and an ``optimistic" case) for each source population. They also reported that for observational counts at frequencies below 40 GHz, the ``optimistic" model is a good fit, especially when the polarized flux $P$ is below 50 mJy, whereas the ``conservative" model fits the observational counts well at high polarized fluxes. Based on this work, we find that the ``conservative" model for all source populations combined fits our data only marginally better with a reduced $\chi^{2}$ of 1.3 compared to 1.4 for the ``optimistic" model, shown in Fig.~\ref{fig:ACTPol_dNdP}. More polarization data are required to better constrain the d$N$/d$P$ models.

\begin{center}
\begin{table*}
\caption{\protect Completeness and purity corrected differential source number counts at 148 GHz for synchrotron-dominated AGN population}
\begin{tabular}{| ll | ll | ll |}
\hline
\multicolumn{1}{|c}{D5+D6:} & Total Flux & \multicolumn{1}{c}{D56:} & Total Flux &  \multicolumn{1}{c}{D56:} & Polarized Flux \\
\hline
\hline
Tot Flux, $S$ & $S^{2.5}$ d$N$/d$S$  & Tot Flux, $S$ & $S^{2.5}$ d$N$/d$S$ & Pol Flux, $P$ & $P^{2.5}$ d$N$/d$P$ \\ 
(mJy) & (Jy$^{1.5}$sr$^{-1}$) & (mJy) & (Jy$^{1.5}$sr$^{-1}$) & (mJy) & (Jy$^{1.5}$sr$^{-1}$) \\
\hline
5.0--7.1 & 2.1$\pm$0.2 & 10.0--16.9 & 3.2$\pm$0.2 & 3.4--9.5 & 0.06$\pm$0.01  \\  
7.1--9.9 & 2.4$\pm$0.3 & 16.9--28.5 & 4.0$\pm$0.3 & 9.5--26.4 & 0.10$\pm$0.03  \\  
9.9--14.0 & 3.2$\pm$0.4 & 28.5--48.0 & 4.9$\pm$0.5 & 26.4--73.2 & 0.04$\pm$0.04  \\   
14.0--19.7 & 3.0$\pm$0.5 & 48.0--80.9 & 7.2$\pm$1.0 & 73.2--203 & 0.19$\pm$0.19 \\    
19.7--27.8 & 3.2$\pm$0.7 & 80.9--136 & 7.2$\pm$1.4 & & \\   
27.8--39.1 & 4.1$\pm$0.9 & 136--230 & 12.0$\pm$2.8 & & \\    
39.1--55.1 & 6.1$\pm$1.5 & 230--388 & 17.9$\pm$5.0 & & \\    
55.1--77.5 & 4.2$\pm$1.6 & 388--654 & 15.1$\pm$6.8 & & \\      
77.5--109 & 10.1$\pm$3.2 & 654--1102 & 19.8$\pm$11.4 & & \\ 
109--154 & 8.4$\pm$3.8 & 1858--3132 & 31.7$\pm$31.7 & & \\      
154--217 & 16.9$\pm$6.9 & & & & \\ 
\hline	                         
\end{tabular}\\
\end{table*}
\end{center}
\label{tab:dNdS}	

\subsection{Astrometry}
\begin{figure}
\begin{center}
\includegraphics[width=0.8\linewidth,keepaspectratio]{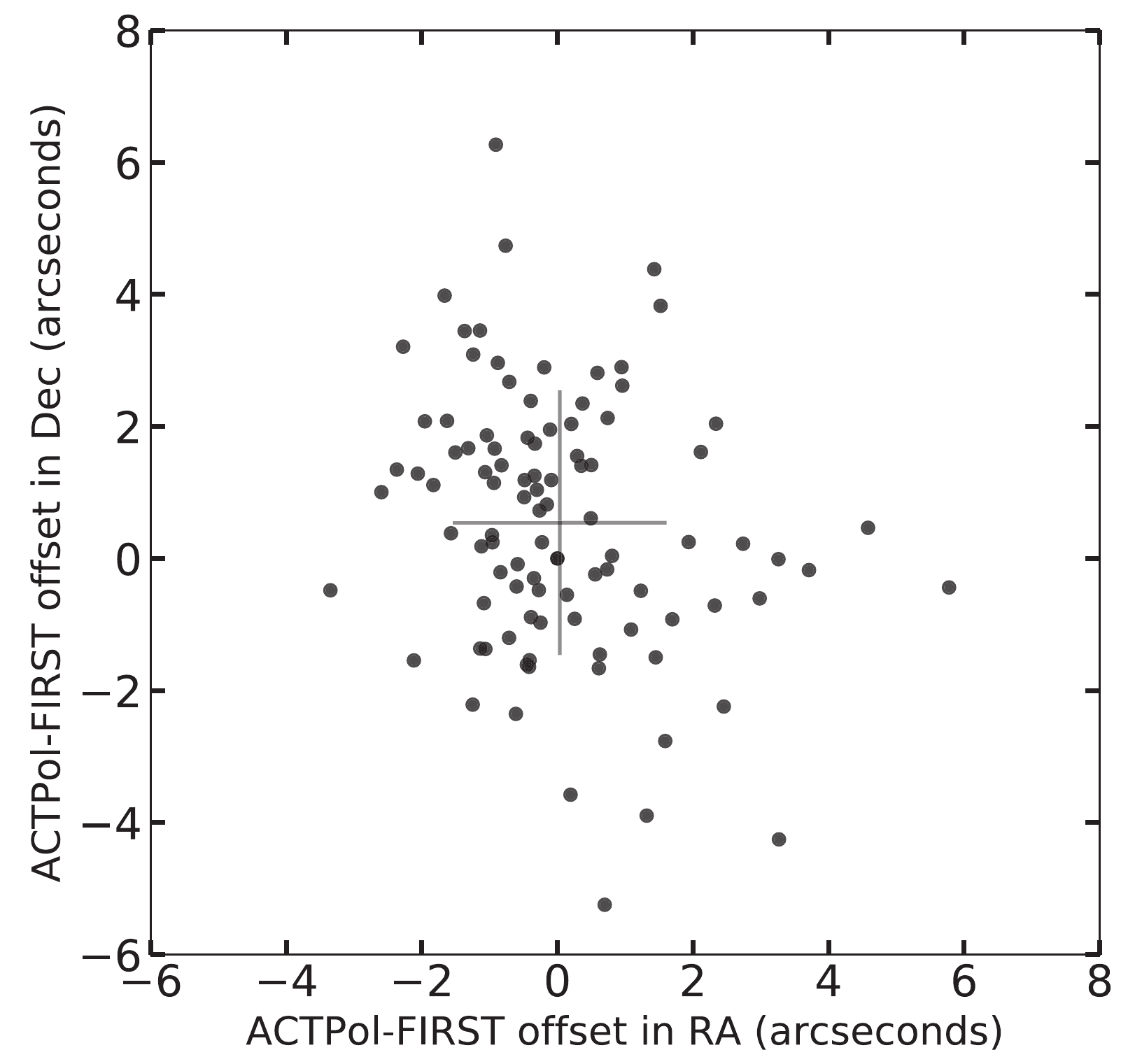}
\caption{\protect \footnotesize Positional offsets between the ACTPol D56 catalog and the VLA-FIRST catalog obtained by comparing the positions of 99 ACTPol sources detected with flux greater than 50 mJy and their VLA-FIRST counterparts. The cross is centered on the mean ACTPol-FIRST offset (0.04$^{\prime}$$^{\prime}$, 0.54$^{\prime}$$^{\prime}$) with lengths along either axis indicating the rms offset (1.55$^{\prime}$$^{\prime}$, 1.98$^{\prime}$$^{\prime}$). }
\label{fig:rms_offset}
\end{center}
\end{figure}
The accuracy of the absolute positions of the sources in the catalog is checked by cross-matching against radio sources listed in the Very Large Array - Faint Images of the Radio Sky at Twenty-centimeters (VLA-FIRST) survey \citep{1995ApJ...450..559B}, in which the astrometry of the maps is accurate to 0.05$^{\prime}$$^{\prime}$, and individual sources at the 3 mJy level have 90$\%$ confidence error circles of radius less than 0.5$^{\prime}$$^{\prime}$. We compare the positions of 99 of the brightest ($S\mathrm{ > 50 mJy}$) sources detected in the D56 intensity map to positions of sources in the VLA-FIRST catalog within a search radius of 10$^{\prime}$$^{\prime}$. The distribution of the positional offsets in right ascension (RA) and declination (Dec) are shown in Fig.~\ref{fig:rms_offset}. The errorbars are centered on the mean offsets, and extend as far as the 1$\sigma$ standard deviation (rms) of the offsets. The mean offsets in RA and Dec are 0.04$\pm$0.16$^{\prime}$$^{\prime}$ and 0.54$\pm$0.20$^{\prime}$$^{\prime}$ respectively (where the quoted uncertainty is the standard error on the mean, i.e., $\sigma$/$\sqrt{N}$, where $N$= 99). The rms offset in RA and Dec are 1.55$^{\prime}$$^{\prime}$ and 1.98$^{\prime}$$^{\prime}$ respectively. The rms astrometry error is thus consistent with zero pointing offset in RA and a less than 1$^{\prime}$$^{\prime}$ pointing offset in Dec. 

\subsection{Cross-matching with other catalogs}
We cross-match the D56 sources with catalogs from observations at other frequencies. Matches are established within a given radius about a source. The radius is chosen based on the predicted positional errors of the catalog being cross-matched with, the beam size of the instruments, and by looking at the distribution of offsets between D56 sources and their closest counterpart in each catalog. We consider the following catalogs from surveys that overlap with the D56 region: NVSS and VLA-FIRST at 1.4 GHz, AT20G at 20 GHz, ACT Equatorial at 148, 218, and 277 GHz, and Herschel SPIRE at 600 and 1199 GHz. 
\begin{figure*}
\begin{center}
\includegraphics[width=0.9\linewidth,keepaspectratio]{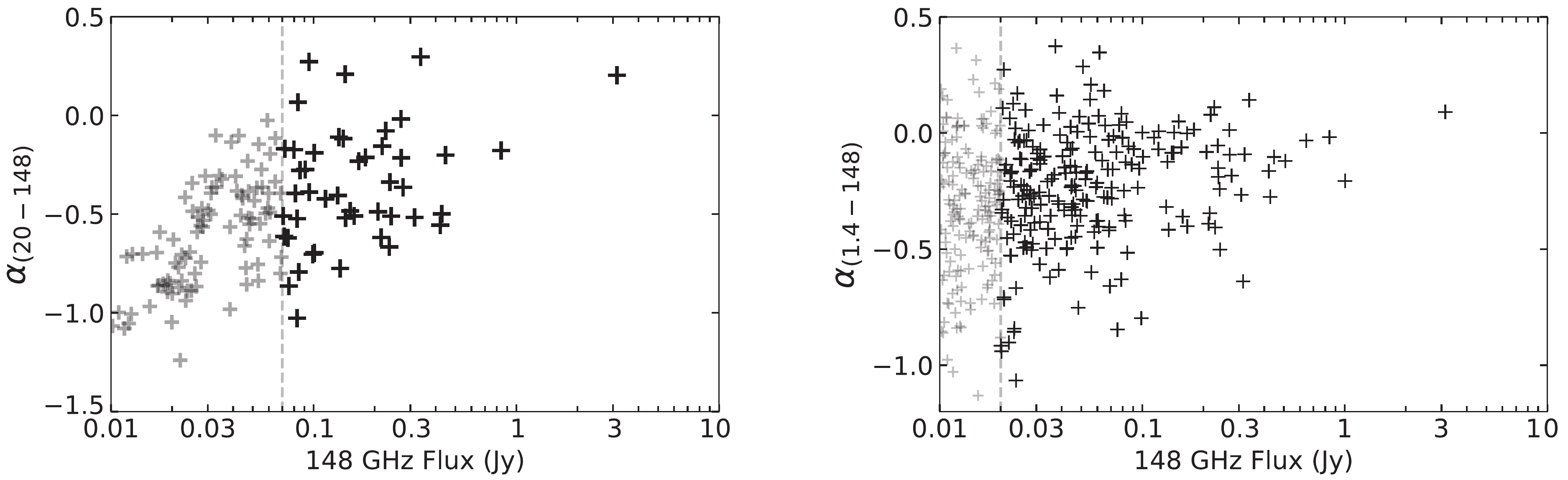}
\caption{\protect \footnotesize Spectral indices between 20--148 GHz (left) and 1.4--148 GHz (right) plotted against measured total flux at 148 GHz for ACTPol-AT20G and ACTPol-VLA FIRST cross-identified sources with 148 GHz flux density greater than 10 mJy. The black pluses represent sources brighter than 70 mJy (left) and 20 mJy (right) at 148 GHz. In the left panel, the vertical gray dashed line is at $S_\mathrm{148}$ = 70 mJy, below which the data suffers from a selection bias owing to the incompleteness of the AT20G catalog that excludes sources with flat or peaked spectra. In the right panel, the vertical gray dashed line is at $S_\mathrm{148}$ = 20 mJy, below which the 148 GHz ACTPol data is incomplete.} 
\label{fig:spec_148}
\end{center}
\end{figure*}

NVSS at 1.4 GHz: The National Radio Astronomy Observatory (NRAO) Very Large Array (VLA) Sky Survey, also known as NVSS, covers the sky north of declination --40$^{\circ}$ at 1.4 GHz. The NVSS catalog consists of almost 2 million discrete sources \citep{1998AJ....115.1693C} brighter than a flux threshold of about 2.5 mJy observed between Sep 1993 and Oct 1996 with a resolution of 17.5$^{\prime}$$^{\prime}$. The root mean square uncertainties in RA and Dec vary from $\lesssim$ 1$^{\prime}$$^{\prime}$ for the 400,000 sources stronger than 15 mJy to 7$^{\prime}$$^{\prime}$ at the survey limit. The catalog also reports measurements of the polarized flux. Among the 103 D56 sources brighter than 50 mJy, 97 matches are found in the NVSS catalog within a matching radius of 10$^{\prime}$$^{\prime}$. We find 456 matches for the 493 D56 sources with deboosted flux greater than 10 mJy within a matching radius of 15$^{\prime}$$^{\prime}$. Based on tests of the catalog purity, less than 15 false detections with flux greater than 10 mJy are expected in the D56 catalog. 

VLA-FIRST at 1.4 GHz: The VLA-FIRST survey \citep{1995ApJ...450..559B}. Data for the FIRST survey were collected from Spring 1993 through Spring 2004 using an elliptical beam with an approximate 6.5$^{\prime}$$^{\prime}$ \ $\times$ \ 5.4$^{\prime}$$^{\prime}$ FWHM. The astrometry of the maps is accurate to 0.05$^{\prime}$$^{\prime}$, and individual sources have 90$\%$ confidence error circles of radius $<0.5^{\prime}$$^{\prime}$ at the 3 mJy level. The catalog contains close to a million sources. Among the 103 D56 sources brighter than 50 mJy, 99 matches are found in the VLA-FIRST catalog within a matching radius of 10$^{\prime}$$^{\prime}$. We find 470 (317) matches for the 493 (337) D56 sources with deboosted flux greater than 10 (15) mJy within a matching radius of 15$^{\prime}$$^{\prime}$. The right panel of Fig.~\ref{fig:spec_148} shows the 1.4 (VLAF) -- 148 (ACTPol) GHz source spectral indices plotted against the measured 148 GHz flux density. The four D56 sources brighter than 50 mJy with no matches in either VLA-FIRST or NVSS are ACTPol-D56 J005735--012334 (an extended source), J233757--023057, J021921--025842, and J010643--031536. Of these, only J233757--023057 has a match in the Herschel SPIRE 600 and 1199 GHz catalogs. 

AT20G at 20 GHz: The Australia Telescope 20 GHz Survey (AT20G) is a blind radio survey carried out at 20 GHz with the Australia Telescope Compact Array (ATCA) from 2004 to 2008, and covers the whole sky south of Dec 0$^{\circ}$. The AT20G source catalog described in \cite{2010MNRAS.402.2403M} includes 5890 sources above a flux-density limit of 40 mJy. All AT20G sources have total intensity and polarization measured at 20 GHz. The declination band 0$^{\circ}>$ Dec $>-15^{\circ}$ was observed during Aug-Sep 2007 with an approximate beam size of 30$^{\prime}$$^{\prime}$. Out of 381 D56 sources south of declination 0$^{\circ}$ that are brighter than 10 mJy, 137 matches were found within 10$^{\prime}$$^{\prime}$ and 147 within 30$^{\prime}$$^{\prime}$ in the AT20G catalog. All of the 46 D56 sources brighter than 70 mJy were cross-identified within AT20G sources within a search radius of 10$^{\prime}$$^{\prime}$. Of these 46 sources in the AT20G catalog, 15 have polarization measurements and 20 sources have upper limits on their polarization. The left panel of Fig.~\ref{fig:spec_148} shows the 20 (AT20G) -- 148 (ACTPol) GHz source spectral indices plotted against the measured 148 GHz flux density. The low flux density ($S_{148} <$ 70 mJy) data suffers from a selection bias owing to the incompleteness of the AT20G catalog that excludes sources with flat or peaked spectra. The sources with $S_{148} >$ 70 mJy represent a complete sample.  

ACT Equatorial Survey at 148, 218, and 277 GHz: ACT conducted surveys of the equatorial sky from 2007 to 2010, observing at three frequencies simultaneously: 148 GHz ($\sim$2.0 mm), 218 GHz ($\sim$1.4 mm) and 277 GHz ($\sim$1.1 mm) with angular resolutions of 1.4$^\prime$,1.0$^\prime$, and 0.9$^\prime$, respectively. The point source catalog from this survey \citep{Gralla:2018} includes 148 and 218 GHz data from the 2009 and 2010 observing seasons, and 277 GHz data from the 2010 observing season. We find 87 cross-matched sources (see Fig.~\ref{fig:Var1}) in the overlapping region between D56 and the ACT equatorial survey that are detected in all three ACT frequency bands by the multifrequency matched filter method described in \cite{Gralla:2018}. Of these, only five (marked with a $\dagger$ symbol in Table~\ref{tab:DSFG}, see below) have dusty spectra identified by $\alpha_{148-217} > 1.0$ in \cite{Gralla:2018}. Of these, two sources J024145+002645 (an extended source, see Fig.~\ref{fig:residuals}) and J020941+001549 have counterparts in the 1.4 GHz NVSS catalog. Dusty sources can have a strong synchrotron component from electrons accelerated in supernovae, the so-called IR-radio correlation \citep{1985ApJ...298L...7H}. The FIRST counterpart could be associated with this synchrotron emission.  

Herschel SPIRE at 600 and 1199 GHz: The Spectral and Photometric Imaging Receiver (SPIRE) instrument onboard Herschel mapped about 9\% of the sky at 250, 350, and 500 $\mu$m (1199, 857, and 600 GHz) with spatial resolutions of 17.9$^{\prime}$$^{\prime}$, 24.2$^{\prime}$$^{\prime}$ and 35.4$^{\prime}$$^{\prime}$, respectively. The SPIRE 1199 and 600 GHz catalogs \citep{Schulz:2017} consist of nearly a million and two hundred thousand objects, respectively. Among the 493 D56 sources brighter than 10 mJy, 32 and 28 matches are found in the SPIRE 1199 and 600 GHz catalogs within a radius of 10$^{\prime}$$^{\prime}$. Table~\ref{tab:DSFG} lists potential DSFG candidates detected at 148 GHz in the D56 intensity map with $\mathrm{S/N}>4$ and $S_\mathrm{148}>10$ mJy with counterparts in the non-contemporaneous SPIRE 250 $\mu$m or 500 $\mu$m catalogs and $\alpha_\mathrm{148-600} \gtrsim$ 1.0 or $\alpha_\mathrm{148-1199} \gtrsim$ 1.0. The brightest of these sources is J020941+001549 which is a gravitationally lensed hyperluminous infrared radio galaxy at redshift $z$ = 2.553 \citep{2015MNRAS.452..502G}; it does not have a detectable polarization signal at 148 GHz. The next brightest source is the extended source J024145+002645; its measured polarization fraction at 148 GHz is 2.8$_\mathrm{-3.8}^\mathrm{+4.9}\%$. The ACTPol-SPIRE cross-matched sources were cross-checked with the SDSS image database \footnote{http://skyserver.sdss.org/dr14/en/tools/chart/listinfo.aspx}. Some of these sources were found to be nearby galaxies and were excluded from the table. Contemporaneous multi-frequency AdvACT data will throw more light on the spectra of these sources. 
\begin{center}
\begin{table*}
\caption{\protect Potential DSFG candidates detected in the ACTPol 148 GHz two-season map and cross-matched in the Herschel SPIRE catalogs. Spectral indices in the table are computed using flux values at 1.4, 600, and 1199 GHz from NVSS, SPIRE 500 $\mu$m, and SPIRE 250 $\mu$m catalogs, respectively, but from observations that are not contemporaneous with the ACTPol D56 survey. The blank entries correspond to no cross-matches. The sources marked with a $\dagger$ symbol have dusty spectra identified by $\alpha_\mathrm{148-217} > 1.0$ in Gralla and those marked with a $^{*}$ symbol are described in further detail there.}
\begin{tabular}{| l| l| l| l| l| l| l| l| l| l}
\hline
\multicolumn{1}{|c|}{\textbf{ID}} & \multicolumn{1}{c|}{\textbf{RA}} & \multicolumn{1}{c|}{\textbf{Dec}} & \multicolumn{1}{c|}{\textbf{$\mathrm{S}$/$\mathrm{N}$}} & \multicolumn{1}{c|}{\textbf{$S_\mathrm{148}$}} & \multicolumn{1}{c|}{\textbf{$\alpha_\mathrm{1.4}^\mathrm{148}$}} & \multicolumn{1}{c|}{\textbf{$\alpha_\mathrm{148}^\mathrm{600}$}} & \multicolumn{1}{c|}{\textbf{$\alpha_\mathrm{148}^\mathrm{1199}$}} \\ 

\multicolumn{1}{|c|}{\textbf{ACTPol-D56}} & \multicolumn{1}{c|}{\textbf{hh:mm:ss}} & \multicolumn{1}{c|}{\textbf{deg:min:sec}} & \multicolumn{1}{c|}{\textbf{}} & \multicolumn{1}{c|}{\textbf{mJy}} & \multicolumn{1}{c|}{\textbf{}} & \multicolumn{1}{c|}{\textbf{}} & \multicolumn{1}{c|}{\textbf{}}  \\ 
\hline
J004533--000115$^{\dagger}$* &     00:45:32.7 &  --00:01:15.0 & 5.4 & 10.8 & & 1.5 & \\ 
J005950--073436 &    00:59:49.9 &  --07:34:36.3 & 8.9 &  17.5  & --0.4  & 2.7 &\\ 
J011248--001721$^{\dagger}$ &     01:12:48.3 &  --00:17:20.6 & 6.8  & 13.9 & --0.1 & 1.3 & 1.7 \\ 
J011255+005904$^{\dagger}$ &     01:12:54.6 &  +00:59:03.9 & 5.4 & 10.2  &  & 2.1 & \\ 
J011531--005147$^{\dagger}$ &     01:15:30.9 &  --00:51:47.0 & 5.1 & 10.3  & & 2.1 & 1.7 \\ 
J020941+001549$^{\dagger}$* &     02:09:41.0 &  +00:15:48.8 & 8.6  & 15.1 & 0.3 & 2.8 & 2.0 \\ 
\hline
\end{tabular}\\
Sources detected at 148 GHz with $\mathrm{S/N}>4$ and $S_{148}<10$ mJy and with counterparts in the SPIRE 500 $\mu$m or 250 $\mu$m catalogs \\
\begin{tabular}{| l| l| l| l| l| l| l| l| l| l}
\hline
\multicolumn{1}{|c|}{\textbf{ID}} & \multicolumn{1}{c|}{\textbf{RA}} & \multicolumn{1}{c|}{\textbf{Dec}} & \multicolumn{1}{c|}{\textbf{$\mathrm{S}$/$\mathrm{N}$}} & \multicolumn{1}{c|}{\textbf{$S_\mathrm{148}$}} & \multicolumn{1}{c|}{\textbf{$\alpha_\mathrm{1.4}^\mathrm{148}$}} & \multicolumn{1}{c|}{\textbf{$\alpha_\mathrm{148}^\mathrm{600}$}} & \multicolumn{1}{c|}{\textbf{$\alpha_\mathrm{148}^\mathrm{1199}$}} \\ 

\multicolumn{1}{|c|}{\textbf{ACTPol-D56}} & \multicolumn{1}{c|}{\textbf{hh:mm:ss}} & \multicolumn{1}{c|}{\textbf{deg:min:sec}} & \multicolumn{1}{c|}{\textbf{}} & \multicolumn{1}{c|}{\textbf{mJy}} & \multicolumn{1}{c|}{\textbf{}} & \multicolumn{1}{c|}{\textbf{}} & \multicolumn{1}{c|}{\textbf{}}  \\ 
\hline
J000202--010457    &      00:02:02.4 &  --01:04:56.5   &  4.1  & 8.7 & --0.2 & & 0.9 \\
J000553--050604    &      00:05:52.7 &  --05:06:03.6   &  4.3  &  8.3 & & 1.4 &  1.2 \\
J002116+013314    &      00:21:15.8 &  +01:33:13.8  &  4.4  &  8.1  & & 1.6 & \\
J003253--034445    &      00:32:53.0 &  --03:44:45.2   &  4.1  &  8.4 & & 1.5 & 1.2 \\
J003815--002254*    &      00:38:14.5 &  --00:22:54.4   &  4.0  &  8.1 & & 2.0 & 1.2\\
J021402--004625    &    02:14:02.2  &   --00:46:24.5   & 5.4  &  3.8  & & 2.5 & 1.6 \\
J021830--053144    &  02:18:30.1    &  --05:31:44.0    & 6.6  &  4.3 & & 2.5 & 1.5\\
J021941--025535    &   02:19:40.7   &   --02:55:34.6    & 4.7  &  3.0  & & & 1.0\\
J022422--044455    &   02:24:21.9   &   --04:44:55.2   & 4.9  &   3.2 & & & 1.2 \\
J022552--045028    &   02:25:52.1   &   --04:50:28.2    & 4.6  &   3.1  & & & 1.0 \\
J022946--034102    &   02:29:45.6   &   --03:41:01.6   & 4.7  &   3.2  & & & 1.6 \\
J233426--042514    &      23:34:26.0 &   --04:25:13.9  &  4.0   &   8.4  & & 1.4 & \\
J234051--041930   &      23:40:50.7   &   --04:19:29.8   &  4.8  &  9.4  & & 2.2 & 1.4\\
\hline
\end{tabular}\\
\end{table*}
\end{center}
\label{tab:DSFG}
\subsection{Spectral Energy Distribution and Color-color diagram}
\begin{figure}
\begin{center}
\includegraphics[width=0.8\linewidth,keepaspectratio]{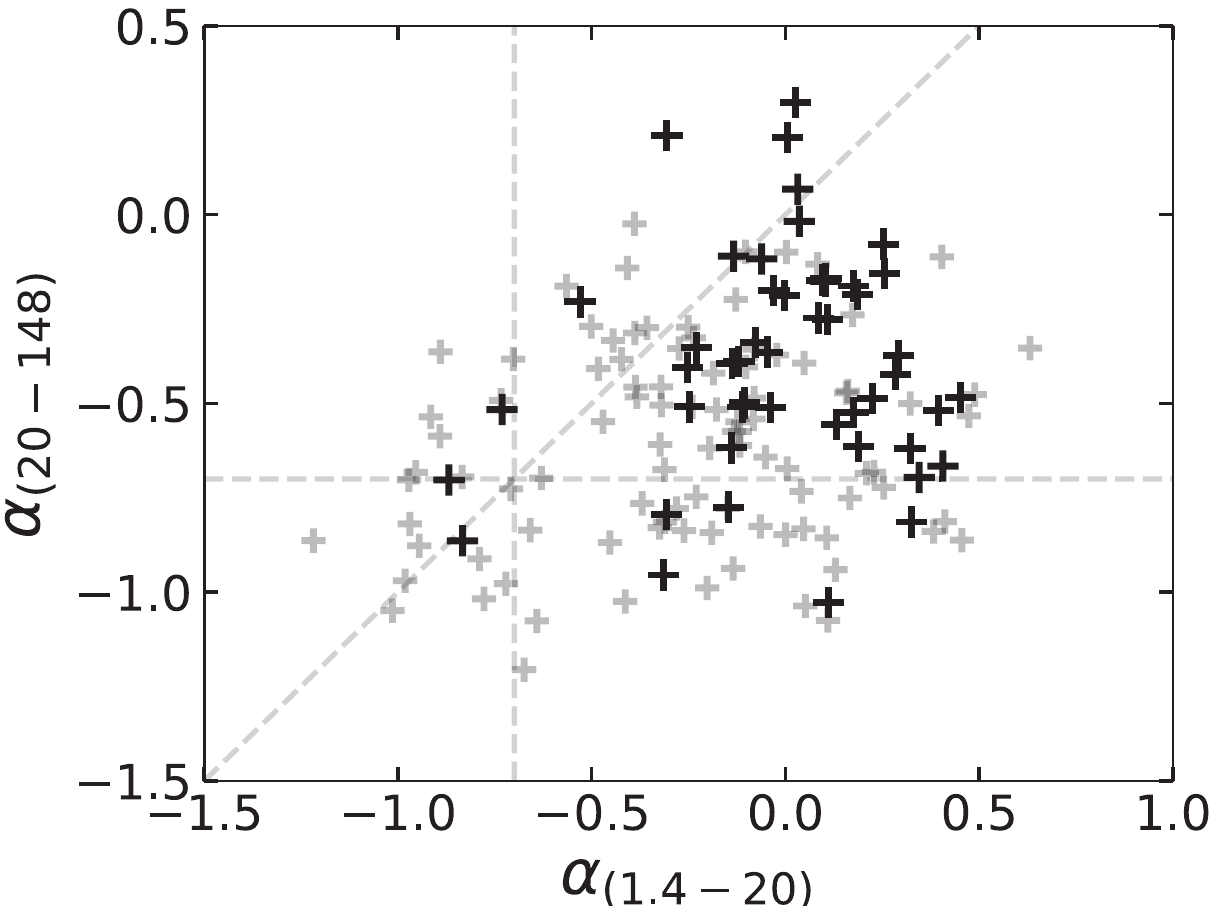}
\caption{\protect \footnotesize Color-color diagram: 1.4--20 GHz spectral indices are plotted against 20--148 GHz spectral indices for 148 ACTPol-AT20G-VLA FIRST cross-identified sources with measured total flux at 148 GHz greater than 20 mJy. The black pluses represent sources that are brighter than 70 mJy at 148 GHz, while the grey pluses represent dimmer sources. The horizontal and vertical gray dashed lines are at $\alpha=-0.7$ (synchrotron sources). The diagonal gray dashed line corresponds to $\alpha_\mathrm{1.4-20}=\alpha_\mathrm{20-148}$.}
\label{fig:spectra}
\end{center}
\end{figure} 
\begin{figure*}
\begin{center}
\includegraphics[width=1.0\linewidth,keepaspectratio]{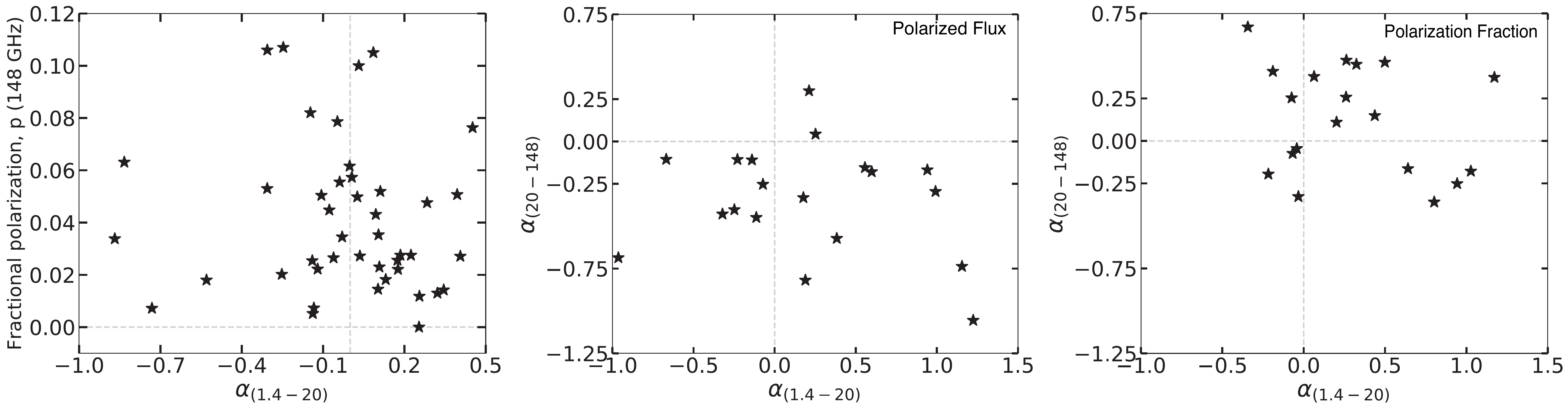} 
\caption{\protect \footnotesize {\it Left:} Fractional polarization at 148 GHz plotted against 1.4--20 GHz spectral index for ACTPol-AT20G-NVSS cross-identified sources with total flux greater than 70 mJy (where the AT20G sample is complete) at 148 GHz. {\it Center and Right:} 1.4--20 GHz and 20--148 GHz spectral indices are shown for ACTPol-AT20G-NVSS cross-identified sources with measurements of polarized flux at all three frequencies. The center panel plots the indices derived from polarized flux $P$ whereas the right panel plots the indices derived from fractional polarization $p$. There is no apparent correlation between the polarized flux spectral indices or the fractional polarization spectral indices in the color-color diagrams.}
\label{fig:ACTPol_NVSS_AT20G_CCD}
\end{center}
\end{figure*}
Spectral energy distributions (SEDs) and color-color comparison of spectral indices have been widely used to classify sources based on their dominant emission mechanisms \citep{2010MNRAS.402.2403M}. The ACTPol 148 GHz catalog is dominated by synchrotron-dominated blazars, which have variable flux densities, often not periodic. Hence, the inferred spectrum of a given source will depend on the epoch of observation, although not biased one way or the other. Here, we are deriving spectral indices from AT20G, VLA-FIRST, and ACTPol surveys which are not contemporaneous, which complicates things further. However, treating the catalog as a whole, a spectral study may give rise to insights about the average spectral behavior of the source population being probed. 

The mean spectral index $\alpha_{20-148}^\mathrm{mean}$ of the complete unbiased sample ($S>$70 mJy) in the left panel of Fig.~\ref{fig:spec_148} is $-0.36\pm0.3$ ($\sigma$/$\sqrt{N}=0.04$). This is consistent with the findings of \cite{2014MNRAS.439.1556M}, who found that synchrotron-classified ACT-selected sources that were followed up with ATCA (Australia Telescope Compact Array) had a mean spectral index $\alpha_{20-148}^\mathrm{mean}$ = $-0.42^{+0. 32}_{-0. 26}$. This ACT sample was completely independent of the ACTPol D56 sample and indicates that the same radio source population is being probed down to the low flux end. There are 5 sources with $\alpha_{20-148} > 0$: J000614--062337, J001736--051243, J020207--055902, J001752--023616, and J005735--012334. Of these, only J001736--051243 has a match in the Herschel SPIRE catalog. This source has also been identified as a DSFG in \cite{Gralla:2018} defined by $\alpha_{1.4-148} > 1.0$. 

The mean spectral index $\alpha_{1.4-148}^\mathrm{mean}$ is --0.21 $\pm$ 0.25 ($\sigma$/$\sqrt{N}$ = 0.03) when only considering sources brighter than 20 mJy at 148 GHz, and $\alpha_\mathrm{1.4-148}^\mathrm{mean}$ is --0.24 $\pm$ 0.26 ($\sigma$/$\sqrt{N}$ = 0.02) when considering all cross-matched sources brighter than 10 mJy at 148 GHz. 

In Fig.~\ref{fig:spectra}, we show a color-color comparison of 1.4--20 GHz versus 20--148 GHz spectral indices for the D56 sources detected above a flux threshold of 20 mJy and cross-identified with VLA-FIRST and AT20G catalogs. The incompleteness of the 20 GHz catalog does not only bias the 20--148 GHz spectral index distribution to steeper spectra, it also affects the 1.4--20 GHz spectral indices. If the 20 GHz catalog threshold were lower, one would see many more sources with steeper 1.4--20 GHz spectral index. The presence of many sources in the $\alpha_{1.4-20} > 0$, $\alpha_{20-148} < 0$ region is indicative of sources with a GHz peaked spectra not characterized by a single power law, as was also argued by \cite{2010A&ARv..18....1D}. It follows that the conventional classification of AGN-powered radio sources into flat and steep spectrum assuming a single power-law spectra is inaccurate at high radio frequencies. One of many factors \citep{2005A&A...432...31T} that could lead to a more complex spectrum is a transition of the synchrotron emission from an optically thick self-absorbing regime characterized by a rising spectrum to an optically thin standard synchrotron regime characterized by a falling spectrum. These sources could have spectra peaking at frequencies of a few to tens of GHz. 

Based on the spectral index $\alpha_{1.4-20}$, we can classify the ACTPol-AT20G-VLAF cross-matched sources into steep ($\alpha_{1.4-20} < -0.5$) and flat ($\alpha_{1.4-20} > -0.5$) spectrum. The population probed here is predominantly comprised of flat spectrum sources, which is expected since sources with very steep spectra are unlikely to be above the detection threshold at a frequency as high as 148 GHz. Further, the AT20G sample is expected to be incomplete for $S_\mathrm{20} < 70$ mJy. We expect that a complete unbiased sample will have almost equal numbers of steep and flat spectrum sources, as predicted by evolutionary models for radio sources in this flux density range \citep{2005A&A...431..893D}. The mean spectral indices $\alpha_{1.4-148}^\mathrm{mean}$ and $\alpha_{20-148}^\mathrm{mean}$ are consistent with that of a source population dominated by synchrotron and free-free emitting AGN-powered radio sources comprised of Flat-Spectrum Radio Quasars (FSRQ) and BL Lacs. These are collectively referred to as blazars. \cite{1989SciAm.260d..20B} argued that whether a source is a FSRQ or a BL Lac depends primarily on its axis of orientation relative to the observer. Powerful flat-spectrum radio quasars (FSRQs) are associated with a line-of-sight aligned with the AGN jet offering a view of the compact, Doppler-boosted, flat-spectrum base of the approaching jet \citep{1995PASP..107..803U}. Low power radio sources are associated with BL Lac objects that have compact jet structures and are known to be variable and relatively highly polarized. They are characterized by flat radio spectra and a steep infrared spectra.  
\begin{figure}
\begin{center}
\includegraphics[width=0.9\linewidth,keepaspectratio]{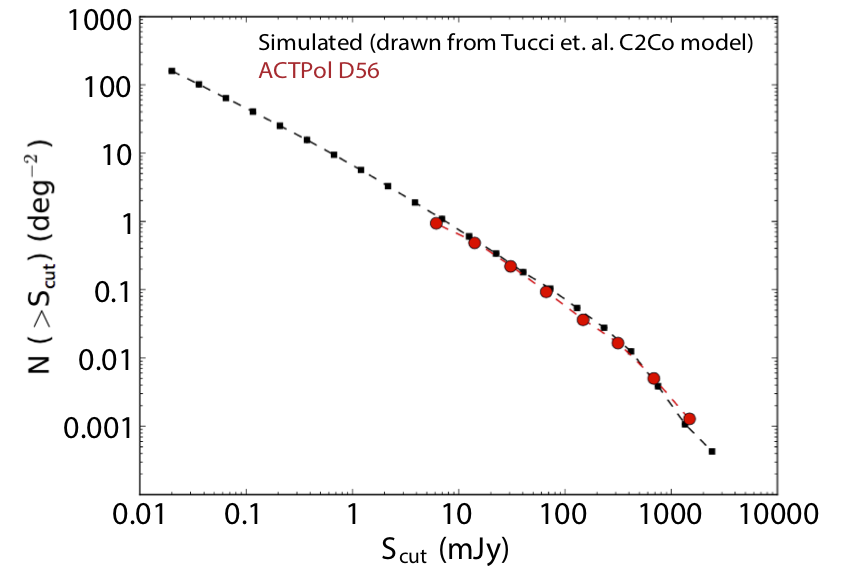}
\caption{\protect \footnotesize The number of synchrotron sources expected per square degree in a 148 GHz map above a given flux threshold, $S_\mathrm{cut}$. Comparison between the number of ACTPol sources (red) and number of simulated sources (black) based on the C2Co model. } 
\label{fig:src_sims}
\end{center}
\end{figure} 

Fig.~\ref{fig:ACTPol_NVSS_AT20G_CCD} shows color-color diagrams derived from measurements of polarized flux and fractional polarization. Only sources detected with total flux density greater than 74 mJy (1000 $\mu$K) were included when making these plots. We do not find any correlation between the spectral indices derived from polarized intensity and polarization fraction. Neither is there any correlation between the spectral indices from total flux and polarized flux. Statistically, the polarized flux densities are on average much more similar between 1.4 and 148 GHz. Hence, the data favors a higher polarization fraction for radio sources at higher frequencies. This is consistent with the findings of \cite{2011ApJ...732...45S} who report that sources on average are more polarized at higher frequencies. 

\section{Impact of Polarized Sources on the CMB EE Measurement}
The dominant foreground contribution to the high-$\ell$ power spectrum at 148 GHz is from synchrotron (radio) and infrared (dusty) extragalactic sources, and the thermal and kinetic SZ effect from galaxy groups and clusters. The power from synchrotron sources is sensitive to the masking level at mm wavelengths being roughly proportional to the total flux threshold above which the sources are masked. This is because most of the power is contributed by the brightest few sources. However, the infrared population is not so sensitive to the masking level since most of the power comes from sources with low total flux at mm wavelengths. The source population we have analyzed here is mostly composed of synchrotron sources, and hence we can predict the synchrotron contribution to the Poisson spectrum for various levels of point source masking. Assuming that polarization angles of the extragalactic point sources are distributed randomly, they are expected to contribute equally on average to the $Q$ and $U$ Stokes parameters, and thus to E- and B-mode power spectra. For a Poisson-distributed population of such sources, the contribution to the polarization power spectrum in terms of the $C_{l}$ is given by
\begin{flalign}
	\begin{gathered}
		C_{\ell,PS}^{EE} = C_{\ell,PS}^{BB} = \frac{1}{2} \bigg(\frac{dB}{dT}\bigg)^{-2} \int_{0}^{P_\mathrm{cut}} \bigg(\frac{dN}{dP}\bigg) P^{2} dP \\
	\end{gathered}	 
\label{eq:ClEE}
\end{flalign} 
\noindent where $dB/dT \approx 100 e^{x}x^{4}/(e^{x}-1)^{2} \mathrm{\mu K}^{-1} \mathrm{Jy}\ \mathrm{sr}^{-1}$ is the multiplicative conversion factor to go from brightness temperature (in $\mu$K) to to flux density (in Jy), $x$ = h$\nu$ / (k$_\mathrm{B}$T$_\mathrm{CMB}$), d$N$/d$P$ is the differential source number counts per steradian with respect to polarized flux, $P_\mathrm{cut}$ is the masking threshold in terms of polarized flux, and the subscript ``$PS$'' refers to point sources. Since the polarized signal is weak, it is difficult to measure polarized source number counts. However, we can express $C_{l,\mathrm{PS}}^\mathrm{EE}$ in terms of the differential source number counts per steradian d$N$/d$S$ with respect to total intensity and statistical estimates of the average fractional polarization of the source population squared $\ev{p^{2}}$ as: 
\begin{flalign}
	\begin{gathered}
		C_{\ell,PS}^{EE} = C_{l,PS}^{BB} = \frac{1}{2} \ev{p^{2}} \ C_{\ell,PS}^{TT} \\
\textnormal{where,\ } \ \  C_{\ell,PS}^{TT} = \bigg(\frac{dB}{dT}\bigg)^{-2} \int_{0}^{S_{cut}} \bigg(\frac{dN}{dS}\bigg) S^{2} dS
	\end{gathered}	 
\label{eq:ClTT}
\end{flalign}  \\

\begin{table}
\begin{center}
\caption{Predicted power from extragalactic synchrotron source population and crossover of its Poisson spectrum with the CMB EE spectrum for different levels of masking }
\begin{tabular}{lllll}
\hline
$S_\mathrm{cut}$ & $l_\mathrm{cross}^\mathrm{PS_\mathrm{EE}}$ & $D_{l\mathrm{=5000}}^\mathrm{PS_\mathrm{EE}}$ \\
(Jy) & & ($\mu$K$^{2}$) \\
\hline
5 &  3100 & 4 \\
1 & 3600 & 0.5 \\
0.1 & 4600 & 0.04 \\
0.01 & 5400 & 0.006 \\
0.002 & $>$6000 & 0.002 \\
\hline	                         
\end{tabular}\\
\end{center}
\end{table}
\label{tab:Dl_PS}

\begin{figure}
\begin{center}
\includegraphics[width=1.0\linewidth,keepaspectratio]{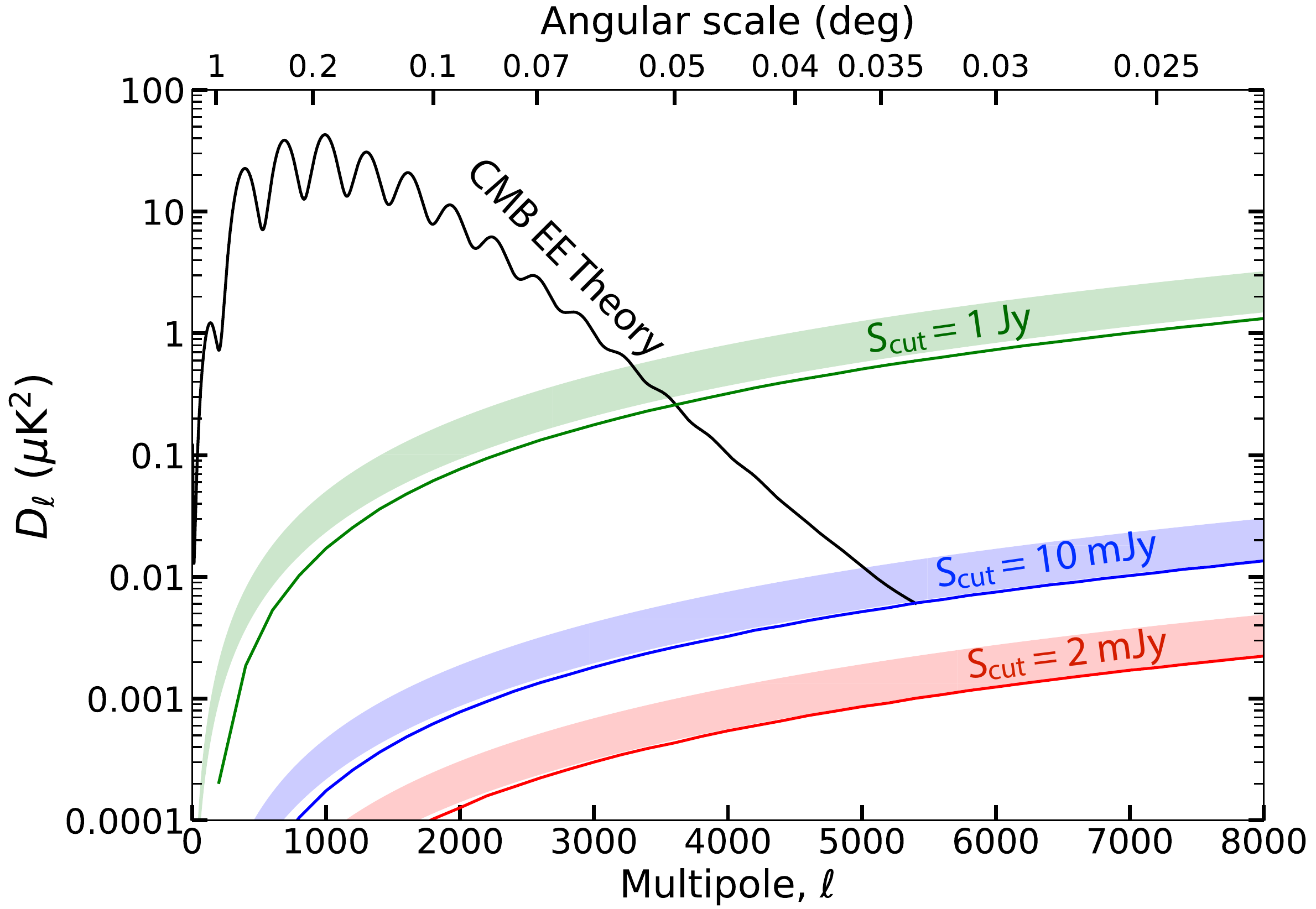} 
\caption{\protect \footnotesize The expected power spectra of the extragalactic synchrotron source population alone (no noise is included in these simulations) with sources above some chosen total flux thresholds $S_\mathrm{cut}$ removed from the Stokes $I$, $Q$, and $U$ maps. The colored lines correspond to a source population whose degree of fractional polarization is described by a truncated (at 0) Gaussian distribution with mean $p_{\mathrm{m}}=0.028$, standard deviation $\sigma_{\mathrm{p}_{\mathrm{m}}}=0.054$ (as found in this work). The black curve shows the CMB theory EE power spectrum. The detection threshold of $S_\mathrm{cut}=2$ mJy is achievable by a future survey with a 1.4$^\prime$ beam and a 1 $\mu$K-arcmin noise level across 10$\%$ of the sky. The lower and upper limits of the colored bands correspond to analytical estimates using equation~(\ref{eq:ClTT}) and assuming $\ev{p^{2}}=0.028^{2}$ and $(0.028^{2}+0.054^{2})$, respectively.} 
\label{fig:pwrspect}
\end{center}
\end{figure} 
\noindent The d$N$/d$S$ we assume for predicting $C_{l,\mathrm{PS}}^\mathrm{EE}$ is based on the C2Co model which best fits the ACTPol source counts. The analytical calculation using equation~(\ref{eq:ClTT}) gives the upper and lower limits of the colored bands in Fig.~\ref{fig:pwrspect} corresponding to $\ev{p^{2}}=(0.028^{2}~+~0.054^{2})$ and $0.028^{2}$, respectively for three different values of $S_{\mathrm{cut}}$. In order to get a more accurate estimate of $C_{l,\mathrm{PS}}^\mathrm{EE}$, we use simulations. First, we estimate the expected number of sources per square degree above a given flux threshold $S_\mathrm{cut}$, at 143 GHz directly from the C2Co d$N$/d$S$. This is consistent with the number of sources detected above a given $S_\mathrm{cut}$ in the D56 map as shown in the left panel of Fig.~\ref{fig:src_sims}. Then, we simulate a $\sim$4600 deg$^{2}$ (2$^{13} \times 2^{13}$ pixels) source-only map by injecting sources drawn from the C2Co d$N$/d$S$ model at random locations in the map. To simulate source-only $Q$ and $U$ maps, these sources are assigned fractional polarizations drawn from the best-fit model described in Section 4.4 which is a Gaussian distribution truncated at $0$ with mean fractional polarization $p_{\mathrm{m}}=0.028$ and standard deviation $\sigma_{\mathrm{p}_{\mathrm{m}}}=0.054$ (see left panel of Fig.~\ref{fig:src_model}), and polarization angles drawn from a flat distribution. The EE power spectrum computed from these maps gives the expected contribution of extragalactic synchrotron sources. Fig.~\ref{fig:pwrspect} shows the residual $D_{l,\mathrm{PS}}^\mathrm{EE}=\ell(\ell+1)C_{l,\mathrm{PS}}^\mathrm{EE}$/2$\pi$ point source spectra for a few different levels of source masking based on total flux threshold $S_\mathrm{cut}$, along with the CMB EE theory power spectrum. This shows that with appropriate point source masking in the $Q$ and $U$ maps, the crossing point between the EE spectrum from extragalactic sources and the measured CMB EE spectrum can be moved to higher $\ell$'s. This represents the level of source subtraction expected to be required in order to measure the faint high-$\ell$ end of the EE spectrum. 

\cite{2014JCAP...10..007N} had placed an upper limit on the contribution from sources $D_{l\mathrm{=3000}}^\mathrm{PS_\mathrm{EE}} < 2.4 \mu$K$^{2}$ based on ACTPol EE spectrum measurement with no sources masked. In \cite{2017JCAP...06..031L}, sources above a total flux threshold $S_\mathrm{cut}=15$ mJy were masked in the total intensity map used for computing the ACTPol TT spectrum but no sources were masked in the Stokes $Q$ and $U$ maps used for computing the EE spectrum. Recent measurements of the EE spectrum reported by the SPTpol experiment in \cite{2018ApJ...852...97H} placed a 95$\%$ confidence upper limit on residual polarized Poisson power in the EE spectrum $D_{l\mathrm{=3000}}^\mathrm{PS_\mathrm{EE}}<0.1 \mu$K$^{2}$ after masking all sources with total flux above 50 mJy. Based on our best fit model, we predict $D_{l\mathrm{=3000}}^\mathrm{PS_\mathrm{EE}}\sim 0.01 \mu$K$^{2}$ after applying a similar level of source masking. Though this prediction is based only on the contribution of synchrotron sources to the Poisson source spectrum, IR sources should not be significantly polarized compared to synchrotron sources (see Section 1). Table~\ref{tab:cat}{\color{blue}4} shows our predictions for $D_{l}^\mathrm{PS_\mathrm{EE}}$ at $\ell=5000$ as well as expected values of $\ell$ at which the Poisson source spectra would cross the CMB theory EE spectrum for different levels of source masking. As we go to higher sensitivity maps, the source detection threshold goes down, and we can apply more aggressive source masks. In Fig.~\ref{fig:pwrspect}, the flux thresholds of 10 mJy and 2 mJy are representative of what is achievable with ongoing and futue CMB surveys \citep{2016arXiv161002743A}, respectively. Assuming a 1.4$^\prime$ FWHM beam (similar to ACTPol) and an average map white noise level of 10 $\mu$K-arcmin and 1 $\mu$K-arcmin for a survey covering $\sim$10$\%$ of the sky, the catalog completeness is expected to be near 100$\%$ at flux thresholds of 10 mJy and 2 mJy, respectively. Therefore, one could get a clean measurement of EE out to $\ell \approx6000$ with a future survey. 

\section{Summary} 
\label{sec:Future}
We have presented the polarization properties of extragalactic sources at 148 GHz from ACTPol observations during the 2013 and 2014 seasons covering 680 deg$^{2}$ of the sky in a region we refer to as ``D56". In our data, out of the 26 sources brighter than 215 mJy detected in total intensity, 14 ($\sim$60$\%$) have a polarization signal with signal-to-noise ratio greater than 3 ($p>3\sigma_{\mathrm{p}}$). The average degree of polarization of these 26 sources, including the ones for which $p<3\sigma_{\mathrm{p}}$, is 2.6$\%$. Using simulations, we find the distribution of the fractional polarization to be consistent with that of a polarized source population having mean fractional polarization $p_{\mathrm{m}}=0.028\pm0.005$ with $\sigma_{\mathrm{p}_{\mathrm{m}}}=0.054$, and consistent with the scenario that the fractional polarization is independent of the total source intensity at least down to $S=100$ mJy. The measured level of average fractional polarization is consistent with typical numbers in the range 1--5$\%$ reported by other mm-wave and radio surveys. We compare the total and polarized flux measurements at 148 GHz with measurements at other frequencies from the VLA-FIRST, NVSS, AT20G, and Herschel SPIRE surveys and found that our source population is predominantly synchrotron with steepening, flat or peaked spectra. We did not find any evidence of correlation between the spectral indices derived from total flux and polarized flux. Our data favors higher levels of fractional polarization at higher frequencies for radio sources on average. The brightest source J000613--062336 in the D56 region, detected at 1.558$^{\circ}$ RA and $\mathrm{-}$6.393$^{\circ}$ Dec, was three times brighter than the next brightest one. This is a previously known and widely studied highly variable BL Lac object by the name QSO B0003-066 at a redshift $z=0.34$ \citep{2014ApJS..215...14D}. It has also been seen in the optical. We also present predictions for the contribution of power from extragalactic sources to the CMB EE spectrum and predict the level of source masking that would be required to enable precision measurements of the faint high-$\ell$ end of the EE spectrum. The lack of foregrounds in EE compared to TT is promising for constraining cosmology with high resolution measurements of the E-mode polarization. With future surveys, it could be possible to measure the EE spectrum out to $\ell\approx6000$. Looking ahead, multi-frequency data from the subsequent seasons of ACTPol and AdvACT observations in five frequency bands spanning 25 to 280 GHz will constrain the spectral signature of thousands of point sources over a large frequency range to unambiguously distinguish between radio and dusty sources and help find new, previously unidentified sources. It will also significantly improve the statistics on the measurements of the fractional polarization of extragalactic sources at millimeter wavelengths and potentially detect hundreds of sources in polarization.  

\section*{Acknowledgements}

This work was supported by the U.S. National Science Foundation through awards AST-1440226, AST-0965625 and AST- 0408698 for the ACT project, as well as awards PHY-1214379 and PHY-0855887. ACT operates in the Parque Astronomico Atacama in northern Chile under the auspices of the Comision Nacional de Investigacion Cientfica y Tecnologica de Chile (CONICYT). R. Datta's research was supported by an appointment to the NASA Postdoctoral Program at the NASA Goddard Space Flight Center, administered by Universities Space Research Association under contract with NASA. R. Datta would also like to acknowledge support from the Balzan foundation in the form of a Center for Cosmological Studies (CCS) Fellowship at the Johns Hopkins University and the Rackham Predoctoral Fellowship awarded by the University of Michigan during the initial phase of this work. CL thanks CONICYT for grant Anillo ACT-1417. LM is funded by CONICYT FONDECYT grant 3170846. R. Dunner acknowledges CONICYT for grants FONDECYT 1141113, Anillo ACT-1417, QUIMAL 160009 and  BASAL PFB-06 CATA. M. Hilton acknowledges financial support from the National Research Foundation, the South African Square Kilometre Array project, and the University of KwaZulu-Natal.

\bibliographystyle{mnras}
\bibliography{adssample}

\begin{appendix}

\section{Fluxes and effective frequencies of compact sources}

Compact extragalactic sources appear in our maps as unresolved hot spots with the profile of the beam. In this appendix, we show how we convert the measured source amplitudes in CMB temperature units to flux densities in Jansky (Jy). A source's emission as a function of frequency is parametrized as $S_\nu\propto(\nu/\nu_e)^{\alpha}$ where $\alpha$ denotes the source spectral index and $\nu_\mathrm{e}$ is a reference frequency. The instrument's frequency response is characterized with a passband, $f(\nu)$, and an antenna gain that also depends on frequency. We parametrize the latter as:
\begin{equation}
G_m(\nu) = \frac {4\pi }{\Omega_B}\bigg(\frac{\nu}{\nu_{\mathrm RJ}}\bigg)^{-2\beta}
\end{equation}
The quantity $\beta$ scales the size of the FWHM and is generally between 0 and $-1$. Because $\beta$ is negative, at higher frequencies the beam is narrower and
 the gain is higher. For ACT, $\beta$ is near $-1$. The gain is scaled to the value measured with planets which are well approximated as Rayleigh-Jeans sources ($\alpha=2$).

Following \cite{0004-637X-585-1-566} we define an effective frequency 
\begin{equation}
\nu_e = \frac{\int \nu f(\nu ) \nu^{-2(1+\beta)+\alpha}d\nu}{\int f(\nu) \nu^{-2(1+\beta)+\alpha}\,d\nu}
\end{equation}
and CMB temperature to flux conversion factor 
\begin{equation}
\Gamma = \frac{c^2}{2 k_B\Omega_B\nu_e^\alpha \nu_{\mathrm RJ}^{-2\beta} } \frac{ \int \nu^{-2(1+\beta)+\alpha} f(\nu)  d\nu}{\int f(\nu) d\nu} \ 10^{-20} ~~[\mu{\mathrm K/ Jy}]
\end{equation}

Let us say the source has a temperature spectrum of $T_\mathrm{src}\propto\nu^{\alpha-2}$.  We convert the measured source amplitude in CMB temperature units to a flux density at its effective central frequency as follows: \\
($i$) Convert to a brightness temperature for the source by dividing by the factor: \\
\begin{flalign}
	\begin{gathered}
\frac{\delta T_{\mathrm CMB}}{\delta T_{\mathrm src}} =  \frac{\delta S_{\nu,src}}{\delta T_{\mathrm src}}\bigg(\frac{\delta S_{\nu,CMB}}{\delta T_{\mathrm CMB}}\bigg)^{-1}= \bigg(\frac{\nu}{\nu_e}\bigg)^{\alpha-2}\frac{(e^x-1)^2}{x^2e^x}\\
{\mathrm with}~~~~x=\frac{h\nu_{e}}{k_BT_{\mathrm CMB}}
	\end{gathered}	 
\end{flalign} 

\noindent Note that this is simply $\delta T_{\mathrm CMB}/\delta T_{\mathrm RJ}$ if $\alpha=2$ and $\nu_e=\nu_{\mathrm RJ}$. \\ 
($ii$) Find the flux by dividing the temperature from step ($i$) by $\Gamma$ as computed above.\\
($iii$) To find the source flux at a different frequency than its effective central frequency, we can use the scaling relations
$S_\nu\propto(\nu/\nu_e)^\alpha$ or $T_\nu\propto(\nu/\nu_e)^{\alpha-2}$. Note that this assumes the source has a simple power law spectrum.\\

Tables~\ref{tab:eff_freq2} and \ref{tab:eff_freq3} give the conversion factors $\delta T_{\mathrm CMB}$/ $\delta T_{\mathrm RJ}$ and $\Gamma$ for $\beta=0$ and --1.0 for different types of sources. The uncertainties in the conversion factors include uncertainties in the measurements of the central frequency and the beam solid angle. These tables update the numbers in Table 1 of \cite{2016ApJS..227...21T}. Since most of the sources in this work are synchrotron dominated, the measured source fluxes in Table~\ref{tab:cat}{\color{blue}4} are reported at $\nu_{e}$ = 147.6 GHz, which is the central frequency to synchrotron sources ($\alpha=-0.7$) assuming $\beta$ = --1. If any of these sources has a known spectral index that is different from --0.7, a multiplicative color correction factor can be used to get its true flux at 147.6 GHz. We provide these factors along with uncertainties for a range of values of $\alpha$ from --2.0 to +4.0 in steps of 0.2 in Table~\ref{tab:cc_Sync}. Note that the flux density uncertainties quoted in Table~\ref{tab:cat}{\color{blue}4} do not include errors in the measurement of the central frequency or the beam solid angle. The complete uncertainty should be obtained by combining the errors in Tables~\ref{tab:eff_freq3}~and~\ref{tab:cat}{\color{blue}4} in quadrature. \\

\begin{table*}
\caption{Central frequencies for compact sources (except CMB) and conversion factors to go from a measured source amplitude in temperature units to flux density assuming $\beta=0$ and solid angle of the beam used in the matched filtering $\Omega_B=182\pm3$ nsr }
\begin{center}
\begin{tabular}{|c|c|c|c|c|c|}
\hline
Source & Sync, $S_\nu\propto\nu^{-0.7}$ & FF, $S_\nu\propto\nu^{-0.1}$&  CMB & RJ, $S_\nu\propto\nu^{2}$ 
& Dust, $S_\nu\propto\nu^{3.7}$ \\
\hline
$\nu_e$ (GHz)  &144.9$\pm$2.4 & 145.7$\pm$2.4  & 148.6$\pm$2.4 & 148.6$\pm$2.4 & 150.8$\pm$2.4 \\
$\frac{\delta T_{CMB}}{\delta T_{RJ}}$ & 1.67$\pm$0.03 & 1.68$\pm$0.03 & 1.71$\pm$0.03 & 1.71$\pm$0.03 & 1.74$\pm$0.03 \\
$\Gamma$ ($\mu$K/Jy)& 8337$\pm$169 & 8331$\pm$138 & & 8109$\pm$294 & 7725$\pm$477 \\
\hline
\end{tabular}
\end{center}
\label{tab:eff_freq2}
\label{default}
\end{table*}

\begin{table*}
\caption{Central frequencies for compact sources (except CMB) and conversion factors to go from a measured source amplitude in temperature units to flux density assuming $\beta=-1$ and solid angle of the beam used in the matched filtering $\Omega_B=182\pm3$ nsr }
\begin{center}
\begin{tabular}{|c|c|c|c|c|c|}
\hline
Source & Sync, $S_\nu\propto\nu^{-0.7}$ & FF, $S_\nu\propto\nu^{-0.1}$&  CMB & RJ, $S_\nu\propto\nu^{2}$ 
& Dust, $S_\nu\propto\nu^{3.7}$ \\
\hline
$\nu_e$ (GHz) & 147.6$\pm$2.4  & 148.4$\pm$2.4 & 151.2$\pm$2.4 & 151.2$\pm$2.4 & 153.3$\pm$2.4 \\
$\frac{\delta T_{CMB}}{\delta T_{RJ}}$ & 1.70$\pm$0.03 & 1.71$\pm$0.03 & 1.75$\pm$0.03 & 1.75$\pm$0.03 & 1.772$\pm$0.03 \\
$\Gamma$ ($\mu$K/Jy)& 7833$\pm$296  &  7829$\pm$282 & & 7634$\pm$502 & 7289$\pm$501 \\
\hline
\end{tabular}
\end{center}
\label{tab:eff_freq3}
\label{default}
\end{table*}

\begin{table*}
\caption{Color corrections: multiplicative factors C($\alpha$) for computing source flux densities at 147.6 GHz for a given source spectral index $\alpha$ and starting with the measured flux densities reported in Table~\ref{tab:cat}{\color{blue}4}}
\begin{center}
\begin{tabular}{|c|c|c||c|c|c|}
\hline
Source Spectral Index $\alpha$ & C($\alpha,\beta=0$) & C($\alpha,\beta=-1$) & Source Spectral Index $\alpha$ & C($\alpha,\beta=0$) & C($\alpha,\beta=-1$)\\
\hline
--2.0 & 0.92$\pm$0.03 & 1.00$\pm$0.03  & +1.2 & 0.95$\pm$0.04 &  0.98$\pm$0.03  \\
--1.8 & 0.93$\pm$0.03 &  1.00$\pm$0.03   & +1.4 & 0.95$\pm$0.04 &  0.97$\pm$0.03 \\
--1.6 & 0.93$\pm$0.04 &  1.00$\pm$0.03  & +1.6 & 0.95$\pm$0.04 &  0.97$\pm$0.03 \\ 
--1.4 & 0.93$\pm$0.04 &  1.00$\pm$0.03  & +1.8 & 0.95$\pm$0.04 &  0.96$\pm$0.03 \\
--1.2 & 0.94$\pm$0.04 &  1.00$\pm$0.03  & +2.0 & 0.95$\pm$0.04 &  0.96$\pm$0.03 \\
--1.0 & 0.94$\pm$0.04 &  1.00$\pm$0.03  & +2.2 & 0.95$\pm$0.04 &  0.95$\pm$0.03 \\
--0.8 & 0.94$\pm$0.04 &  1.00$\pm$0.03  & +2.4 & 0.94$\pm$0.04 &  0.94$\pm$0.03 \\
--0.6 & 0.95$\pm$0.04 &  1.00$\pm$0.03  & +2.6 & 0.94$\pm$0.04 &  0.94$\pm$0.03 \\
--0.4 & 0.95$\pm$0.04 &  1.00$\pm$0.03  & +2.8 & 0.94$\pm$0.04 &  0.93$\pm$0.03 \\
--0.2 & 0.95$\pm$0.04 &  1.00$\pm$0.03  & +3.0 & 0.93$\pm$0.04 &  0.93$\pm$0.03 \\
0.0 & 0.95$\pm$0.04 &  1.00$\pm$0.03   & +3.2 & 0.93$\pm$0.04 &  0.92$\pm$0.03\\
+0.2 & 0.95$\pm$0.04 &  0.99$\pm$0.03 & +3.4 & 0.93$\pm$0.03 &  0.91$\pm$0.03  \\
+0.4 & 0.95$\pm$0.04 &  0.99$\pm$0.03  & +3.6 & 0.92$\pm$0.03 &  0.90$\pm$0.03 \\
+0.6 & 0.95$\pm$0.04 &  0.99$\pm$0.03  & +3.8 & 0.92$\pm$0.03 &  0.90$\pm$0.03 \\
+0.8 & 0.95$\pm$0.04 &  0.98$\pm$0.03  & +4.0 & 0.91$\pm$0.03 &  0.89$\pm$0.03\\
+1.0 & 0.95$\pm$0.04 &  0.98$\pm$0.03  & & & \\ 
\hline
\end{tabular}
\end{center}
\label{tab:cc_Sync}
\end{table*}
Below is a description of the columns in Table~\ref{tab:cat}{\color{blue}4}: \\
1. ID: Source ID in the format ``ACTPol-D56 J[hh mm ss][deg min sec]" ($^{*}$ indicates extended source) \\
2. RA: J2000 right ascension of the source in decimal degrees \\
3. dec: J2000 declination of the source in decimal degrees \\
4. S/N: Signal-to-noise ratio of the detection in total intensity \\
5. T$_\mathrm{CMB}$: Source amplitude in CMB temperature units ($\mu$K) along with the 1$\sigma$ uncertainty \\
6. Stokes $Q$: Stokes Q amplitude in CMB temperature units ($\mu$K) for sources brighter than 30 mJy in total flux density along with the 1$\sigma$ uncertainty \\
7. Stokes $U$: Stokes U amplitude in CMB temperature units ($\mu$K, following the IAU standard) for sources brighter than 30 mJy in total flux density along with the 1$\sigma$ uncertainty \\
8. S$_\mathrm{raw}$: Raw measured flux density in mJy units at 147.6 GHz (assumes a source spectral index of -0.7) along with the 1$\sigma$ uncertainty \\
9. S$_\mathrm{deb}$: Deboosted flux density in mJy units (only sources dimmer than 100 mJy are deboosted) along with the negative and positive 1$\sigma$ uncertainties \\
10. S/N$_\mathrm{p}$: Signal-to-noise ratio in polarization \\
11. p$_\mathrm{WK}$: Percentage fractional polarization estimated using the Wardle-Kronberg estimator along with the negative and positive 1$\sigma$ uncertainties \\
12. p$_\mathrm{x}$: Percentage fractional polarization estimated using the 2-way data splits along with the negative and positive 1$\sigma$ uncertainties \\
13. pol ang: Polarization angle of the source following the IAU standard (only for sources detected in polarization at a significance larger than 3-sigma) along with the 1$\sigma$ uncertainty \\
14. var?: for sources observed during both seasons, checkmark indicates that the variation in the measured source flux density between the 2013 (S13) and 2014 (S14) observing seasons is greater than 40$\%$ \\
15. V?: checkmark indicates that the source has a cross-match in the VLA FIRST 1.4 GHz catalog \\ 
16. A?: checkmark indicates that the source has a cross-match in the AT20G 20 GHz catalog \\ 
17. S?: checkmark indicates that the source has a cross-match in the Herschel SPIRE 600 GHz catalog \\ 

\newpage 

\clearpage
\onecolumn


\begin{center}

\footnotesize
{
\setlength{\tabcolsep}{1pt} 
\begin{longtable}{| l| l| l| l| l| l| l| l| l| l| l| l| l| l| l| l| l| l| l}
\caption{Catalog of point sources from the ACTPol D56 survey at 148 GHz with total flux density $S_\mathrm{148}$ greater than 30 mJy. The full catalog is available on LAMBDA. The flux densities reported here are at 147.6 GHz, and assume a source spectral index of --0.7. }
\label{tab:cat}\\
\hline \multicolumn{1}{|c|}{\textbf{ID}} & \multicolumn{1}{c|}{\textbf{RA}} & \multicolumn{1}{c|}{\textbf{dec}} & \multicolumn{1}{c|}{\textbf{$\mathrm{S}$/$\mathrm{N}$}} & \multicolumn{1}{c|}{\textbf{T$_\mathrm{CMB}$}} & \multicolumn{1}{c|}{\textbf{Stokes Q}} & \multicolumn{1}{c|}{\textbf{Stokes U}} & \multicolumn{1}{c|}{\textbf{S$_\mathrm{raw}$}} & \multicolumn{1}{c|}{\textbf{S$_\mathrm{deb}$}} & \multicolumn{1}{c|}{\textbf{$\mathrm{S}$/$\mathrm{N}_\mathrm{p}$}}  & \multicolumn{1}{c|}{\textbf{p$_\mathrm{WK}$}} & \multicolumn{1}{c|}{\textbf{p$_\mathrm{\times}$}} & \multicolumn{1}{c|}{\textbf{pol ang}} & \multicolumn{1}{c|}{\textbf{var?}} & \multicolumn{1}{c|}{\textbf{V?}} & \multicolumn{1}{c|}{\textbf{A?}} & \multicolumn{1}{c|}{\textbf{S?}}\\ 
\multicolumn{1}{|c|}{\textbf{ACTPol-D56}} & \multicolumn{1}{c|}{\textbf{(deg)}} & \multicolumn{1}{c|}{\textbf{(deg)}} & \multicolumn{1}{c|}{\textbf{}} & \multicolumn{1}{c|}{\textbf{($\mu$K)}} & \multicolumn{1}{c|}{\textbf{($\mu$K)}} & \multicolumn{1}{c|}{\textbf{($\mu$K)}} & \multicolumn{1}{c|}{\textbf{(mJy)}} & \multicolumn{1}{c|}{\textbf{(mJy)}} & \multicolumn{1}{c|}{\textbf{}} & \multicolumn{1}{c|}{\textbf{($\%$)}} & \multicolumn{1}{c|}{\textbf{($\%$)}} & \multicolumn{1}{c|}{\textbf{(deg)}} & \multicolumn{1}{c|}{\textbf{}} & \multicolumn{1}{c|}{\textbf{}} & \multicolumn{1}{c|}{\textbf{}} & \multicolumn{1}{c|}{\textbf{}} \\ \hline 
\endfirsthead

\multicolumn{17}{c}
{{\bfseries \tablename\ \thetable{} -- continued from previous page}} \\
\hline \multicolumn{1}{|c|}{\textbf{ID}} & \multicolumn{1}{c|}{\textbf{RA}} & \multicolumn{1}{c|}{\textbf{dec}} & \multicolumn{1}{c|}{\textbf{$\mathrm{S}$/$\mathrm{N}$}} & \multicolumn{1}{c|}{\textbf{T$_\mathrm{CMB}$}} & \multicolumn{1}{c|}{\textbf{Stokes Q}} & \multicolumn{1}{c|}{\textbf{Stokes U}} & \multicolumn{1}{c|}{\textbf{S$_\mathrm{raw}$}} & \multicolumn{1}{c|}{\textbf{S$_\mathrm{deb}$}} & \multicolumn{1}{c|}{\textbf{$\mathrm{S}$/$\mathrm{N}_\mathrm{p}$}} & \multicolumn{1}{c|}{\textbf{p$_\mathrm{WK}$}} & \multicolumn{1}{c|}{\textbf{p$_\mathrm{\times}$}} & \multicolumn{1}{c|}{\textbf{pol ang}} & \multicolumn{1}{c|}{\textbf{var?}} & \multicolumn{1}{c|}{\textbf{V?}} & \multicolumn{1}{c|}{\textbf{A?}} & \multicolumn{1}{c|}{\textbf{S?}} \\  
\multicolumn{1}{|c|}{\textbf{(ACTPol-D56)}} & \multicolumn{1}{c|}{\textbf{(deg)}} & \multicolumn{1}{c|}{\textbf{(deg)}} & \multicolumn{1}{c|}{\textbf{}} & \multicolumn{1}{c|}{\textbf{($\mu$K)}} & \multicolumn{1}{c|}{\textbf{($\mu$K)}} & \multicolumn{1}{c|}{\textbf{($\mu$K)}} & \multicolumn{1}{c|}{\textbf{(mJy)}} & \multicolumn{1}{c|}{\textbf{(mJy)}} & \multicolumn{1}{c|}{\textbf{}} & \multicolumn{1}{c|}{\textbf{($\%$)}} & \multicolumn{1}{c|}{\textbf{($\%$)}} & \multicolumn{1}{c|}{\textbf{(deg)}} & \multicolumn{1}{c|}{\textbf{}} & \multicolumn{1}{c|}{\textbf{}} & \multicolumn{1}{c|}{\textbf{}} & \multicolumn{1}{c|}{\textbf{}} \\ \hline 
\endhead
\hline 
\endfoot

J000111--002015 & 0.2969 & --0.3375 & 21.0 & 616$\pm$29 & .. & .. & 46.3$\pm$2.1 & 47.3$^{+7.7}_{-7.7}$ & .. & .. & .. & .. & .. & $\checkmark$ & .. & .. \\
J000118--074627 & 0.3250 & --7.7744 & 60.9 & 1938$\pm$32 & 38$\pm$33 & 68$\pm$33 & 145.6$\pm$2.3 & .. & 2.4 & 3.7$^{+1.7}_{-0.9}$ & 2.7$^{+2.4}_{-2.4}$ & .. & .. & $\checkmark$ & $\checkmark$ & .. \\
J000257--002445 & 0.7391 & --0.4125 & 13.9 & 403$\pm$29 & .. & .. & 30.3$\pm$2.1 & 29.3$^{+5.6}_{-5.6}$ & .. & .. & .. & .. & $\checkmark$ & $\checkmark$ & .. & .. \\
J000613--062336* & 1.5578 & --6.3935 & 1238 & 43452$\pm$35 & 2060$\pm$36 & 1438$\pm$36 & 3261.4$\pm$2.6 & .. & 68.7 & 5.8$^{+0.2}_{-0.2}$ & 5.7$^{+0.1}_{-0.1}$ & +17$\pm$1 & .. & $\checkmark$ & $\checkmark$ & .. \\
J000622--000422 & 1.5943 & --0.0729 & 37.6 & 1046$\pm$28 & --54$\pm$33 & --88$\pm$33 & 78.5$\pm$2.1 & 77.6$^{+13.8}_{-12.8}$ & 3.1 & 9.4$^{+3.0}_{-2.9}$ & 6.3$^{+4.5}_{-4.5}$ & --61$\pm$10 & .. & $\checkmark$ & $\checkmark$ & .. \\
J001000--043347 & 2.5016 & --4.5632 & 78.2 & 2490$\pm$32 & --45$\pm$33 & 93$\pm$33 & 186.9$\pm$2.4 & .. & 3.1 & 4.0$^{+1.3}_{-1.3}$ & 2.8$^{+1.9}_{-1.9}$ & +58$\pm$10 & .. & $\checkmark$ & $\checkmark$ & .. \\
J001130+005756 & 2.8771 & +0.9657 & 46.8 & 1230$\pm$26 & --71$\pm$31 & 2$\pm$31 & 92.3$\pm$2.0 & 93.2$^{+15.7}_{-15.5}$ & 2.3 & 5.3$^{+2.6}_{-1.3}$ & 3.7$^{+3.6}_{-3.6}$ & .. & .. & $\checkmark$ & .. & .. \\
J001354--042351 & 3.4755 & --4.3976 & 87.9 & 2873$\pm$33 & 49$\pm$34 & --62$\pm$34 & 215.7$\pm$2.5 & .. & 2.3 & 2.5$^{+1.2}_{-0.6}$ & 2.8$^{+1.7}_{-1.7}$ & .. & .. & $\checkmark$ & $\checkmark$ & .. \\
J001611--001509 & 4.0458 & --0.2526 & 72.5 & 2309$\pm$32 & --7$\pm$33 & 65$\pm$33 & 173.3$\pm$2.4 & .. & 2.0 & 2.5$^{+1.5}_{-0.7}$ & 1.8$^{+2.0}_{-1.8}$ & .. & .. & $\checkmark$ & $\checkmark$ & .. \\
J001709--065036 & 4.2896 & --6.8433 & 19.3 & 580$\pm$30 & .. & .. & 43.6$\pm$2.3 & 45.5$^{+7.1}_{-10.5}$ & .. & .. & .. & .. & .. & $\checkmark$ & $\checkmark$ & .. \\
J001735--051242* & 4.3990 & --5.2119 & 140.7 & 4670$\pm$33 & 106$\pm$34 & 199$\pm$34 & 350.5$\pm$2.5 & .. & 6.5 & 4.8$^{+0.7}_{-0.8}$ & 5.0$^{+1.0}_{-1.0}$ & +31$\pm$5 & .. & $\checkmark$ & $\checkmark$ & $\checkmark$ \\
J001752--023616 & 4.4677 & --2.6045 & 41.6 & 1160$\pm$28 & 6$\pm$33 & 119$\pm$33 & 87.0$\pm$2.1 & 85.5$^{+14.3}_{-13.9}$ & 3.6 & 10.0$^{+2.6}_{-2.7}$ & 10.0$^{+4.1}_{-4.1}$ & +44$\pm$8 & .. & $\checkmark$ & $\checkmark$ & .. \\
J001834--062838 & 4.6422 & --6.4773 & 15.4 & 474$\pm$31 & .. & .. & 35.6$\pm$2.3 & 34.3$^{+6.3}_{-7.6}$ & .. & .. & .. & .. & .. & $\checkmark$ & $\checkmark$ & .. \\
J001917--010356 & 4.8229 & --1.0657 & 23.3 & 651$\pm$28 & .. & .. & 48.8$\pm$2.1 & 48.6$^{+8.4}_{-8.0}$ & .. & .. & .. & .. & .. & $\checkmark$ & $\checkmark$ & .. \\
J002238+014704 & 5.6594 & +1.7847 & 41.7 & 1069$\pm$26 & .. & .. & 80.2$\pm$1.9 & 78.2$^{+13.9}_{-12.9}$ & .. & 0.0$^{+0.4}_{-0.0}$ & 0.0$^{+4.1}_{-0.0}$ & .. & .. & $\checkmark$ & .. & .. \\
J002400--081114 & 6.0037 & --8.1872 & 93.2 & 3148$\pm$34 & 25$\pm$35 & 15$\pm$35 & 236.3$\pm$2.5 & .. & .. & 0.0$^{+1.3}_{-0.0}$ & 0.0$^{+1.6}_{-0.0}$ & .. & .. & $\checkmark$ & $\checkmark$ & .. \\
J002446--041204 & 6.1917 & --4.2012 & 24.2 & 687$\pm$28 & .. & .. & 51.6$\pm$2.1 & 51.7$^{+9.1}_{-8.6}$ & .. & .. & .. & .. & .. & $\checkmark$ & $\checkmark$ & .. \\
J002733+045915 & 6.8885 & +4.9876 & 25.5 & 681$\pm$27 & .. & .. & 51.1$\pm$2.0 & 50.8$^{+9.0}_{-8.4}$ & .. & .. & .. & .. & .. & $\checkmark$ & .. & .. \\
J002901--011339 & 7.2542 & --1.2277 & 47.5 & 1320$\pm$28 & 16$\pm$33 & 28$\pm$33 & 99.1$\pm$2.1 & 98.4$^{+16.6}_{-15.8}$ & .. & 0.0$^{+3.2}_{-0.0}$ & 2.2$^{+0.0}_{-2.2}$ & .. & .. & $\checkmark$ & $\checkmark$ & .. \\
J003007--000015 & 7.5328 & --0.0042 & 15.0 & 418$\pm$28 & .. & .. & 31.4$\pm$2.1 & 30.3$^{+6.1}_{-6.0}$ & .. & .. & .. & .. & .. & $\checkmark$ & $\checkmark$ & .. \\
J003031--021154 & 7.6323 & --2.1985 & 94.4 & 3024$\pm$32 & --38$\pm$33 & 12$\pm$33 & 227.0$\pm$2.4 & .. & 1.2 & 0.9$^{+1.1}_{-0.6}$ & 1.2$^{+1.6}_{-1.2}$ & .. & .. & $\checkmark$ & $\checkmark$ & $\checkmark$ \\
J003259+051010 & 8.2474 & +5.1695 & 18.3 & 509$\pm$28 & .. & .. & 38.2$\pm$2.1 & 37.7$^{+7.3}_{-8.3}$ & .. & .. & .. & .. & .. & $\checkmark$ & .. & .. \\
J003307--042933 & 8.2802 & --4.4926 & 15.6 & 457$\pm$29 & .. & .. & 34.3$\pm$2.2 & 33.7$^{+5.9}_{-7.3}$ & .. & .. & .. & .. & .. & $\checkmark$ & $\checkmark$ & .. \\
J003443--005411 & 8.6828 & --0.9032 & 35.4 & 1003$\pm$28 & --16$\pm$34 & 41$\pm$34 & 75.3$\pm$2.1 & 73.9$^{+12.4}_{-12.1}$ & 1.3 & .. & 3.5$^{+4.8}_{-3.5}$ & .. & .. & $\checkmark$ & $\checkmark$ & .. \\
J003546--083552 & 8.9427 & --8.5979 & 57.2 & 2103$\pm$37 & 61$\pm$44 & --146$\pm$44 & 157.9$\pm$2.8 & .. & 3.6 & 7.3$^{+1.9}_{-2.0}$ & 7.6$^{+3.0}_{-3.0}$ & --34$\pm$8 & .. & $\checkmark$ & $\checkmark$ & .. \\
J003635--031812 & 9.1474 & --3.3034 & 15.8 & 437$\pm$28 & .. & .. & 32.8$\pm$2.1 & 32.4$^{+5.9}_{-6.2}$ & .. & .. & .. & .. & .. & $\checkmark$ & $\checkmark$ & .. \\
J003637--024007 & 9.1568 & --2.6692 & 14.8 & 406$\pm$27 & .. & .. & 30.5$\pm$2.0 & 29.5$^{+5.6}_{-5.6}$ & .. & .. & .. & .. & .. & $\checkmark$ & $\checkmark$ & .. \\
J003704--010905 & 9.2667 & --1.1516 & 48.3 & 1375$\pm$28 & --5$\pm$34 & 42$\pm$34 & 103.2$\pm$2.1 & .. & 1.3 & 2.4$^{+2.4}_{-1.3}$ & 3.4$^{+3.5}_{-3.4}$ & .. & .. & $\checkmark$ & $\checkmark$ & .. \\
J003816--012204 & 9.5688 & --1.3678 & 27.8 & 774$\pm$28 & .. & .. & 58.1$\pm$2.1 & 59.5$^{+9.6}_{-9.6}$ & .. & .. & .. & .. & .. & $\checkmark$ & $\checkmark$ & .. \\
J003820--020741 & 9.5854 & --2.1281 & 133.7 & 4369$\pm$33 & --24$\pm$34 & 3$\pm$34 & 327.9$\pm$2.5 & .. & .. & 0.0$^{+0.8}_{-0.0}$ & 0.7$^{+0.0}_{-0.7}$ & .. & .. & $\checkmark$ & $\checkmark$ & $\checkmark$ \\
J003849--011719 & 9.7057 & --1.2887 & 21.2 & 593$\pm$28 & .. & .. & 44.5$\pm$2.1 & 46.0$^{+7.4}_{-10.5}$ & .. & .. & .. & .. & .. & $\checkmark$ & $\checkmark$ & .. \\
J003852--012156 & 9.7182 & --1.3658 & 20.5 & 571$\pm$28 & .. & .. & 42.8$\pm$2.1 & 44.9$^{+7.2}_{-10.4}$ & .. & .. & .. & .. & .. & $\checkmark$ & $\checkmark$ & .. \\
J003913--025708 & 9.8057 & --2.9524 & 30.0 & 821$\pm$27 & .. & .. & 61.6$\pm$2.1 & 61.9$^{+9.8}_{-9.8}$ & .. & .. & .. & .. & .. & $\checkmark$ & $\checkmark$ & .. \\
J003919+031955 & 9.8292 & +3.3321 & 21.1 & 630$\pm$30 & .. & .. & 47.3$\pm$2.2 & 47.3$^{+7.9}_{-7.8}$ & .. & .. & .. & .. & .. & $\checkmark$ & .. & .. \\
J004013+012549 & 10.0563 & +1.4304 & 72.3 & 2164$\pm$30 & --32$\pm$31 & 97$\pm$31 & 162.4$\pm$2.2 & .. & 3.3 & 4.5$^{+1.4}_{-1.4}$ & 4.7$^{+2.0}_{-2.0}$ & +54$\pm$9 & .. & $\checkmark$ & .. & $\checkmark$ \\
J004020--004031* & 10.0839 & --0.6755 & 29.5 & 847$\pm$29 & .. & .. & 63.5$\pm$2.2 & 63.0$^{+10.8}_{-10.4}$ & .. & .. & .. & .. & .. & $\checkmark$ & $\checkmark$ & .. \\
J004057--014632 & 10.2401 & --1.7758 & 102.5 & 3295$\pm$32 & --145$\pm$33 & --5$\pm$33 & 247.3$\pm$2.4 & .. & 4.3 & 4.3$^{+0.9}_{-1.0}$ & 4.5$^{+1.4}_{-1.4}$ & --79$\pm$4 & .. & $\checkmark$ & $\checkmark$ & $\checkmark$ \\
J004300--075354 & 10.7500 & --7.8984 & 21.2 & 663$\pm$31 & .. & .. & 49.8$\pm$2.3 & 49.7$^{+8.5}_{-8.2}$ & .. & .. & .. & .. & .. & $\checkmark$ & $\checkmark$ & .. \\
J004332+002452 & 10.8870 & +0.4146 & 17.5 & 476$\pm$27 & .. & .. & 35.7$\pm$2.0 & 34.3$^{+6.3}_{-7.7}$ & .. & .. & .. & .. & .. & $\checkmark$ & .. & .. \\
J004345--025108 & 10.9406 & --2.8522 & 20.2 & 533$\pm$26 & .. & .. & 40.0$\pm$2.0 & 40.7$^{+7.1}_{-9.2}$ & .. & .. & .. & .. & .. & $\checkmark$ & $\checkmark$ & .. \\
J004346+050259 & 10.9438 & +5.0498 & 24.1 & 643$\pm$27 & .. & .. & 48.3$\pm$2.0 & 48.2$^{+8.2}_{-7.9}$ & .. & .. & .. & .. & .. & $\checkmark$ & .. & $\checkmark$ \\
J004404+010152 & 11.0188 & +1.0313 & 24.8 & 661$\pm$27 & .. & .. & 49.6$\pm$2.0 & 49.6$^{+8.5}_{-8.2}$ & .. & .. & .. & .. & .. & $\checkmark$ & .. & .. \\
J004705+031959 & 11.7745 & +3.3331 & 60.9 & 1665$\pm$27 & 27$\pm$32 & 18$\pm$32 & 125.0$\pm$2.1 & .. & .. & 0.4$^{+2.3}_{-0.4}$ & 0.0$^{+0.0}_{-0.0}$ & .. & .. & $\checkmark$ & .. & .. \\
J004943+023705 & 12.4302 & +2.6181 & 46.3 & 1133$\pm$24 & --60$\pm$29 & 22$\pm$29 & 85.0$\pm$1.8 & 82.9$^{+14.7}_{-13.9}$ & 2.2 & 5.1$^{+2.6}_{-1.3}$ & 4.6$^{+3.7}_{-3.7}$ & .. & .. & $\checkmark$ & .. & $\checkmark$ \\
J005021--045223 & 12.5896 & --4.8731 & 114.8 & 3733$\pm$33 & --3$\pm$34 & --100$\pm$34 & 280.2$\pm$2.4 & .. & 2.9 & 2.5$^{+0.9}_{-0.9}$ & 2.7$^{+1.3}_{-1.3}$ & .. & .. & $\checkmark$ & $\checkmark$ & .. \\
J005108--065003 & 12.7844 & --6.8344 & 165.9 & 5837$\pm$35 & 47$\pm$36 & --101$\pm$36 & 438.1$\pm$2.6 & .. & 3.1 & 1.8$^{+0.6}_{-0.6}$ & 1.8$^{+0.9}_{-0.9}$ & --33$\pm$10 & .. & $\checkmark$ & $\checkmark$ & .. \\
J005159+042742 & 12.9964 & +4.4618 & 24.5 & 737$\pm$30 & .. & .. & 55.3$\pm$2.3 & 56.6$^{+9.6}_{-9.6}$ & .. & .. & .. & .. & .. & $\checkmark$ & .. & .. \\
J005205+003537 & 13.0240 & +0.5938 & 23.1 & 620$\pm$27 & .. & .. & 46.5$\pm$2.0 & 47.3$^{+7.7}_{-7.7}$ & .. & .. & .. & .. & .. & $\checkmark$ & .. & .. \\
J005717--002435 & 14.3208 & --0.4099 & 23.3 & 641$\pm$28 & .. & .. & 48.1$\pm$2.1 & 48.1$^{+8.2}_{-8.0}$ & .. & .. & .. & .. & .. & $\checkmark$ & $\checkmark$ & .. \\
J005734--012334* & 14.3953 & --1.3929 & 47.7 & 1315$\pm$28 & 71$\pm$33 & 19$\pm$33 & 98.7$\pm$2.1 & 98.4$^{+16.5}_{-15.7}$ & 2.2 & 5.1$^{+2.5}_{-1.3}$ & 5.6$^{+3.5}_{-3.5}$ & .. & .. & .. & $\checkmark$ & .. \\
J005805--053953 & 14.5208 & --5.6650 & 106.8 & 3342$\pm$31 & 134$\pm$32 & 92$\pm$32 & 250.9$\pm$2.3 & .. & 5.0 & 4.8$^{+0.7}_{-1.0}$ & 4.8$^{+1.4}_{-1.4}$ & +17$\pm$6 & .. & $\checkmark$ & $\checkmark$ & .. \\
J005826+032911 & 14.6099 & +3.4865 & 19.0 & 562$\pm$30 & .. & .. & 42.2$\pm$2.2 & 43.3$^{+7.2}_{-9.9}$ & .. & .. & .. & .. & .. & $\checkmark$ & .. & .. \\
J005905+000652 & 14.7729 & +0.1146 & 106.5 & 3355$\pm$32 & --32$\pm$32 & 94$\pm$32 & 251.8$\pm$2.4 & .. & 3.1 & 2.8$^{+0.9}_{-0.9}$ & 2.7$^{+1.4}_{-1.4}$ & +54$\pm$10 & .. & $\checkmark$ & .. & $\checkmark$ \\
J010230--080432 & 15.6271 & --8.0757 & 22.4 & 620$\pm$28 & .. & .. & 46.5$\pm$2.1 & 47.3$^{+7.7}_{-7.7}$ & .. & .. & .. & .. & .. & $\checkmark$ & .. & .. \\
J010237+042107 & 15.6547 & +4.3521 & 16.7 & 488$\pm$29 & .. & .. & 36.6$\pm$2.2 & 35.9$^{+6.8}_{-7.8}$ & .. & .. & .. & .. & .. & $\checkmark$ & .. & .. \\
J010622--015536 & 16.5953 & --1.9269 & 19.5 & 538$\pm$28 & .. & .. & 40.4$\pm$2.1 & 41.1$^{+7.2}_{-9.4}$ & .. & .. & .. & .. & .. & $\checkmark$ & $\checkmark$ & .. \\
J010643--031536 & 16.6797 & --3.2601 & 60.6 & 1583$\pm$26 & --36$\pm$31 & 78$\pm$31 & 118.8$\pm$2.0 & .. & 2.8 & 5.1$^{+2.0}_{-2.0}$ & 5.3$^{+2.8}_{-2.8}$ & .. & .. & .. & $\checkmark$ & .. \\
J010728+033351 & 16.8693 & +3.5643 & 36.9 & 1096$\pm$30 & --8$\pm$35 & --87$\pm$35 & 82.3$\pm$2.2 & 79.3$^{+14.4}_{-13.0}$ & 2.5 & 7.5$^{+3.2}_{-1.7}$ & 8.2$^{+4.6}_{-4.6}$ & .. & .. & $\checkmark$ & .. & .. \\
J010826--003722 & 17.1115 & --0.6229 & 41.2 & 1168$\pm$28 & --125$\pm$34 & 5$\pm$34 & 87.7$\pm$2.1 & 85.5$^{+14.4}_{-13.9}$ & 3.7 & 10.3$^{+2.6}_{-2.8}$ & 10.7$^{+4.1}_{-4.1}$ & +77$\pm$5 & .. & $\checkmark$ & $\checkmark$ & .. \\
J010838+013500 & 17.1615 & +1.5835 & 449.7 & 13901$\pm$31 & --197$\pm$32 & --416$\pm$32 & 1043.4$\pm$2.3 & .. & 14.3 & 3.3$^{+0.3}_{-0.3}$ & 3.2$^{+0.3}_{-0.3}$ & --58$\pm$2 & .. & $\checkmark$ & .. & $\checkmark$ \\
J011030--041532 & 17.6287 & --4.2591 & 30.8 & 832$\pm$27 & .. & .. & 62.4$\pm$2.0 & 62.1$^{+10.1}_{-10.0}$ & .. & .. & .. & .. & .. & $\checkmark$ & $\checkmark$ & .. \\
J011050--074143 & 17.7089 & --7.6955 & 69.9 & 2200$\pm$31 & --58$\pm$33 & 103$\pm$33 & 165.2$\pm$2.4 & .. & 3.6 & 5.2$^{+1.4}_{-1.4}$ & 5.1$^{+2.1}_{-2.1}$ & +60$\pm$8 & .. & $\checkmark$ & $\checkmark$ & .. \\
J011239--032843 & 18.1630 & --3.4787 & 65.2 & 1994$\pm$31 & --42$\pm$32 & --27$\pm$32 & 149.6$\pm$2.3 & .. & 1.6 & 2.1$^{+1.6}_{-0.8}$ & 2.5$^{+2.3}_{-2.3}$ & .. & .. & $\checkmark$ & $\checkmark$ & .. \\
J011316--063052 & 18.3182 & --6.5146 & 27.0 & 753$\pm$28 & .. & .. & 56.5$\pm$2.1 & 58.0$^{+9.5}_{-9.5}$ & .. & .. & .. & .. & .. & $\checkmark$ & $\checkmark$ & .. \\
J011343+022217 & 18.4297 & +2.3715 & 155.4 & 4430$\pm$29 & --5$\pm$29 & --1$\pm$29 & 332.5$\pm$2.1 & .. & .. & 0.0$^{+0.3}_{-0.0}$ & 0.0$^{+1.0}_{-0.0}$ & .. & .. & $\checkmark$ & .. & .. \\
J011407+020809 & 18.5323 & +2.1359 & 26.4 & 654$\pm$25 & .. & .. & 49.1$\pm$1.9 & 48.7$^{+8.4}_{-8.0}$ & .. & .. & .. & .. & .. & $\checkmark$ & .. & .. \\
J011429+000037 & 18.6229 & +0.0104 & 16.4 & 462$\pm$28 & .. & .. & 34.6$\pm$2.1 & 33.8$^{+6.0}_{-7.4}$ & .. & .. & .. & .. & .. & $\checkmark$ & .. & .. \\
J011517--012700 & 18.8208 & --1.4502 & 92.8 & 2979$\pm$32 & --5$\pm$33 & 76$\pm$33 & 223.6$\pm$2.4 & .. & 2.3 & 2.3$^{+1.1}_{-0.6}$ & 2.6$^{+1.6}_{-1.6}$ & .. & .. & $\checkmark$ & $\checkmark$ & .. \\
J011540+035645 & 18.9188 & +3.9458 & 46.8 & 1392$\pm$30 & 25$\pm$35 & --81$\pm$35 & 104.4$\pm$2.2 & .. & 2.4 & 5.7$^{+2.5}_{-1.3}$ & 5.7$^{+3.6}_{-3.6}$ & .. & .. & $\checkmark$ & .. & .. \\
J011727--042513 & 19.3662 & --4.4205 & 30.5 & 803$\pm$26 & .. & .. & 60.3$\pm$2.0 & 60.5$^{+9.8}_{-9.8}$ & .. & .. & .. & .. & .. & $\checkmark$ & $\checkmark$ & .. \\
J011818+025806 & 19.5771 & +2.9685 & 112.0 & 3173$\pm$28 & 27$\pm$29 & --20$\pm$29 & 238.1$\pm$2.1 & .. & 1.2 & 0.7$^{+1.0}_{-0.5}$ & 0.7$^{+1.3}_{-0.7}$ & .. & .. & $\checkmark$ & .. & .. \\
J011844--085059 & 19.6844 & --8.8492 & 36.2 & 1615$\pm$45 & 9$\pm$53 & --43$\pm$53 & 121.2$\pm$3.3 & .. & .. & 0.0$^{+3.8}_{-0.0}$ & 0.0$^{+4.7}_{-0.0}$ & .. & .. & .. & .. & .. \\
J012014--062652 & 20.0599 & --6.4480 & 15.3 & 431$\pm$28 & .. & .. & 32.3$\pm$2.1 & 31.7$^{+5.9}_{-6.2}$ & .. & .. & .. & .. & .. & $\checkmark$ & $\checkmark$ & .. \\
J012156+042226 & 20.4865 & +4.3740 & 137.3 & 4270$\pm$31 & 66$\pm$32 & --94$\pm$32 & 320.5$\pm$2.3 & .. & 3.5 & 2.6$^{+0.7}_{-0.7}$ & 1.9$^{+1.1}_{-1.1}$ & --27$\pm$9 & .. & $\checkmark$ & .. & .. \\
J012213--001758 & 20.5578 & --0.2995 & 34.9 & 948$\pm$27 & .. & .. & 71.1$\pm$2.0 & 70.2$^{+12.1}_{-11.8}$ & .. & .. & .. & .. & .. & $\checkmark$ & $\checkmark$ & $\checkmark$ \\
J012217--005615 & 20.5729 & --0.9375 & 23.6 & 633$\pm$27 & .. & .. & 47.5$\pm$2.0 & 48.1$^{+8.2}_{-8.0}$ & .. & .. & .. & .. & .. & $\checkmark$ & $\checkmark$ & .. \\
J012335--034838 & 20.8990 & --3.8106 & 44.6 & 1148$\pm$26 & --58$\pm$31 & --20$\pm$31 & 86.2$\pm$1.9 & 84.8$^{+14.4}_{-13.8}$ & 2.0 & 4.8$^{+2.7}_{-1.4}$ & 5.0$^{+3.8}_{-3.8}$ & .. & .. & $\checkmark$ & $\checkmark$ & .. \\
J012450--062501 & 21.2104 & --6.4171 & 19.3 & 546$\pm$28 & .. & .. & 41.0$\pm$2.1 & 41.6$^{+7.3}_{-9.4}$ & .. & .. & .. & .. & .. & $\checkmark$ & $\checkmark$ & .. \\
J012505+014632 & 21.2729 & +1.7758 & 16.7 & 433$\pm$26 & .. & .. & 32.5$\pm$1.9 & 31.9$^{+5.9}_{-6.1}$ & .. & .. & .. & .. & .. & $\checkmark$ & .. & .. \\
J012528--000554 & 21.3698 & --0.0984 & 194.1 & 5934$\pm$31 & --77$\pm$31 & --119$\pm$31 & 445.4$\pm$2.3 & .. & 4.5 & 2.3$^{+0.5}_{-0.5}$ & 2.2$^{+0.8}_{-0.8}$ & --61$\pm$7 & .. & $\checkmark$ & $\checkmark$ & $\checkmark$ \\
J012600--012041* & 21.5026 & --1.3449 & 52.3 & 1400$\pm$27 & --4$\pm$32 & 32$\pm$32 & 105.1$\pm$2.0 & .. & 1.1 & 0.7$^{+2.6}_{-0.7}$ & 2.3$^{+0.0}_{-2.3}$ & .. & .. & $\checkmark$ & $\checkmark$ & $\checkmark$ \\
J012716--082129 & 21.8188 & --8.3583 & 42.2 & 1180$\pm$28 & 36$\pm$33 & --118$\pm$33 & 88.6$\pm$2.1 & 87.6$^{+15.0}_{-14.6}$ & 3.7 & 10.1$^{+2.5}_{-2.7}$ & 10.5$^{+4.0}_{-4.0}$ & --36$\pm$8 & .. & $\checkmark$ & $\checkmark$ & .. \\
J013241--080404 & 23.1719 & --8.0678 & 25.0 & 664$\pm$27 & .. & .. & 49.8$\pm$2.0 & 49.6$^{+8.5}_{-8.2}$ & .. & .. & .. & .. & .. & $\checkmark$ & $\checkmark$ & .. \\
J013412+000348 & 23.5526 & +0.0635 & 29.1 & 779$\pm$27 & .. & .. & 58.5$\pm$2.0 & 59.5$^{+9.7}_{-9.7}$ & .. & .. & .. & .. & .. & $\checkmark$ & .. & .. \\
J013514--000701 & 23.8094 & --0.1172 & 36.0 & 951$\pm$26 & .. & .. & 71.4$\pm$2.0 & 70.2$^{+12.2}_{-11.9}$ & .. & .. & .. & .. & .. & $\checkmark$ & .. & .. \\
J013852--054008 & 24.7167 & --5.6691 & 37.9 & 979$\pm$26 & .. & .. & 73.5$\pm$1.9 & 72.1$^{+12.5}_{-11.9}$ & .. & .. & .. & .. & .. & $\checkmark$ & $\checkmark$ & .. \\
J013957+013149 & 24.9885 & +1.5304 & 104.9 & 2952$\pm$28 & 33$\pm$29 & 2$\pm$29 & 221.6$\pm$2.1 & .. & 1.1 & 0.7$^{+1.0}_{-0.5}$ & 0.7$^{+1.4}_{-0.7}$ & .. & .. & $\checkmark$ & .. & $\checkmark$ \\
J014043+025445 & 25.1813 & +2.9127 & 40.9 & 978$\pm$24 & .. & .. & 73.4$\pm$1.8 & 72.1$^{+12.5}_{-11.9}$ & .. & .. & .. & .. & .. & $\checkmark$ & .. & .. \\
J014133--020220 & 25.3906 & --2.0390 & 49.4 & 1263$\pm$26 & --22$\pm$30 & 36$\pm$30 & 94.8$\pm$1.9 & 95.7$^{+16.1}_{-15.4}$ & 1.4 & 2.7$^{+2.4}_{-1.2}$ & 1.3$^{+3.4}_{-1.3}$ & .. & .. & $\checkmark$ & $\checkmark$ & .. \\
J014239--054402 & 25.6625 & --5.7341 & 37.8 & 955$\pm$25 & .. & .. & 71.7$\pm$1.9 & 70.6$^{+12.1}_{-11.9}$ & .. & .. & .. & .. & .. & $\checkmark$ & $\checkmark$ & .. \\
J014316--011858 & 25.8193 & --1.3163 & 23.2 & 605$\pm$26 & .. & .. & 45.4$\pm$2.0 & 47.1$^{+7.6}_{-7.7}$ & .. & .. & .. & .. & .. & $\checkmark$ & $\checkmark$ & .. \\
J014609+004324 & 26.5406 & +0.7235 & 34.8 & 849$\pm$24 & .. & .. & 63.7$\pm$1.8 & 63.2$^{+11.0}_{-10.5}$ & .. & .. & .. & .. & .. & $\checkmark$ & .. & .. \\
J014632+025627 & 26.6354 & +2.9409 & 23.9 & 572$\pm$24 & .. & .. & 42.9$\pm$1.8 & 44.9$^{+7.1}_{-10.4}$ & .. & .. & .. & .. & .. & $\checkmark$ & .. & .. \\
J014721--044542 & 26.8391 & --4.7617 & 33.6 & 841$\pm$25 & .. & .. & 63.1$\pm$1.9 & 63.0$^{+10.8}_{-10.4}$ & .. & .. & .. & .. & .. & $\checkmark$ & $\checkmark$ & .. \\
J014738--082718 & 26.9099 & --8.4551 & 35.9 & 1038$\pm$29 & --41$\pm$34 & --103$\pm$34 & 77.9$\pm$2.2 & 77.6$^{+13.6}_{-12.7}$ & 3.2 & 10.3$^{+3.2}_{-3.1}$ & 10.6$^{+4.8}_{-4.8}$ & --56$\pm$10 & .. & $\checkmark$ & $\checkmark$ & .. \\
J014816+001946 & 27.0677 & +0.3297 & 33.2 & 832$\pm$25 & .. & .. & 62.4$\pm$1.9 & 62.1$^{+10.1}_{-10.0}$ & .. & .. & .. & .. & .. & $\checkmark$ & .. & $\checkmark$ \\
J014934--073318 & 27.3922 & --7.5552 & 36.7 & 946$\pm$26 & .. & .. & 71.0$\pm$1.9 & 70.2$^{+12.1}_{-11.8}$ & .. & .. & .. & .. & .. & $\checkmark$ & .. & .. \\
J015239+014719 & 28.1651 & +1.7888 & 26.3 & 628$\pm$24 & .. & .. & 47.1$\pm$1.8 & 47.3$^{+7.8}_{-7.8}$ & .. & .. & .. & .. & .. & $\checkmark$ & .. & .. \\
J015243+002041 & 28.1797 & +0.3448 & 53.3 & 1341$\pm$25 & --16$\pm$30 & --48$\pm$30 & 100.7$\pm$1.9 & .. & 1.7 & 3.3$^{+2.2}_{-1.2}$ & 3.2$^{+3.2}_{-3.2}$ & .. & .. & $\checkmark$ & .. & $\checkmark$ \\
J015422--023453 & 28.5948 & --2.5816 & 29.0 & 706$\pm$24 & .. & .. & 53.0$\pm$1.8 & 53.3$^{+9.5}_{-8.9}$ & .. & .. & .. & .. & .. & $\checkmark$ & $\checkmark$ & .. \\
J015500+004952 & 28.7510 & +0.8313 & 18.1 & 452$\pm$25 & .. & .. & 33.9$\pm$1.9 & 33.6$^{+5.9}_{-7.2}$ & .. & .. & .. & .. & .. & $\checkmark$ & .. & .. \\
J015553--062135 & 28.9724 & --6.3599 & 17.2 & 471$\pm$27 & .. & .. & 35.4$\pm$2.1 & 34.3$^{+6.3}_{-7.6}$ & .. & .. & .. & .. & .. & $\checkmark$ & .. & .. \\
J015710+001126 & 29.2938 & +0.1906 & 24.8 & 627$\pm$25 & .. & .. & 47.1$\pm$1.9 & 47.3$^{+7.8}_{-7.8}$ & .. & .. & .. & .. & .. & $\checkmark$ & .. & .. \\
J020004+012513 & 30.0167 & +1.4205 & 20.3 & 486$\pm$24 & .. & .. & 36.5$\pm$1.8 & 35.6$^{+6.7}_{-7.8}$ & .. & .. & .. & .. & .. & $\checkmark$ & .. & .. \\
J020040+032248 & 30.1703 & +3.3801 & 78.0 & 2304$\pm$30 & 28$\pm$30 & --79$\pm$30 & 173.0$\pm$2.2 & .. & 2.7 & 3.4$^{+1.3}_{-1.3}$ & 2.2$^{+1.9}_{-1.9}$ & .. & .. & $\checkmark$ & .. & .. \\
J020137--002346 & 30.4073 & --0.3964 & 20.7 & 530$\pm$26 & .. & .. & 39.8$\pm$1.9 & 40.1$^{+7.2}_{-9.3}$ & .. & .. & .. & .. & .. & $\checkmark$ & .. & .. \\
J020151+034307 & 30.4651 & +3.7188 & 21.8 & 589$\pm$27 & .. & .. & 44.2$\pm$2.0 & 45.8$^{+7.3}_{-10.4}$ & .. & .. & .. & .. & .. & $\checkmark$ & .. & .. \\
J020207--055901 & 30.5292 & --5.9838 & 67.5 & 1984$\pm$29 & 18$\pm$30 & --38$\pm$30 & 148.9$\pm$2.2 & .. & 1.4 & 1.7$^{+1.5}_{-0.8}$ & 2.0$^{+2.2}_{-2.0}$ & .. & .. & $\checkmark$ & $\checkmark$ & .. \\
J020214--001748 & 30.5599 & --0.2969 & 29.8 & 763$\pm$26 & .. & .. & 57.3$\pm$1.9 & 58.6$^{+9.6}_{-9.6}$ & .. & .. & .. & .. & $\checkmark$ & $\checkmark$ & $\checkmark$ & .. \\
J020224--014506 & 30.6037 & --1.7518 & 24.6 & 622$\pm$25 & .. & .. & 46.7$\pm$1.9 & 47.3$^{+7.8}_{-7.8}$ & .. & .. & .. & .. & .. & $\checkmark$ & .. & .. \\
J020234+000301 & 30.6427 & +0.0505 & 33.0 & 828$\pm$25 & .. & .. & 62.2$\pm$1.9 & 62.1$^{+10.1}_{-10.0}$ & .. & .. & .. & .. & .. & $\checkmark$ & .. & .. \\
J020354+023555 & 30.9781 & +2.5988 & 40.4 & 935$\pm$23 & .. & .. & 70.2$\pm$1.7 & 69.3$^{+12.0}_{-11.5}$ & .. & .. & .. & .. & .. & $\checkmark$ & .. & .. \\
J020529+040321 & 31.3729 & +4.0560 & 18.4 & 514$\pm$28 & .. & .. & 38.6$\pm$2.1 & 38.1$^{+7.2}_{-8.4}$ & .. & .. & .. & .. & .. & $\checkmark$ & .. & .. \\
J020826--003228 & 32.1089 & --0.5412 & 17.1 & 439$\pm$26 & .. & .. & 32.9$\pm$1.9 & 32.4$^{+5.9}_{-6.2}$ & .. & .. & .. & .. & .. & $\checkmark$ & .. & .. \\
J020826--004741 & 32.1099 & --0.7948 & 60.8 & 1820$\pm$30 & --75$\pm$31 & 55$\pm$31 & 136.6$\pm$2.2 & .. & 3.0 & 4.8$^{+1.6}_{-1.6}$ & 5.2$^{+2.4}_{-2.4}$ & +71$\pm$10 & .. & $\checkmark$ & $\checkmark$ & $\checkmark$ \\
J020931--043827 & 32.3792 & --4.6410 & 26.4 & 649$\pm$25 & .. & .. & 48.7$\pm$1.8 & 48.4$^{+8.3}_{-8.0}$ & .. & .. & .. & .. & .. & $\checkmark$ & $\checkmark$ & .. \\
J020931--073648 & 32.3807 & --7.6135 & 24.0 & 618$\pm$26 & .. & .. & 46.4$\pm$1.9 & 47.3$^{+7.7}_{-7.7}$ & .. & .. & .. & .. & .. & $\checkmark$ & $\checkmark$ & .. \\
J021053--063336 & 32.7229 & --6.5602 & 15.4 & 411$\pm$27 & .. & .. & 30.8$\pm$2.0 & 30.1$^{+6.0}_{-5.9}$ & .. & .. & .. & .. & .. & $\checkmark$ & .. & .. \\
J021316--071932 & 33.3203 & --7.3257 & 32.5 & 824$\pm$25 & .. & .. & 61.8$\pm$1.9 & 61.9$^{+9.9}_{-9.8}$ & .. & .. & .. & .. & .. & $\checkmark$ & $\checkmark$ & .. \\
J021541--022256 & 33.9245 & --2.3825 & 47.5 & 1148$\pm$24 & 69$\pm$29 & --57$\pm$29 & 86.1$\pm$1.8 & 84.9$^{+14.3}_{-13.9}$ & 3.1 & 7.4$^{+2.4}_{-2.4}$ & 6.2$^{+3.6}_{-3.6}$ & --20$\pm$10 & .. & $\checkmark$ & $\checkmark$ & .. \\
J021553+001826 & 33.9740 & +0.3073 & 17.8 & 437$\pm$25 & .. & .. & 32.8$\pm$1.8 & 32.4$^{+5.9}_{-6.2}$ & .. & .. & .. & .. & .. & $\checkmark$ & .. & .. \\
J021555+052427 & 33.9800 & +5.4075 & 27.0 & 933$\pm$35 & .. & .. & 70.0$\pm$2.6 & 68.9$^{+12.1}_{-11.4}$ & .. & .. & .. & .. & .. & .. & .. & .. \\
J021702--082051 & 34.2609 & --8.3477 & 118.7 & 3746$\pm$32 & 297$\pm$32 & --66$\pm$32 & 281.2$\pm$2.4 & .. & 9.3 & 8.1$^{+0.9}_{-1.0}$ & 8.2$^{+1.2}_{-1.2}$ & --6$\pm$3 & .. & $\checkmark$ & $\checkmark$ & .. \\
J021749+014449 & 34.4542 & +1.7472 & 338.7 & 8943$\pm$26 & --148$\pm$28 & --101$\pm$28 & 671.2$\pm$2.0 & .. & 6.4 & 2.0$^{+0.3}_{-0.3}$ & 2.0$^{+0.4}_{-0.4}$ & --73$\pm$5 & .. & $\checkmark$ & .. & $\checkmark$ \\
J021835+004039 & 34.6484 & +0.6776 & 34.2 & 817$\pm$24 & .. & .. & 61.3$\pm$1.8 & 61.4$^{+9.8}_{-9.8}$ & .. & .. & .. & .. & .. & $\checkmark$ & .. & .. \\
J021907+012100 & 34.7792 & +1.3501 & 123.8 & 3287$\pm$27 & .. & --167$\pm$27 & 246.7$\pm$2.0 & .. & 6.1 & 5.0$^{+0.8}_{-0.9}$ & 4.7$^{+1.2}_{-1.2}$ & --45$\pm$5 & .. & $\checkmark$ & .. & .. \\
J021920--025842 & 34.8370 & --2.9784 & 39.9 & 957$\pm$24 & .. & .. & 71.9$\pm$1.8 & 70.9$^{+12.1}_{-12.0}$ & .. & .. & .. & .. & .. & .. & .. & .. \\
J022146--013306 & 35.4443 & --1.5518 & 28.7 & 708$\pm$25 & .. & .. & 53.1$\pm$1.9 & 53.9$^{+9.5}_{-9.1}$ & .. & .. & .. & .. & $\checkmark$ & $\checkmark$ & .. & .. \\
J022152+023614 & 35.4698 & +2.6040 & 23.0 & 526$\pm$23 & .. & .. & 39.5$\pm$1.7 & 39.3$^{+7.4}_{-8.9}$ & .. & .. & .. & .. & .. & $\checkmark$ & .. & .. \\
J022313--020505 & 35.8042 & --2.0848 & 29.7 & 731$\pm$25 & .. & .. & 54.9$\pm$1.8 & 56.2$^{+9.6}_{-9.5}$ & .. & .. & .. & .. & .. & $\checkmark$ & $\checkmark$ & .. \\
J022430--021932 & 36.1260 & --2.3256 & 17.7 & 433$\pm$24 & .. & .. & 32.5$\pm$1.8 & 31.9$^{+5.9}_{-6.1}$ & .. & .. & .. & .. & $\checkmark$ & $\checkmark$ & .. & .. \\
J022640--055239 & 36.6667 & --5.8775 & 34.7 & 881$\pm$25 & .. & .. & 66.1$\pm$1.9 & 64.6$^{+11.3}_{-10.7}$ & .. & .. & .. & .. & .. & $\checkmark$ & .. & $\checkmark$ \\
J022853--033736 & 37.2219 & --3.6269 & 66.5 & 1872$\pm$28 & 45$\pm$29 & --63$\pm$29 & 140.5$\pm$2.1 & .. & 2.7 & 3.9$^{+1.6}_{-1.6}$ & 2.7$^{+2.2}_{-2.2}$ & .. & .. & $\checkmark$ & $\checkmark$ & $\checkmark$ \\
J023322--045504 & 38.3422 & --4.9180 & 29.3 & 742$\pm$25 & .. & .. & 55.7$\pm$1.9 & 56.9$^{+9.6}_{-9.5}$ & .. & .. & .. & .. & $\checkmark$ & $\checkmark$ & $\checkmark$ & .. \\
J023330--020320 & 38.3766 & --2.0557 & 21.4 & 517$\pm$24 & .. & .. & 38.8$\pm$1.8 & 38.1$^{+7.2}_{-8.4}$ & .. & .. & .. & .. & $\checkmark$ & $\checkmark$ & .. & .. \\
J023507--040206 & 38.7807 & --4.0351 & 39.3 & 961$\pm$24 & .. & .. & 72.1$\pm$1.8 & 71.0$^{+12.2}_{-12.0}$ & .. & .. & .. & .. & .. & $\checkmark$ & $\checkmark$ & .. \\
J023919+025109 & 39.8297 & +2.8527 & 19.6 & 496$\pm$25 & .. & .. & 37.2$\pm$1.9 & 36.3$^{+6.9}_{-7.9}$ & .. & .. & .. & .. & .. & $\checkmark$ & .. & .. \\
J023945--023440 & 39.9391 & --2.5780 & 116.6 & 3270$\pm$28 & --24$\pm$29 & --32$\pm$29 & 245.4$\pm$2.1 & .. & 1.4 & 1.0$^{+0.9}_{-0.5}$ & 0.7$^{+1.3}_{-0.7}$ & .. & .. & $\checkmark$ & $\checkmark$ & .. \\
J023951+041623 & 39.9630 & +4.2732 & 156.5 & 7035$\pm$45 & 96$\pm$47 & 252$\pm$47 & 528.0$\pm$3.4 & .. & 5.8 & 3.8$^{+0.7}_{-0.7}$ & 3.7$^{+0.9}_{-0.9}$ & +34$\pm$5 & .. & $\checkmark$ & .. & .. \\
J024056--050440 & 40.2333 & --5.0780 & 53.6 & 1847$\pm$34 & --7$\pm$41 & --28$\pm$41 & 138.6$\pm$2.6 & .. & .. & 0.0$^{+2.3}_{-0.0}$ & 1.5$^{+0.0}_{-1.5}$ & .. & .. & $\checkmark$ & $\checkmark$ & .. \\
J024104--081514 & 40.2698 & --8.2540 & 274.7 & 11652$\pm$42 & 69$\pm$44 & --39$\pm$44 & 874.6$\pm$3.2 & .. & 1.8 & 0.6$^{+0.4}_{-0.2}$ & 0.5$^{+0.5}_{-0.5}$ & .. & .. & $\checkmark$ & $\checkmark$ & .. \\
J024145+002645* & 40.4385 & +0.4458 & 24.6 & 619$\pm$25 & .. & .. & 46.5$\pm$1.9 & 47.3$^{+7.7}_{-7.7}$ & .. & .. & .. & .. & .. & $\checkmark$ & .. & .. \\
J232839--063433 & 352.1635 & --6.5759 & 29.0 & 991$\pm$34 & .. & .. & 74.4$\pm$2.6 & 73.4$^{+12.3}_{-12.0}$ & .. & .. & .. & .. & .. & $\checkmark$ & $\checkmark$ & .. \\
J232925+050310 & 352.3557 & +5.0529 & 16.1 & 906$\pm$56 & .. & .. & 68.0$\pm$4.2 & 67.1$^{+11.9}_{-10.9}$ & .. & .. & .. & .. & .. & $\checkmark$ & .. & .. \\
J233316--013108 & 353.3193 & --1.5189 & 39.3 & 1128$\pm$29 & --92$\pm$34 & 16$\pm$34 & 84.7$\pm$2.2 & 82.4$^{+14.8}_{-13.8}$ & 2.7 & 7.8$^{+3.0}_{-3.1}$ & 7.9$^{+4.4}_{-4.4}$ & .. & $\checkmark$ & $\checkmark$ & $\checkmark$ & .. \\
J233520--013110 & 353.8349 & --1.5195 & 186.9 & 6206$\pm$33 & --35$\pm$34 & 92$\pm$34 & 465.8$\pm$2.5 & .. & 2.9 & 1.5$^{+0.6}_{-0.6}$ & 1.4$^{+0.8}_{-0.8}$ & .. & .. & $\checkmark$ & $\checkmark$ & $\checkmark$ \\
J233552--003458 & 353.9672 & --0.5828 & 16.8 & 486$\pm$29 & .. & .. & 36.5$\pm$2.2 & 35.6$^{+6.7}_{-7.8}$ & .. & .. & .. & .. & .. & $\checkmark$ & $\checkmark$ & .. \\
J233757--023057 & 354.4891 & --2.5159 & 122.4 & 3829$\pm$31 & --132$\pm$32 & 12$\pm$32 & 287.4$\pm$2.3 & .. & 4.1 & 3.4$^{+0.8}_{-0.8}$ & 3.5$^{+1.2}_{-1.2}$ & +80$\pm$5 & $\checkmark$ & .. & $\checkmark$ & $\checkmark$ \\
J233807+032652 & 354.5323 & +3.4479 & 34.4 & 1091$\pm$32 & 32$\pm$38 & .. & 81.9$\pm$2.4 & 78.8$^{+14.0}_{-12.9}$ & .. & 0.0$^{+4.0}_{-0.0}$ & 2.6$^{+0.0}_{-2.6}$ & .. & .. & $\checkmark$ & .. & .. \\
J233853--072103 & 354.7245 & --7.3509 & 17.8 & 538$\pm$30 & .. & .. & 40.4$\pm$2.3 & 41.1$^{+7.2}_{-9.4}$ & .. & .. & .. & .. & .. & $\checkmark$ & .. & .. \\
J233929+024407 & 354.8740 & +2.7354 & 55.4 & 1672$\pm$30 & --8$\pm$31 & --36$\pm$31 & 125.5$\pm$2.3 & .. & 1.2 & 1.5$^{+1.9}_{-0.8}$ & 2.0$^{+2.7}_{-1.0}$ & .. & .. & $\checkmark$ & .. & .. \\
J234106+001830 & 355.2787 & +0.3083 & 21.0 & 588$\pm$28 & .. & .. & 44.2$\pm$2.1 & 45.8$^{+7.3}_{-10.4}$ & .. & .. & .. & .. & .. & $\checkmark$ & .. & $\checkmark$ \\
J234110--034156 & 355.2927 & --3.6989 & 16.7 & 459$\pm$27 & .. & .. & 34.4$\pm$2.1 & 33.8$^{+5.9}_{-7.4}$ & .. & .. & .. & .. & $\checkmark$ & $\checkmark$ & $\checkmark$ & .. \\
J234357--052150 & 355.9875 & --5.3641 & 31.5 & 898$\pm$29 & .. & .. & 67.4$\pm$2.1 & 66.2$^{+11.8}_{-11.0}$ & .. & .. & .. & .. & .. & $\checkmark$ & $\checkmark$ & .. \\
J234656--020344 & 356.7333 & --2.0624 & 28.0 & 770$\pm$28 & .. & .. & 57.8$\pm$2.1 & 59.2$^{+9.7}_{-9.7}$ & .. & .. & .. & .. & $\checkmark$ & $\checkmark$ & .. & .. \\
J234811--042557 & 357.0490 & --4.4326 & 52.5 & 1403$\pm$27 & --31$\pm$32 & 59$\pm$32 & 105.3$\pm$2.0 & .. & 2.1 & 4.3$^{+2.3}_{-1.2}$ & 4.3$^{+3.2}_{-3.2}$ & .. & .. & $\checkmark$ & $\checkmark$ & .. \\
J234910--043805 & 357.2917 & --4.6347 & 27.2 & 733$\pm$27 & .. & .. & 55.0$\pm$2.0 & 56.2$^{+9.6}_{-9.5}$ & .. & .. & .. & .. & .. & $\checkmark$ & $\checkmark$ & .. \\
J235050--002846 & 357.7120 & --0.4797 & 14.9 & 420$\pm$28 & .. & .. & 31.5$\pm$2.1 & 30.5$^{+6.1}_{-6.1}$ & .. & .. & .. & .. & .. & $\checkmark$ & .. & .. \\
J235054--042701 & 357.7260 & --4.4503 & 32.2 & 870$\pm$27 & .. & .. & 65.3$\pm$2.0 & 64.1$^{+11.3}_{-10.7}$ & .. & .. & .. & .. & .. & $\checkmark$ & .. & .. \\
J235156--010917 & 357.9844 & --1.1548 & 23.4 & 671$\pm$29 & .. & .. & 50.4$\pm$2.2 & 50.2$^{+8.5}_{-8.3}$ & .. & .. & .. & .. & .. & $\checkmark$ & $\checkmark$ & .. \\
J235409--001946 & 358.5375 & --0.3297 & 29.0 & 834$\pm$29 & .. & .. & 62.6$\pm$2.2 & 62.0$^{+10.2}_{-10.0}$ & .. & .. & .. & .. & .. & $\checkmark$ & $\checkmark$ & .. \\
J235451--040503 & 358.7162 & --4.0842 & 24.0 & 647$\pm$27 & .. & .. & 48.5$\pm$2.0 & 48.2$^{+8.2}_{-7.9}$ & .. & .. & .. & .. & .. & $\checkmark$ & $\checkmark$ & .. \\
J235524--072946 & 358.8526 & --7.4964 & 22.6 & 616$\pm$27 & .. & .. & 46.2$\pm$2.0 & 47.3$^{+7.7}_{-7.7}$ & .. & .. & .. & .. & .. & $\checkmark$ & $\checkmark$ & .. \\
J235706+044859 & 359.2787 & +4.8166 & 22.5 & 621$\pm$28 & .. & .. & 46.6$\pm$2.1 & 47.3$^{+7.8}_{-7.8}$ & .. & .. & .. & .. & .. & $\checkmark$ & .. & .. \\
J235725--015214 & 359.3542 & --1.8706 & 32.1 & 892$\pm$28 & .. & .. & 67.0$\pm$2.1 & 65.9$^{+11.7}_{-10.9}$ & .. & .. & .. & .. & $\checkmark$ & $\checkmark$ & $\checkmark$ & .. \\
J235809--080005 & 359.5406 & --8.0015 & 16.0 & 436$\pm$27 & .. & .. & 32.7$\pm$2.0 & 32.3$^{+5.8}_{-6.2}$ & .. & .. & .. & .. & .. & $\checkmark$ & .. & .. \\
J235828+043024 & 359.6193 & +4.5067 & 30.8 & 939$\pm$30 & .. & .. & 70.5$\pm$2.3 & 69.5$^{+12.0}_{-11.6}$ & .. & .. & .. & .. & $\checkmark$ & $\checkmark$ & .. & .. \\
J235931--063940 & 359.8828 & --6.6614 & 38.8 & 1108$\pm$29 & --26$\pm$34 & --26$\pm$34 & 83.2$\pm$2.1 & 81.7$^{+14.7}_{-13.6}$ & 1.1 & 1.7$^{+3.2}_{-1.6}$ & 2.1$^{+4.4}_{-2.1}$ & .. & .. & $\checkmark$ & $\checkmark$ & .. \\
J235936--003115 & 359.9037 & --0.5208 & 25.2 & 736$\pm$29 & .. & .. & 55.3$\pm$2.2 & 56.4$^{+9.6}_{-9.6}$ & .. & .. & .. & .. & $\checkmark$ & $\checkmark$ & $\checkmark$ & .. \\

\hline \hline
\end{longtable}
}
\end{center}
\label{tab:cat}

\clearpage
\twocolumn

\end{appendix}


\end{document}